\documentclass[journal]{IEEEtran}
\IEEEsettopmargin{t}{1.92cm}

\IEEEoverridecommandlockouts
\usepackage{cite}
\usepackage{amsmath,amssymb,amsfonts}
\usepackage{algorithmic}
\usepackage{graphicx}
\usepackage{textcomp}
\usepackage{multicol}
\usepackage{import}
\usepackage[table, svgnames, dvipsnames]{xcolor}
\usepackage{multirow}
\usepackage{makecell}
\usepackage[letterspace=-13]{microtype}
\usepackage{breakcites}
\usepackage{dsfont} 
\usepackage{pdfrender}
\usepackage{bm}
\makeatletter
\let\MYcaption\@makecaption
\makeatother
\usepackage[font=footnotesize,labelformat=simple]{subcaption}
\renewcommand\thesubfigure{(\alph{subfigure})}
\makeatletter
\let\@makecaption\MYcaption
\makeatother
\usepackage{tikz}
\DeclareMathOperator*{\argmax}{argmax}
\usepackage{comment}
\usepackage{pgfplots}
\usetikzlibrary{shapes,arrows,positioning,calc, matrix,spy,patterns,backgrounds}
\usepackage[nolist]{acronym}
\usepackage{fontawesome5}
\pgfplotsset{compat=1.18}

\usepackage{booktabs}
\usepackage{xfrac}
\usepackage{comment}
\usepackage{url} 
\usepackage{layouts} 
\usepackage{standalone}

\usepackage{lipsum,adjustbox}
\usetikzlibrary{mindmap,shadows}
\newcommand*{\info}[4][16.3]{%
  \node [ annotation, #3, scale=0.65, text width = #1em,
          inner sep = 2mm ] at (#2) {%
  \list{$\bullet$}{\topsep=0pt\itemsep=0pt\parsep=0pt
    \parskip=0pt\labelwidth=8pt\leftmargin=8pt
    \itemindent=0pt\labelsep=2pt}%
    #4
  \endlist
  };
}

\newcommand*{\figures}{figures}

\definecolor{cb-1}{HTML}{4477AA}
\definecolor{cb-2}{HTML}{EE6677}
\definecolor{cb-3}{HTML}{228833}
\definecolor{cb-4}{HTML}{CCBB44}
\definecolor{cb-5}{HTML}{66CCEE}
\definecolor{cb-6}{HTML}{AA3377}
\definecolor{cb-7}{HTML}{BBBBBB}

\pgfplotscreateplotcyclelist{cb list}{
	cb-1,cb-2,cb-3,cb-4,cb-5,cb-6,cb-7
}

\definecolor{cb-1}{HTML}{4477AA}
\definecolor{cb-2}{HTML}{EE6677}
\definecolor{cb-3}{HTML}{228833}
\definecolor{cb-4}{HTML}{CCBB44}
\definecolor{cb-5}{HTML}{66CCEE}
\definecolor{cb-6}{HTML}{AA3377}
\definecolor{cb-7}{HTML}{BBBBBB}

\pgfplotscreateplotcyclelist{cb list}{
	cb-1,cb-2,cb-3,cb-4,cb-5,cb-6,cb-7
}

\definecolor{KITgreen}{rgb}{0,.59,.51}

\definecolor{KITpalegreen}{RGB}{130,190,60}

\definecolor{KITblue}{rgb}{.27,.39,.66}

\definecolor{KITyellow}{rgb}{.98,.89,0}

\definecolor{KITorange}{rgb}{.87,.60,.10}

\definecolor{KITred}{rgb}{.63,.13,.13}

\definecolor{KITpurple}{RGB}{160,0,120}

\definecolor{KITcyanblue}{RGB}{80,170,230}

\let\j\relax
\newcommand{\j}{\mathrm{j}}
\let\Re\relax
\let\Im\relax
\DeclareMathOperator{\Re}{Re}
\DeclareMathOperator{\Im}{Im}

\newcommand*{\vect}[1]{\boldsymbol{#1}}
\newcommand*{\mat}[1]{\MakeUppercase{\boldsymbol{#1}}}

\DeclareDocumentCommand{\complexset}{o o}{%
    \mathbb{C}\IfValueT{#1}{\IfValueTF{#2}{^{#1\times#2}}{^{#1}}}
}

\newcommand{\hermit}{\mathsf{H}}

\newcommand{\bmRY}{\bm{R}_{\text{Y}}}

\usepackage{ellipsis}
\usetikzlibrary{calc, patterns,fit,shapes,positioning}
\usetikzlibrary{decorations.pathreplacing,decorations.markings,shapes.geometric}
\tikzset{naming/.style={align=center,font=\small}}
\tikzset{antenna/.style={insert path={-- coordinate (ant#1) ++(0,0.25) -- +(135:0.25) + (0,0) -- +(45:0.25)}}}
\tikzset{station/.style={naming,draw,shape=dart,shape border rotate=90, minimum width=10mm, minimum height=10mm,outer sep=0pt,inner sep=3pt}}
\tikzset{mobile/.style={naming,draw,shape=rectangle,minimum width=12mm,minimum height=6mm, outer sep=0pt,inner sep=3pt}}
\tikzset{radiation/.style={{decorate,decoration={expanding waves,angle=90,segment length=4pt}}}}
\usetikzlibrary{decorations.pathreplacing}

\tikzset{pics/MUE/.style args={#1}
{code={
\node[mobile,label={[inner ysep=+.3333em]\dots}] (box){};
\draw ([xshift=.25cm] box.south west) circle (4pt)
      ([xshift=-.25cm]box.south east) circle (4pt);

\fill ([xshift=.25cm] box.south west) circle (1pt)
      ([xshift=-.25cm]box.south east) circle (1pt);

\draw ([xshift=.25cm] box.north west) [antenna=1];
\draw ([xshift=-.25cm]box.north east) [antenna=2];
}}}

\tikzset{pics/UE/.style args={#1}
{code={
\node[mobile] (box) {#1};

\draw ([xshift=.25cm] box.south west) circle (4pt)
      ([xshift=-.25cm]box.south east) circle (4pt);

\fill ([xshift=.25cm] box.south west) circle (1pt)
      ([xshift=-.25cm]box.south east) circle (1pt);

\draw (box.north) [antenna=1];
}}}

\tikzset{pics/MBS/.style args={#1}{code={
\node[coordinate] (-base) at (0,0)  {};
\node[coordinate] (-l) at (-2,-1) {};
\node[coordinate] (-r) at (2,-1) {};
\node[coordinate] (-t) at (0,4) {};
\draw[line join=bevel] (-base) -- (-l) -- (-t) -- (-r) -- cycle;
\draw[line join=bevel] ($(-l)!0.5!(-t)$) -- ($(-r)!0.5!(-t)$) -- ($(-l)!0.7!(-t)$) -- ($(-r)!0.7!(-t)$) -- cycle;
\draw[line join=bevel] ($(-l)!0.7!(-t)$) -- ($(-r)!0.8!(-t)$) -- ($(-l)!0.8!(-t)$) -- ($(-r)!0.7!(-t)$) -- cycle;
\node[coordinate, label={[label distance=0.5]above:\dots}] (-d) at (0,4) {};
\draw[line cap=rect,-] ($ (-t)+(-1,0)$) [antenna=1];
\draw[line cap=rect,-] ($ (-t)+(-1,0)$) -- ($(-t)+(1,0)$) [antenna=2];
\node[coordinate] (-a1) at ($ (-t)+(-1,0)$) {};
\node[coordinate] (-a2) at ($ (-t)+(1,0)$) {};
}}}

\tikzset{pics/BS/.style args={#1}{code={

\node[station] (base) {#1};

\draw[line join=bevel] (base.100) -- (base.80) -- (base.110) -- (base.70) -- (base.north west) -- (base.north east);
\draw[line join=bevel] (base.100) -- (base.70) (base.110) -- (base.north east);

\draw[line cap=rect] ([yshift=0pt]base.north) [antenna=1];

}}}

\tikzset{pics/car/.style args={#1}{code={
    \node (-base) at (4.2,2){};
    \draw[very thick,#1, rounded corners=0.18ex,fill=black!20!blue!20!white]  (2.7,1.8) -- ++(1,0.7) -- ++(1.6,0) -- ++(0.6,-0.7) -- (2.7,1.8);
        \draw[thick,#1, fill=#1,thick] (4.2,2.5) -- (5.3,2.5)  -- ++(0.2,-0.23) -- (4.4,2.27) -- (4.4,1.8) --(4.2,1.8) -- cycle;
      \draw[thick]  (4.2,1.8) -- (4.2,2.5);
      \fill[draw=#1,fill=#1,rounded corners=0.6ex,very thick] (1.5,.5) -- ++(0,1) -- ++(1.2,0.3) --  ++(3,0) -- ++(1,0) -- ++(0,-1.3) -- (1.5,.5) -- cycle;
      \draw[draw=black,fill=gray!50,thick] (2.75,.5) circle (.5);
      \draw[draw=black,fill=gray!50,thick] (5.5,.5) circle (.5);
      \draw[draw=black,fill=gray!80,semithick] (2.75,.5) circle (.4);
      \draw[draw=black,fill=gray!80,semithick] (5.5,.5) circle (.4);
      \filldraw[draw=#1, fill=KITorange, semithick] (2.1,1.3) -- ++(150:.5) arc (155:205:0.5) -- cycle;
  
}}}

\tikzset{
  pobl/.style={
    inner sep=0pt, outer sep=0pt, fill=#1,
  },
  pobl gron/.style n args={2}{
    pobl=#1, rounded corners=#2,
  },
  pics/person/.style n args={3}{
    code={
      \node (-corff) [pobl=#1, minimum width=.25*#2, minimum height=.375*#2, rotate=#3, pic actions] {};
      \node (-pen) [minimum width=.3*#2, circle, pobl=#1, outer sep=.01*#2, anchor=south, rotate=#3, pic actions] at (-corff.north) {};
      \node (-coes dde) [pobl gron={#1}{1pt}, anchor=north west, minimum width=.12125*#2, minimum height=.25*#2, rotate=#3, pic actions] at (-corff.south west) {};
      \node [pobl=#1, anchor=north, minimum width=.12125*#2, minimum height=.15*#2, rotate=#3, pic actions] at (-coes dde.north) {};
      \node (-coes chwith) [pobl gron={#1}{1pt}, anchor=north east, minimum width=.12125*#2, minimum height=.25*#2, rotate=#3, pic actions] at (-corff.south east) {};
      \node [pobl=#1, anchor=north, minimum width=.12125*#2, minimum height=.15*#2, rotate=#3, pic actions] at (-coes chwith.north) {};
      \node (-braich dde) [pobl gron={#1}{.75pt}, minimum width=.075*#2, minimum height=.325*#2, outer sep=.0064*#2, anchor=north west, rotate=#3, pic actions] at (-corff.north east)  {};
      \node [pobl=#1, minimum width=.05*#2, minimum height=.2*#2, outer sep=.0064*#2, anchor=north west, rotate=#3, pic actions] at (-corff.north east) {};
      \node (-braich chwith) [pobl gron={#1}{.75pt}, minimum width=.075*#2, minimum height=.325*#2, outer sep=.0064*#2, anchor=north east, rotate=#3, pic actions] at (-corff.north west) {};
      \node [pobl=#1, minimum width=.0375*#2, minimum height=.2*#2, outer sep=.0064*#2, anchor=north east, rotate=#3, pic actions] at (-corff.north west) {};
      \node (-fit person) [fit={(-pen.north) (-braich dde.east) (-coes chwith.south) (-braich chwith.west)}] {};
      \node (-pwy) [below=25pt of -fit person, every pin] {\tikzpictext};
      \draw [every pin edge] (-fit person) -- (-pwy);
    };
  };
}

\tikzset{pics/phone/.style args={#1}{code={
    \node (-base) at (1,1.75) {};
      \draw[draw=black,fill=black!80,rounded corners=0.2ex,very thick] (0,0) -- ++(0,3.5) -- ++(2,0) --  ++(0,-3.5) -- ++(-2,0) -- cycle;
      \fill[thick, fill=KITblue] (0.2,0.2) -- ++(0,3.1)  -- ++(1.6,0) -- ++(0,-3.1) -- ++(-1.6,0) -- cycle;
      \fill[fill=#1] (0.65,2) rectangle (0.75,2.1);
      \fill[fill=#1] (0.85,2) rectangle (0.95,2.3);
      \fill[fill=#1] (1.05,2) rectangle (1.15,2.5);
      \fill[fill=#1] (1.25,2) rectangle (1.35,2.7);
  
}}}

\def\BibTeX{{\rm B\kern-.05em{\sc i\kern-.025em b}\kern-.08em
    T\kern-.1667em\lower.7ex\hbox{E}\kern-.125emX}}

\begin{document}
\title{Integrated Radio Sensing Capabilities for 6G Networks: AI/ML Perspective

\author{Victor Shatov,~\IEEEmembership{Graduate Student Member,~IEEE}, 
Steffen Schieler,~\IEEEmembership{Graduate Student Member,~IEEE},\\ 
Charlotte Muth,~\IEEEmembership{Graduate Student Member,~IEEE}, 
Jos\'{e} Miguel Mateos-Ramos,~\IEEEmembership{Graduate Student Member,~IEEE},\\
Ivo Bizon,~\IEEEmembership{Graduate Student Member,~IEEE},
Florian Euchner,~\IEEEmembership{Graduate Student Member,~IEEE},\\
Sebastian Semper,
Stephan ten Brink,~\IEEEmembership{Fellow,~IEEE},
Gerhard Fettweis,~\IEEEmembership{Fellow,~IEEE},\\
Christian H\"{a}ger,~\IEEEmembership{Member,~IEEE},
Henk Wymeersch,~\IEEEmembership{Fellow,~IEEE},
Laurent Schmalen,~\IEEEmembership{Fellow,~IEEE},\\
Reiner Thomä,~\IEEEmembership{Life Fellow,~IEEE},
and Norman Franchi,~\IEEEmembership{Member,~IEEE}

\thanks{Manuscript received XX April, 2025; revised XX XXXXX, 2025; date of current version XX XXXXX, 2025.}
\thanks{This work was supported, in part, by the German
	Federal Ministry of Education and Research (BMBF) within the projects Open6GHub (Grant Agreements 16KISK005, 16KISK010) and KOMSENS-6G (Grant Agreement 16KISK123); in part, by Hexa-X-II, part of the European Union’s Horizon Europe research and innovation program under Grant Agreement No 101095759.}

\thanks{Victor Shatov and Norman Franchi are with the Institute for Smart Electronics and Systems (LITES), Friedrich-Alexander Universität Erlangen-Nürnberg, Erlangen, 91058, Germany (e-mail: victor.shatov@fau.de; norman.franchi@fau.de). Their work contributes to research within the 6G-Valley innovation cluster.}
\thanks{Steffen Schieler, Sebastian Semper, and Reiner Thom\"{a} are with the Electronic Measurements and Signal Processing Group, Technische Universitat Ilmenau, Germany (e-mail: steffen.schieler@tu-ilmenau.de; reiner.thomae@tu-ilmenau.de).}
\thanks{Charlotte Muth and Laurent Schmalen are with the Communications Engineering Lab (CEL), Karlsruhe Institute of Technology (KIT) Hertzstr. 16, 76187 Karlsruhe, Germany (e-mail: charlotte.muth@kit.edu; laurent.schmalen@kit.edu).}
\thanks{Jos\'{e} Miguel Mateos-Ramos, Christian H\"{a}ger, and Henk Wymeersch are with the Department of Electrical Engineering, Chalmers University of Technology, Sweden (e-mail: josemi@chalmers.se; christian.haeger@chalmers.se; henkw@chalmers.se).}
\thanks{Ivo Bizon and Gerhard Fettweis are with the Vodafone Chair Mobile Communications Systems, Technische Universit{\"a}t Dresden (TUD), Germany (e-mail: ivo.bizon@ifn.et.tu-dresden.de; fettweis@ifn.et.tu-dresden.de).}
\thanks{Florian Euchner and Stephan ten Brink are with the Institute of Telecommunications, University of Stuttgart, 70569 Stuttgart, Germany (e-mail: florian.euchner@inue.uni-stuttgart.de; tenbrink@inue.uni-stuttgart.de).}
}


}


\markboth{IEEE XXXXX,~Vol.~XX, No.~XX, MM~2025}%
{Shell \MakeLowercase{\textit{et al.}}: A Sample Article Using IEEEtran.cls for IEEE Journals}


\maketitle

\begin{abstract}
The \ac{6G} is often labeled as "connected intelligence".
Radio sensing, aligned with \ac{ML} and \ac{AI}, promises, among other benefits, breakthroughs in the system's ability to perceive the environment and effectively utilize this awareness.
This article offers a tutorial-style survey of \ac{AI} and \ac{ML} approaches to enhance the sensing capabilities of next-generation wireless networks.
To this end, while staying in the framework of \ac{ISAC}, we expand the term "sensing" from radar, via spectrum sensing, to miscellaneous applications of radio sensing like non-cooperative transmitter localization.
We formulate the problems, explain the state-of-the-art approaches, and detail \ac{AI}-based techniques to tackle various objectives in the context of wireless sensing.
We discuss the advantages, enablers, and challenges of integrating various sensing capabilities into an envisioned \ac{AI}-powered multi-modal multi-task network.
In addition to the tutorial-style core of this work based on direct authors' involvement in \ac{6G} research problems, we review the related literature, and provide both a good start for those entering this field of research, and a topical overview for a general reader with a background in wireless communications.
\end{abstract}


\bstctlcite{IEEEexample:BSTcontrol}
\acresetall
\section{Introduction}\label{sec:intro}

It is generally accepted that wireless networks, once designed solely for transmitting voice calls, will continue to evolve, becoming significantly more capable and intelligent.
Similarly to how the five human senses perceive raw data from our surroundings for subsequent processing and interpretation by the brain, radio sensors can feed the wireless network with diverse environmental data, enabling the network's capability to adapt and optimize autonomously.
Nowadays, the research community is exploring data-driven solutions as promising alternatives to traditional algorithms based on domain knowledge, statistics and mathematical models.

\subsection{Motivation and Background}

Since around a decade ago, algorithms based on \ac{AI} and \ac{ML} have prevailed in speech recognition, visual object recognition, and object detection~\cite{LeCun2015,yolo,speech}.
Meanwhile, a relatively small number of enthusiasts tried to leverage \ac{ML}-based techniques for wireless communications.
Though \ac{ML} was long envisioned as a natural way to adapt wireless networks based on spectrum sensing in the context of \ac{CR}~\cite{ThesisMitola2000,CR_overview2005}, its potential for wireless communication systems was clearly and systematically articulated only in the second half of the 2010s~\cite{OSheaHoydis2017,tenBrinkHoydis2018,Bjornson2019}.
Moreover, around that time, wireless networks were predicted to go well beyond mobile Internet and be \ac{ML}-powered in early roadmaps for \ac{6G}~\cite{Letaief2019}. 
As of 2025, \ac{ML}-based algorithms have become mainstream in wireless research on the physical layer, while practical implementations are mostly still in their infancy.

Although the \ac{ML} origins can be traced back to 1950s, the recent exponential advancement in data-driven methods is mainly due to the following three factors:
(\textit{i}) improvement in algorithms and model architectures;
(\textit{ii}) breakthrough in computational capabilities, notably with the development of \acp{GPU} and specialized hardware like Google's \acp{TPU}; (\textit{iii}) availability of huge amounts of data.
In the wireless domain, (\textit{i}) and (\textit{ii}) can be expected and attributed to natural technological progress over time, while (\textit{iii}) seems to be the bottleneck. Compared to audio or image signals, high quality wireless raw data is more difficult to obtain. 
For instance, such data frames might need to include details like interference patterns, channel conditions, and user behavior, which are not trivial to collect or label in plenty.
The data shortage problem is understood and currently being tackled by the \ac{3GPP}.
In particular, enhanced data collection is part of the recently frozen \ac{3GPP} Release 18~\cite{3GPP_data}.

\begin {figure}
\centering
\begin{adjustbox}{width=0.49\textwidth}

\begin{tikzpicture}
[ every annotation/.style = {draw,
                     fill = white, font = \Large}]
  \path[mindmap,concept color=black!40,text=white,
    every node/.style={concept,circular drop shadow},
    root/.style    = {concept color=black!40,
      font=\large\bfseries,text width=10em},
    level 1 concept/.append style={font=\Large\bfseries,
      sibling angle=120,text width=7.7em,
    level distance=15em,inner sep=0pt},
    level 2 concept/.append style={font=\bfseries,level distance=9em},
  ]
  node[root] {Radio\\Sensing} [clockwise from=90]
    child[concept color=blue!50] {
      node {Physical\\Perception\\(Radar)} [clockwise from=240]
        child { node[concept] (ObjectDetection)
        {Object Detection}}
        child {node(ObjectLocalization){Object Positioning}}
        child {node(ObjectOrientation){Object Orientation}}
        child {node(ShapeDetection){Shape Detection}}
        child {node(ObjectTracking){Object Tracking}}
        child {node(Environment){Environ-\\mental\\ Sensing}}
    }
    child[concept color=green!50!black] {
      node[concept] {Channel\\ Measure-\\ments}
        [clockwise from=-10]
      child { node[concept] (ChannelEstimation){Channel Estimation}}
      child { node[concept] (ChannelSounding){Channel Sounding}}
    }
    child[concept color=red!50!black] {
      node[concept] {Emitter\\Analysis}
        [counterclockwise from=160]
      child { node[concept] (SpectrumSensing){Spectrum Usage}}
      child { node[concept] (SignalClassification){Signal Classification}}
      child { node[concept] (EmitterLocalization){Emitter Localization}} 
    };
    \info[20]{ObjectDetection.west}{above,anchor=west,xshift=-14em, yshift=0.5em}{%
      \item Identifies the presence of objects in the radar's field of view
      \item Initiates further analysis and decision-making in autonomous systems, security, and surveillance applications
    }
    \info[20]{ObjectLocalization.north west}{above,anchor=south,xshift=-5em}{%
      \item Determines the precise position (coordinates) of a detected object
      \item Navigation, collision avoidance, and situational awareness for autonomous vehicles, air traffic control, and military operations
    }
    \info[17.5]{ObjectOrientation.north}{below,anchor=south west,xshift=-5em,yshift=-0.25em}{%
      \item Estimates the orientation (angle or tilt) of an object relative to the radar
      \item Understanding the object's alignment and motion in robotics, for AGVs and UAVs
    }
    \info[20]{ShapeDetection.north east}{below,anchor=south,xshift=7.7em,yshift=-3em}{%
      \item Helps in determining the approximate shape and structure of a detected object
      \item Used to classify objects (e.g., distinguish vehicles from pedestrians) and improving safety. Applications: automated systems and smart surveillance
    }
    \info[22]{ObjectTracking.east}{below,anchor=west}{%
      \item Continuously tracks and predicts the movement of an object over time
      \item To maintain situational awareness, prevent collisions, and optimize paths in autonomous systems, robotics, and defense applications
      \item Beamforming enhancement
    }
    \info[14]{Environment.east}{below,anchor=west}{%
      \item Material Parameters
      \item Particle distribution
      \item Weather prediction
    }
    \info[14]{ChannelEstimation.north}{below,anchor=south,xshift=0.0em,yshift=-0.5em}{%
      \item CSI feedback for real-time network optimization
      \item Compensate impairments for communications
      \item Adaptive modulation and coding
      \item Adaptive beamforming
    }
    \info[19]{ChannelSounding.south}{anchor=north,xshift=-0.0em, yshift=0.5em}{%
      \item Radio propagation environment characterization
      \item Facilitates network planning
    }
    \info[22]{EmitterLocalization.east}{anchor=west,xshift=-0.5em,yshift=-1.5em}{%
      \item Localization of known/cooperative TX using AoA, ToA, TDoA
      \item Localization of unknown/non-cooperative TX
      \item TX localization via channel charting
      \item Helps to enhance beamforming
      \item Helps to initiate actions against non-legitimate TXs
    }
    \info[22]{SpectrumSensing.north}{anchor=south,xshift = 0.5em,yshift = -0.5em}{%
      \item Interference avoidance and mitigation
      \item Dynamic spectrum allocation
      \item Opportunistic access via 'white spaces'
    }
    \info[17]{SignalClassification.south}{anchor=north,xshift=-0.5em, yshift=0.5em}{%
      \item Spectrum anomaly detection
      \item Jamming detection
      \item Waveform recognition\\ (Comm., radar, ISAC...)
      \item Automatic modulation classification
    }
\end{tikzpicture}
\end{adjustbox}
\caption{An illustration of the various radio sensing capabilities discussed in this article, along with some of their potential applications.}
\label{sensing_outlook}
\end{figure}
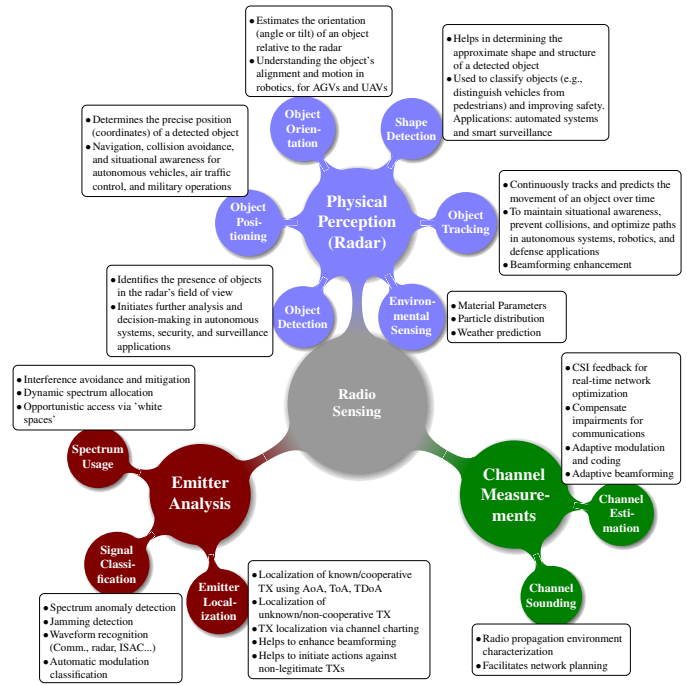

Sensing ubiquitous wireless signals will facilitate the extraction of abundant data and help to achieve complete environmental awareness.
Following~\cite{FNWF2023}, we roughly categorize radio sensing as follows: (\textit{i}) radar sensing to extract the information about the physical world; (\textit{ii}) channel measurements to understand the propagation characteristics and parameters, and (\textit{iii}) \ac{EM} analysis of the environment, or spectrum monitoring.
Fig.~\ref{sensing_outlook} illustrates this classification with a few details on each type of radio sensing.
In this work, we roughly group the first two as spatial awareness enablers with their own \ac{TX}, with the underlying goal of extracting information regarding physical objects, while the last one is to obtain \ac{EM} awareness that answers the questions who and where transmits, is purely passive.

Processing large amounts of diverse sensing data is challenging, and mathematical models often cannot capture the dependencies that may occur due to their dynamic nature or the peculiarities of an environment.
However, it is worth noting that via radio sensing, the increased number of devices and overall complexity in wireless environments can be turned from a challenge into an opportunity.
For example, localization algorithms might exploit signals broadcasted by \acp{TX} and captured by \acp{RX} to find the coordinates of objects or emitters. 
At the same time, \ac{ML}-based algorithms can refine this process, increasing accuracy and robustness in complex environments.
Furthermore, expansion of available frequency bands for \ac{6G} have been under consideration.
This added diversity can also be used to better understand the propagation environment by leveraging multi-band channel sounding, which provides a frequency-dependent view of how radio waves interact with their surroundings.

\ac{3GPP} is examining the integration of \ac{ML} from the 5G foundation to the emergence of \ac{6G}, which is promised to be the first \ac{AI}-native generation~\cite{Lin2023_1}, \cite{Lin2023_2}.
In Release 18, finalized in 2024, \ac{3GPP}, for the first time, investigated the potential of \ac{AI} for a 5G air interface, while the ongoing Release 19, scheduled for completion at the end of 2025, continues this work with the focus on beam management and positioning.
In parallel, \ac{6G} is envisioned to be the first generation that enables \ac{ISAC} technology~\cite{3GPP_ISAC}, with \ac{ML} being a natural choice to tackle various associated problems~\cite{AI_ISAC_2024}.
Enhanced with \ac{ML}, \ac{ISAC} will provide a foundation towards building a \ac{DT} of the radio environment~\cite{ISAC_DT_2024}, where the complete information about the physical and electromagnetic environment can open a new era in automated communications performance optimization, spectrum management, detection and localization of threats, as well as unveil new use cases for customers.
Finally, yet equally significant, the trend towards \ac{ML} in wireless networks is accepted and driven by traditional telecommunication stakeholders from all over the globe like Nokia~\cite{Nokia_24}, Ericsson~\cite{Ericsson_24}, and Huawei~\cite{Huawei_24}.
At this point, the topic has attracted significant attention and investments from other domain big tech players like Nvidia~\cite{Nvidia_24}, prominent for accelerated computing hardware, currently also offering tools for designing, simulating, training, and deploying \ac{AI}-based wireless networks~\cite{cammerer2025sionnaresearchkitgpuaccelerated}.

\begin{table*}
\caption{Comparison of relevant surveys and tutorials on sensing for wireless networks. Symbols \checkmark and \pmb{\checkmark} denote partially and detailed covered topics, respectively. Lozenge symbol indicates whether the article discusses leveraging \ac{AI} for a certain task, where $\lozenge$ and $\blacklozenge$ denote a glance (e.g., mentioning only use cases and opportunities) and strong focus (e.g., overview of concrete \ac{AI}/\ac{ML} algorithms), respectively.}
\centering
\scalebox{0.89}{
\begin{tabular}{c|c|l|cc|ccccc|ccc|c}
\hline\hline
\multirow{2}{*}{\textbf{Ref.}} & \multirow{2}{*}{\textbf{Year}} & \multirow{2}{*}{\textbf{\phantom{to center}Key contributions}} & \multicolumn{2}{c|}{\textbf{Basics}}    & \multicolumn{5}{c|}{\textbf{Physical perception of objects}}                                                                            & \multicolumn{3}{l|}{\textbf{Emitter analysis}}                            & \multirow{2}{*}{\makecell{\textbf{Vision}\\\textbf{Aspect}\\\textbf{B5G}}} \\ \cline{4-13}
                  &                   &                   & \multicolumn{1}{c|}{\textbf{\rotatebox{90}{Wireless}}}& \textbf{\rotatebox{90}{AI/ML}} & \multicolumn{1}{l|}{\textbf{\rotatebox{90}{\makecell{Object\\ detection}}}} & \multicolumn{1}{l|}{\textbf{\rotatebox{90}{\makecell{Object\\ positioning}}}} & \multicolumn{1}{l|}{\textbf{\rotatebox{90}{\makecell{Object\\ orientation}}}} & \multicolumn{1}{l|}{\textbf{\rotatebox{90}{\makecell{Shape\\ detection}}}} & \textbf{\rotatebox{90}{\makecell{Object\\ tracking}}} & \multicolumn{1}{l|}{\textbf{\rotatebox{90}{\makecell{Spectrum\\ sensing}}}} & \multicolumn{1}{l|}{\textbf{\rotatebox{90}{\makecell{Signal\\ classification}}}} & \textbf{\rotatebox{90}{\makecell{Transmitter\\ Localization}}} &                   \\ \hline

\cite{CR_AI_Survey2010}& 2010 & \makecell[l]{One of the first known surveys with \\ a clear focus on \ac{AI} for CR}   & \multicolumn{1}{l|}{} & \pmb{\checkmark} & \multicolumn{1}{l|}{} & \multicolumn{1}{l|}{} & \multicolumn{1}{l|}{} & \multicolumn{1}{l|}{} &  & \multicolumn{1}{l|}{\pmb{\checkmark}$\blacklozenge$} & \multicolumn{1}{l|}{} &  &    
\\ \hline       

\cite{CR_AI_Survey2013}& 2013 & \makecell[l]{Overviews supervised and unsupervised\\ \ac{ML} techniques for both\\ spectrum sensing and signal classification}   & \multicolumn{1}{l|}{} & \pmb{\checkmark} & \multicolumn{1}{l|}{} & \multicolumn{1}{l|}{} & \multicolumn{1}{l|}{} & \multicolumn{1}{l|}{} &  & \multicolumn{1}{l|}{\pmb{\checkmark}$\blacklozenge$} & \multicolumn{1}{l|}{\pmb{\checkmark}$\blacklozenge$} &  &  
\\ \hline

\cite{DeLima2021}&2021 & \makecell[l]{Identifies key localization and sensing\\ enablers for B5G} & \multicolumn{1}{l|}{} &  & \multicolumn{1}{l|}{\checkmark$\lozenge$} & \multicolumn{1}{l|}{\checkmark$\lozenge$} & \multicolumn{1}{l|}{\checkmark$\lozenge$} & \multicolumn{1}{l|}{\checkmark$\lozenge$} & \checkmark$\lozenge$ & \multicolumn{1}{l|}{} & \multicolumn{1}{l|}{} & \checkmark$\lozenge$ &\pmb{\checkmark}            \\ \hline

\cite{ISAC_vision_Liu_2022}& 2022 & \makecell[l]{Historical overview, performance trade-\\offs, waveform and receiver design\\ of ISAC systems}   & \multicolumn{1}{l|}{\pmb{\checkmark}} &  & \multicolumn{1}{l|}{\pmb{\checkmark}} & \multicolumn{1}{l|}{\pmb{\checkmark}} & \multicolumn{1}{l|}{} & \multicolumn{1}{l|}{} & \checkmark & \multicolumn{1}{l|}{} & \multicolumn{1}{l|}{\checkmark$\lozenge$} &  &\pmb{\checkmark}   
\\ \hline

\cite{Demirhan2023}& 2023 & \makecell[l]{Focus on a vision of \ac{ML} roles\\ for ISAC in 6G}   & \multicolumn{1}{l|}{} &  & \multicolumn{1}{l|}{\checkmark$\lozenge$} & \multicolumn{1}{l|}{\checkmark$\lozenge$} & \multicolumn{1}{l|}{\checkmark$\lozenge$} & \multicolumn{1}{l|}{} &  & \multicolumn{1}{l|}{} & \multicolumn{1}{l|}{} &  &\pmb{\checkmark}    
\\ \hline

\cite{JRC_Shatov}& 2024 & \makecell[l]{Architectures, use cases, signal processing\\ and hardware aspects of ISAC for B5G}   & \multicolumn{1}{l|}{} &  & \multicolumn{1}{l|}{\pmb{\checkmark}} & \multicolumn{1}{l|}{\pmb{\checkmark}} & \multicolumn{1}{l|}{} & \multicolumn{1}{l|}{} &  & \multicolumn{1}{l|}{} & \multicolumn{1}{l|}{} &  & \checkmark   
\\ \hline

\cite{respati2024}& 2024 & \makecell[l]{Overview of the principles in
ML and DL\\and their potential applications in ISAC,\\ \ac{ML}-enhanced ISAC}   & \multicolumn{1}{l|}{} & \pmb{\checkmark} & \multicolumn{1}{l|}{{\checkmark$\lozenge$}} & \multicolumn{1}{l|}{\checkmark$\lozenge$} & \multicolumn{1}{l|}{} & \multicolumn{1}{l|}{\checkmark$\lozenge$} & \checkmark$\lozenge$ & \multicolumn{1}{l|}{} & \multicolumn{1}{l|}{} & \checkmark$\lozenge$ & \checkmark   
\\ \hline

\cite{CognitiveRadio_ML_Survey2024}& 2024 & \makecell[l]{Comprehensive survey on \ac{ML}-driven\\ cognitive radio for wireless networks}   & \multicolumn{1}{l|}{} & \pmb{\checkmark} & \multicolumn{1}{l|}{} & \multicolumn{1}{l|}{} & \multicolumn{1}{l|}{} & \multicolumn{1}{l|}{} &  & \multicolumn{1}{l|}{\pmb{\checkmark}$\blacklozenge$} & \multicolumn{1}{l|}{\pmb{\checkmark}$\blacklozenge$} &  &\checkmark   
\\ \hline

\cite{JammerDetection2024}& 2024 & \makecell[l]{Review of \ac{AI}/\ac{ML}-driven jamming \\detection and interference management}  & \multicolumn{1}{l|}{\checkmark} & \checkmark & \multicolumn{1}{l|}{} & \multicolumn{1}{l|}{} & \multicolumn{1}{l|}{} & \multicolumn{1}{l|}{} &  & \multicolumn{1}{l|}{\pmb{\checkmark}$\blacklozenge$} & \multicolumn{1}{l|}{\pmb{\checkmark}$\blacklozenge$} &  & {\pmb{\checkmark}}   
\\ \hline

\cite{Prelcic2024}& 2024 & \makecell[l]{Signal processing, applications \\and vision of ISAC systems}   & \multicolumn{1}{l|}{\pmb{\checkmark}} &  & \multicolumn{1}{l|}{\pmb{\checkmark}$\lozenge$} & \multicolumn{1}{l|}{\pmb{\checkmark}$\lozenge$} & \multicolumn{1}{l|}{\pmb{\checkmark}$\lozenge$} & \multicolumn{1}{l|}{\pmb{\checkmark}$\lozenge$} & \pmb{\checkmark}$\lozenge$ & \multicolumn{1}{l|}{} & \multicolumn{1}{l|}{} & \pmb{\checkmark}$\lozenge$ &\pmb{\checkmark}    
\\ \hline

\cite{Lu_ISAC_2024}& 2024 & \makecell[l]{Discussion on synchronization and \\overview of the other problems of ISAC}   & \multicolumn{1}{l|}{\pmb{\checkmark}} &  & \multicolumn{1}{l|}{\pmb{\checkmark}$\lozenge$} & \multicolumn{1}{l|}{\pmb{\checkmark}$\lozenge$} & \multicolumn{1}{l|}{} & \multicolumn{1}{l|}{} & \checkmark & \multicolumn{1}{l|}{} & \multicolumn{1}{l|}{} &  &\checkmark    
\\ \hline

\cite{SS_DL_Surv2023}& 2023 & \makecell[l]{A survey on \ac{ML} for spectrum sensing \\and signal classification}   & \multicolumn{1}{l|}{} & \pmb{\checkmark} & \multicolumn{1}{l|}{} & \multicolumn{1}{l|}{} & \multicolumn{1}{l|}{} & \multicolumn{1}{l|}{} &  & \multicolumn{1}{l|}{\pmb{\checkmark}$\blacklozenge$} & \multicolumn{1}{l|}{\pmb{\checkmark}$\blacklozenge$} &  &    
\\ \hline

\cite{SpectrumSensingTrends2025}& 2024 & \makecell[l]{Comprehensive survey on \ac{ML} for\\ spectrum sensing, sharing, and access}   & \multicolumn{1}{l|}{} & \pmb{\checkmark} & \multicolumn{1}{l|}{} & \multicolumn{1}{l|}{} & \multicolumn{1}{l|}{} & \multicolumn{1}{l|}{} &  & \multicolumn{1}{l|}{\pmb{\checkmark}$\blacklozenge$} & \multicolumn{1}{l|}{} &  & \checkmark    
\\ \hline

\cite{AI_ISAC_2024}& 2024 & \makecell[l]{Bird's eye overview of \ac{AI}-enhanced ISAC,\\ including key challenges and perspectives }   & \multicolumn{1}{l|}{} &  & \multicolumn{1}{l|}{\checkmark$\lozenge$} & \multicolumn{1}{l|}{\checkmark$\lozenge$} & \multicolumn{1}{l|}{\checkmark$\lozenge$} & \multicolumn{1}{l|}{} &  & \multicolumn{1}{l|}{} & \multicolumn{1}{l|}{} &  &\pmb{\checkmark}    
\\ \hline

\cite{SurvISACin6G2024}& 2024 & \makecell[l]{Discussion on multimodal sensing and sen-\\ sing and computation convergence in 6G}   & \multicolumn{1}{l|}{} & \pmb{\checkmark} & \multicolumn{1}{l|}{\pmb{\checkmark}$\blacklozenge$} & \multicolumn{1}{l|}{\pmb{\checkmark}$\blacklozenge$} & \multicolumn{1}{l|}{\pmb{\checkmark}$\blacklozenge$} & \multicolumn{1}{l|}{} & \pmb{\checkmark}$\blacklozenge$ & \multicolumn{1}{l|}{} & \multicolumn{1}{l|}{} & \ &\pmb{\checkmark}    
\\ \hline

\cite{DiSAC2025}& 2025 & \makecell[l]{Highlights principles, architecture, \\and potential applications of\\ distributed ISAC (DISAC)}   & \multicolumn{1}{l|}{} &  & \multicolumn{1}{l|}{\checkmark$\lozenge$} & \multicolumn{1}{l|}{\checkmark$\lozenge$} & \multicolumn{1}{l|}{\checkmark$\lozenge$} & \multicolumn{1}{l|}{\checkmark$\lozenge$} & \checkmark$\lozenge$ & \multicolumn{1}{l|}{} & \multicolumn{1}{l|}{} & \checkmark &\pmb{\checkmark}    
\\ \hline

\hline\hline
\rowcolor{blue!20!white}
Our work & 2025 & \makecell[l]{A tutorial on various aspects of \ac{AI}-Enhanced\\ radio sensing for 6G}  & \multicolumn{1}{l|}{\pmb{\checkmark}} & \pmb{\checkmark} & \multicolumn{1}{l|}{\pmb{\checkmark}$\blacklozenge$} & \multicolumn{1}{l|}{\pmb{\checkmark}$\blacklozenge$} & \multicolumn{1}{l|}{\pmb{\checkmark}$\blacklozenge$} & \multicolumn{1}{l|}{\pmb{\checkmark}$\blacklozenge$} & \pmb{\checkmark}$\blacklozenge$ & \multicolumn{1}{l|}{\pmb{\checkmark}$\blacklozenge$} & \multicolumn{1}{l|}{\pmb{\checkmark}$\blacklozenge$} &\pmb{\checkmark}$\blacklozenge$  &\pmb{\checkmark}    
\\ \hline\hline

\end{tabular}
}
\label{Tab:SurveyComparison}
\end{table*}

\subsection{Related Works and Key Contributions}
In the context of \ac{6G}, integrating sensing and \ac{ML} capabilities into wireless networks is one of the hottest research topics, regardless of whether sensing and \ac{ML} are considered together or separately.
Here, we familiarize the reader with key and recent works that fall into the review, tutorial, or perspective article categories.

Two early surveys of \ac{ML} for spectrum sensing in the context of \ac{CR}, a concept formalized in the late 1990s, can be found in~\cite{CR_AI_Survey2010} and~\cite{CR_AI_Survey2013}.
In~\cite{CR_AI_Survey2010}, the authors formulate the \ac{ML}-empowered \ac{CR} observation through spectrum sensing with subsequent system reconfiguration and learning the impact of these actions on the performance of the radio.
While~\cite{CR_AI_Survey2010} focuses on spectrum sensing to implement a practical \ac{CR} and the relative merits of various proposed techniques in different \ac{CR} applications, a later study~\cite{CR_AI_Survey2013} addresses supervised and unsupervised \ac{ML} algorithms for both spectrum sensing and signal classification.
A recent work~\cite{CognitiveRadio_ML_Survey2024} provided a survey on \ac{ML}-driven \ac{CR} wireless networks, including \ac{IoT}, mobile (namely vehicular and railway), and unmanned aerial vehicle (UAV) networks.
This comprehensive survey also goes beyond spectrum occupancy detection techniques, reviewing various aspects of non-cooperative TX signal classification, which is denoted here as 'intelligent spectrum sensing'.
A review of \ac{ML}-driven jamming detection and interference management presents a slightly different angle of view on spectrum sensing in~\cite{JammerDetection2024}, where the authors presented a detailed taxonomy of intentional and unintentional spectrum-related troubles and \ac{ML}-based mitigation techniques in various wireless communication scenarios, including proper resource allocation through spectrum sensing.
In~\cite{SS_DL_Surv2023} and~\cite{SpectrumSensingTrends2025}, surveys on \ac{ML} methods for standard spectrum sensing, sharing, and access are provided.
The authors of both reviews treat sensing as part of the classical \ac{CR} technology scenario, in which secondary users detect the unutilized bands through sensing, thus facilitating better spectrum utilization.

Narrowing down the term sensing purely to radar,~\cite{ISAC_vision_Liu_2022} reviewed the historical development of \ac{ISAC}, its performance gains and trade-offs, signal processing, potential waveforms, and the \ac{RX} design.
Interestingly, as one of the open challenges, the authors mentioned the importance of \ac{ML}-based signal recognition in the context of radar-communication convergent systems.
Similarly, the perspective paper~\cite{JRC_Shatov} considers only radar sensing integration into \ac{6G} networks.
The authors introduced the architectures of \ac{ISAC} and reviewed use cases, waveforms, signal processing, and hardware aspects of \ac{ISAC}, only mentioning \ac{ML} as a potential performance enhancer.
Another recent milestone work~\cite{Prelcic2024} provides a detailed overview and vision of \ac{ISAC} in cellular networks, focusing on signal processing aspects for localization of sources and passive targets.
Furthermore, this article discusses sensing-aided communications, discussing beam training for overhead reduction and blockage prediction in practical scenarios.
In~\cite{Lu_ISAC_2024}, additionally to the state-of-the-art \ac{ISAC} overview as of 2024, the authors provided design guidelines and ten open challenges for \ac{ISAC} systems.
An early pre-standardization vision of \ac{6G} networks as convergent communication, localization, and sensing systems is formulated in~\cite{DeLima2021}.
That paper identifies key localization and sensing enablers for beyond 5G (B5G).
This work focuses on futuristic integrated radar sensing solutions with fine-range, Doppler, and angular resolutions. 
It also notes the vital role of the flexible use of the \ac{RF} spectrum and \ac{AI} in \ac{6G}.
A more recent study ~\cite{DISAC} took a step further and highlighted the importance of distributed observations of the environment, discussing principles, architecture, and potential applications of the distributed \ac{ISAC} (DISAC).

While some papers mention the value of the convergence of various sensing capabilities in \ac{6G}, their focus is often limited to only one capability. Hence, there is a need for a broader perspective on the different modalities of radio sensing.
To the best of our knowledge, no reviews consider sensing of physical objects and \ac{EM} sensing together, and no clear formulation of the importance of \ac{AI}-empowered radio sensing exists.
Meanwhile, to realize the expectations of \ac{6G}, e.g., build a \ac{DT} of the wireless network, knowledge of both physical and \ac{EM} worlds is equally essential.
Our paper's primary goal is to provide the research community with a tutorial based on hands-on experience while providing an overview of the latest works, datasets, and research gaps.
The main contributions of this paper are as follows:
\begin{itemize}
    \item We offer a guidance on \ac{ML}-enhanced radar sensing for physical object perception and its possible integration into wireless networks, one of the critical topics in the context of \ac{6G}. 
    \item Similarly, we formulate the problems in the context of what we roughly categorize as emitter analysis, including \ac{CR}-like standard spectrum sensing to detect underutilized parts of the spectrum and waveform classification, while providing \ac{ML}-based existing solutions.
    We add insights on \ac{ML}-empowered \ac{TX} localization and channel charting.
    \item Finally, we consider the potential integration of various radio sensing functions enhanced by \ac{AI} algorithms into future wireless networks, highlighting potential advantages and challenges to overcome.
\end{itemize}

Table~\ref{Tab:SurveyComparison} structures the scope of this work and compares it with other existing surveys, tutorials, and vision articles. 
This table focuses on recent articles, while a few selected older publications are introduced for completeness.

\subsection{Organization of the Paper}

\begin{figure*}
\centering
    \includegraphics[width=0.98\linewidth]{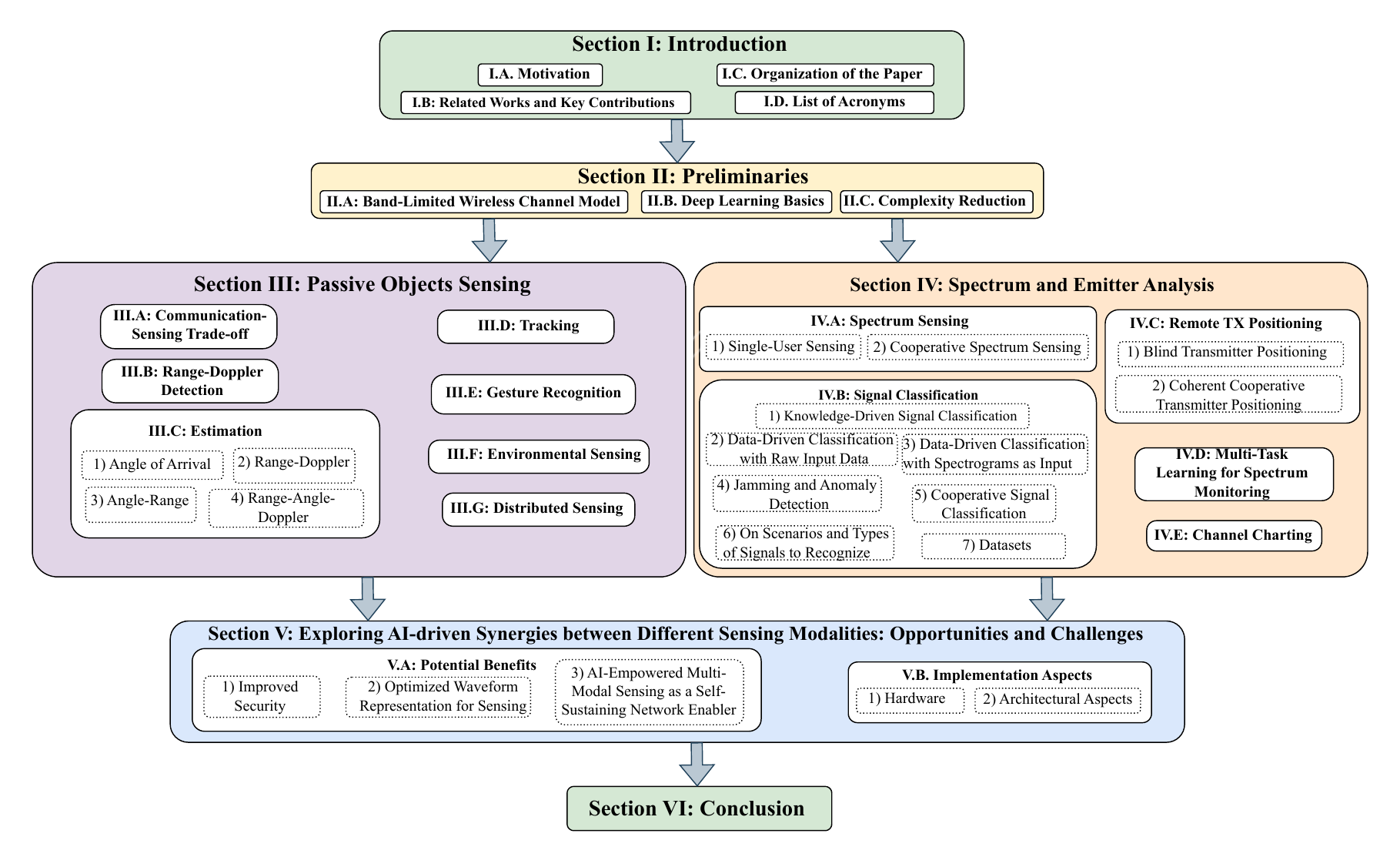}
\caption{The organization of the article.} \label{structure}
\end{figure*}

Fig.~\ref{structure} illustrates the article’s organization.
In section~\ref{sec:prelim}, we introduce the background in \ac{AI} and wireless basics that make it easier for a general reader to follow the content of this work.
Section~\ref{sec:passive_object_sensing} provides guidance to the \ac{AI}-empowered physical perception of passive objects, focusing on radar capabilities integrated into the wireless network.
Section~\ref{sec:spectral_sens} aims to help navigate in the broad area of passive radio sensing, including spectrum sensing, signal classification, and TX localization.
Section~\ref{sec:integr} highlights the importance of different types of radio sensing for the envisioned 6G wireless applications, discussing potential advantages and challenges.
Finally, section~\ref{sec:conclusion} concludes the article with final remarks. 

\begin{acronym}[TROLOLO]
\acro{6G}{sixth-generation wireless communications}
  \acro{ACM}{auto correlation matrix}
  \acro{ADC}{analog to digital converter}
  \acro{AE}{autoencoder}
  \acro{AI}{artificial intelligence}
  \acro{AMC}{automatic modulation classification}
  \acro{ASK}{amplitude shift keying}
  \acro{AoA}{angle of arrival}
  \acro{AWGN}{additive white Gaussian noise}
  \acro{BER}{bit error rate}
  \acro{BCE}{binary cross entropy}
  \acro{BD}{Bhattacharyya distance}
  \acro{BMI}{bit-wise mutual information}
  \acro{BPTT}{backpropagation through time}
  \acro{BPSK}{binary phase shift keying}
  \acro{BP}{backpropagation}
  \acro{BSC}{binary symmetric channel}
  \acro{CAZAC}{constant amplitude zero autocorrelation waveform}
  \acro{CDF}{cumulative distribution function}
  \acro{CE}{cross entropy}
  \acro{CFAR}{constant false alarm rate}
  \acro{CIR}{channel impulse response}
  \acro{CNN}{convolutional neural network}
  \acro{CP}{cyclic prefix}
  \acro{CR}{cognitive radio}
  \acro{CRB}{Cramér-Rao bound}
  \acro{CRC}{cyclic redundancy check}
  \acro{CS}{compressive sensing}
  \acro{CSI}{channel state information}
  \acro{CU}{central unit}
  \acro{CZT}{chirp z transform}
  \acro{DFT}{discrete Fourier transform}
  \acro{DNN}{deep neural network}
  \acro{DDD}{delay-Doppler domain}
  \acro{DL}{deep learning}
  \acro{DoA}{direction of arrival}  
  \acro{DoD}{direction of departure}  
  \acro{DOCSIS}{data over cable services}
  \acro{DPSK}{differential phase shift keying}
  \acro{DSL}{digital subscriber line}
  \acro{DSP}{digital signal processing}
  \acro{DT}{digital twin}
  \acro{DTFT}{discrete-time Fourier transform}
  \acro{DVB}{digital video broadcasting}
  \acro{ELM}{extreme learning machine}
  \acro{ELU}{exponential linear unit}
  \acro{EDC}{efficient detection criterion}
  \acro{EM}{electromagnetic}
  \acro{ESPRIT}{estimation of signal parameter via rotational invariance techniques}
  \acro{FCF}{forward charting function}
  \acro{FEM}{finite element method}
  \acro{FFNN}{feed-forward neural network}
  \acro{FFT}{fast Fourier transform}
  \acro{FIR}{finite impulse response} 
  \acro{FL}{federated learning}
  \acro{FMCW}{frequency-modulated continuous-wave}
  \acro{FPGA}{field-programmable gate array}
  \acro{GD}{gradient descent}
  \acro{GF}{Galois field}
  \acro{GMM}{Gaussian mixture model}
  \acro{GMI}{generalized mutual information}
  \acro{GPU}{graphics processing unit}
  \acro{ICI}{inter-channel interference}
  \acro{ICZT}{inverse chirp z transform}
  \acro{IDE}{integrated development environment}
  \acro{IDFT}{inverse discrete Fourier transform}
  \acro{IFFT}{inverse fast Fourier transform}
  \acro{IIR}{infinite impulse response}
  \acro{IoT}{Internet of Things}
  \acro{IQ}{in-phase and quadrature}
  \acro{ISAC}{integrated sensing and communication}
  \acro{ISI}{inter-symbol interference}
  \acro{JCAS}{joint communication and sensing}
  \acro{KKT}{Karush-Kuhn-Tucker}
  \acro{kldiv}{Kullback-Leibler divergence}
  \acro{KPI}{key performance indicator}
  \acro{LDPC}{low-density parity-check}
  \acro{LFMW}{linear frequency modulated waveform}
  \acro{LLR}{log-likelihood ratio}
  \acro{LoS}{line of sight}
  \acro{LSTM}{long short-term memory}
  \acro{LTE}{long-term evolution}
  \acro{LTI}{linear time-invariant}
  \acro{LR}{logistic regression}
  \acro{MAC}{medium access control}
  \acro{MAP}{maximum a posteriori}
  \acro{MBL}{model-based learning}
  \acro{MI}{mutual information}
  \acro{MIMO}{multiple-input and multiple-output}
  \acro{MLP}{multilayer perceptron}
  \acro{ML}{machine learning}
  \acro{MLE}{maximum likelihood estimation}
  \acro{MSE}{mean squared error}
  \acro{MLSE}{maximum-likelihood sequence estimation}
  \acro{MTL}{multi-task learning}
  \acro{MM-MTL}{multi-modal multi-task learning}
  \acro{MMSE}{miminum mean squared error}
  \acro{MUSIC}{multiple signal classification}
  \acro{NIST}{National Institute of Standards and Technology}
  \acro{NLoS}{non-line of sight}
  \acro{NN}{neural network}
  \acro{OFDM}{orthogonal frequency-division multiplexing}
  \acro{OFDMA}{orthogonal frequency-division multiple access}
  \acro{OLA}{overlap-add}
  \acro{OMP}{orthogonal matching pursuit}
  \acro{PAPR}{peak-to-average-power ratio}
  \acro{PDF}{probability density function}
  \acro{pmf}{probability mass function}
  \acro{PMCW}{phase-modulated continuous waveform}
  \acro{PSD}{power spectral density}
  \acro{PSK}{phase shift keying}
  \acro{PLM}{path loss model}
  \acro{PICM}{position information correlation matrix}
  \acro{QAM}{quadrature amplitude modulation}
  \acro{QPSK}{quadrature phase shift keying}
  \acro{radar}{radio detection and ranging}
  \acro{RAN}{radio access network}
  \acro{RAT}{radio access technology}
  \acro{RC}{raised cosine}
  \acro{RCS}{radar cross section}
  \acro{RDM}{range-Doppler matrix}
  \acro{RF}{radio frequency}
  \acro{RIS}{reflective intelligent surface}
  \acro{RL}{reinforcement learning}
  \acro{RMSE}{root mean squared error}
  \acro{RNN}{recurrent neural network}
  \acro{ROM}{read-only memory}
  \acro{RRC}{root raised cosine}
  \acro{RSS}{received signal strength}
  \acro{RV}{random variable}
  \acro{RX}{receiver}
  \acro{SAR}{synthetic aperture radar}
  \acro{SER}{symbol error rate}
  \acro{SDR}{software defined radio}
  \acro{SNR}{signal-to-noise ratio}
  \acro{SINR}{signal-to-noise-and-interference ratio}
  \acro{SISO}{single-input single-output}
  \acro{SPA}{sum-product algorithm}
  \acro{SSN}{self-sustaining network}
  \acro{SU}{sensing unit}
  \acro{SVM}{support vector machine}
  \acro{TDoA}{time-difference-of-arrival}
  \acro{ToA}{time-of-arrival}
  \acro{TPU}{tensor processing unit}
  \acro{TX}{transmitter}
  \acro{UE}{user equipment}
  \acro{ULA}{uniform linear array}
  \acro{URA}{uniform rectangular array}
  \acro{VCS}{version control system}
  \acro{WLAN}{wireless local area network}
  \acro{WSN}{wireless sensor network}
  \acro{WSS}{wide-sense stationary}
  \acro{3GPP}{3rd Generation Partnership Project}
\end{acronym}

\section{Preliminaries}\label{sec:prelim}

In this section, we first review the fundamental concepts and analytical expressions that are vital for understanding different types of wireless systems. 
Then, we summarize some \ac{ML} basics and related terminology used throughout the article.

\subsection{Band-Limited Wireless Channel Model}\label{sec:channel-model} 
Knowledge of the channel is crucial in wireless networks because it enables efficient transmission and reception of signals, as well as effective use of resources.
Various signal processing algorithms are based on this information.
Since real-world systems can only transmit signals within a specific frequency range defined by hardware and regulations, wireless channels are typically limited in bandwidth.

The wireless propagation channel can be modeled as a superposition of incoming signal components that travel through multiple propagation paths. 
Therefore, they arrive at the \ac{RX} with different delays, amplitudes, and phases.
The combination of the signal components at the receiving antenna results in a signal that varies extensively both in amplitude and phase. 
Moreover, each signal component experiences a frequency shift that arises due to the relative motion between transmitter and receiver, i.e., the Doppler shift. 
The complex baseband \ac{CIR} at time $t$, with relative channel delay $\tau$, after mixing the signal down from the carrier frequency $f_{\mathrm{c}}$ is expressed by
\begin{equation} \label{continuous_time_channel}
	h(t, \tau) = \sum_{\ell=0}^{L(t)-1} \alpha_{\ell}(t) \delta\left(\tau-\tau_{\ell}(t)\right) \mathrm{e}^{-\j2\pi f_{\mathrm{D},\ell}t},
\end{equation}
where $L(t)$ represents the number of propagation paths, $\alpha_{\ell}(t)$, $\tau_{\ell}(t)$ represent the complex path weight and delay associated with the $\ell$-th path, respectively, $f_{\mathrm{D},\ell}$ denotes the Doppler shift, and $\delta(\cdot)$ the Dirac delta function \cite{principles_of_mobile_communication, goldsmith, rappaport}.
Equation (\ref{continuous_time_channel}) models the channels as a function of two time variables by assuming the channel is underspread such that $B_{f_D} \cdot \tau_{\text{rms}} \ll 1$, such that the Doppler shift changes only with $t$.
The two time variables are commonly referred to as fast-time $\tau$ and slow-time $t$ in radar sensing.

In digital receivers, the channel is sampled at a given sampling frequency $f_{\mathrm{s}} = 1/T_{\mathrm{s}}$ Hz, where $T_{\mathrm{s}}$ is the sampling time interval in seconds. When the difference among delays from propagation paths is smaller than $T_{\mathrm{s}}$, these are vectorially combined into a single tap. The time dependency $t$ can be dropped if the symbol time is smaller than the coherence time $T_{\mathrm{c}}$.
During the coherence time, channel responses are highly correlated, and the desired correlation level in a given time interval defines the relation between the coherence time $T_{\mathrm{c}}$ and the maximum Doppler shift $f_{\mathrm{D}\max}$.

If multiple \acp{RX} are present, and they are spaced closely together, we can assume that similar signal components are received by all \acp{RX}.
There is a small residual time delay between the different receiver nodes for each different path
which can be modeled by rephrasing~\eqref{continuous_time_channel} for an antenna element $i$ as
\begin{align} 
	h_i(\tau) &= \sum_{\ell=0}^{L-1} \alpha_{\ell} \delta\left(\tau-\tau_{\ell}-\tau_{i,\ell}\right) \mathrm{e}^{\j\Delta\varphi_{i,\ell}}
    \label{continuous_time_channel_node}
\end{align}
with a small residual time delay $\tau_{i,l}$ between antenna $i$ and a reference point that causes a phase shift $\Delta \varphi_{i,\ell}$ depending on the delay difference with respect to the antenna elements. This description allows for \ac{AoA} estimation from the different receive signals with a suitable antenna array.   

The \ac{CIR} describes how the radio channel affects an impulse signal.
When the transmitted signal is $s_{\text{Tx}}(t)$, the received signal $s_{\text{Rx}}(t)$ is given by

\begin{equation}\label{ReceiverSignal}
    s_{\text{Rx}}(t) = \int_{-\infty}^{+\infty}s_{\text{Tx}}(\tau) h(t,\tau) d \tau + n(t),
\end{equation}
where $n(t)~\sim~\mathcal{CN}~(0, N_0)$ is a Gaussian process corresponding to the \ac{AWGN} with the noise power spectral density $N_0$.\footnote{The noise is often assumed to be \ac{AWGN} after sampling and filtering}

Equation~(\ref{ReceiverSignal}) represents the received signal as the sum of delayed, attenuated, and phase-shifted versions of the transmit signal, determined by different propagation paths.
For communication systems,
accurate channel estimation is essential to mitigate the effects of fading and multipath propagation. 
Furthermore, since the \ac{CIR} provides valuable information on how radio waves interact with the environment, its knowledge is a critical component in radio sensing.
For example, it helps to detect, track, and characterize objects, as well as map the environment.

\subsection{Machine Learning Basics}\label{sec:ml-basics}
\Ac{ML} is a branch of \ac{AI} that focuses on enabling computers to learn patterns from data and make predictions or decisions without being explicitly programmed~\cite{LeCun2015}. 
Instead, a function $f(\mat{D},\mat{\Theta})$ with input data $\mat{D}$ and parameters $\mat{\Theta}$ is used to approximate the prediction or decision task. 
The parameters $\mat{\Theta}$ of the function are then obtained using a dataset $\mathcal{D}$ with $\mat{D} \in \mathcal{D}$ and a performance measure $P(f(\mat{D},\mat{\Theta}),T(\mat{D}))$, where the task to be learned is represented by $T(\mat{D})$.
The performance measure of numerous \ac{ML} approaches is expressed as a \textit{loss function}.

There are three primary learning paradigms: 
\textit{Supervised learning} is based on a labeled dataset, where $T(\mat{D})$ is the known intended output, e.g., the ground truth of a parameter that $f(\mat{D},\mat{\Theta})$ estimates. 
\textit{Unsupervised or self-supervised learning} finds patterns and structures in unlabeled data. In this case, $T(\mat{D})$ can include the variance of found patterns or different structures for clustering.
\textit{\Ac{RL}} teaches an agent to make decisions by rewarding or penalizing it based on its actions in an environment, potentially changing the environment with each action. The environment replaces $\mathcal{D}$, as each input $\mat{D}$ is dependent on previous actions taken by the agent.
\ac{RL} is of particular interest in the context of wireless networks, since it is researched as a candidate for complex decision-making in dynamic environments~\cite{Pradankla2022}.

Focusing on supervised learning, random subsets $\mathcal{B}$ of $\mathcal{D}$ called \textit{batches} (or mini-batches), with $|\mathcal{B}|\ll|\mathcal{D}|$ are usually propagated through $f(\mat{D},\mat{\Theta})$. Then, the parameters $\mat{\Theta}$ are optimized iteratively using different $\mathcal{B}$.
In each iteration, the loss function evaluates the mean performance over the current data batch.
The \ac{MSE} is a common choice for regression problems, i.e., tasks where a value dependent on $\mat{D}$ should be predicted such as estimating the delay from a received signal. 
For a batch ${\mathcal{B}}$, the \ac{MSE} creates a criterion that measures the mean squared error between each element of the estimated value $f(\mat{D}_{i},\mat{\Theta}) = \hat{x}_i(\mat{\Theta})$ and the corresponding ground truth value $T(\mat{D}_i)=x_i$ as follows:

    \begin{equation}\label{MSE_eq}
    \mathcal{L}_{\text{MSE}}(\mat{\Theta}) = \frac{1}{|{\mathcal{B}}|} \sum_{i =1}^{|\mathcal{B}|} (x_i - \hat{x}_i(\mat{\Theta}))^2.
    \end{equation}

The categorical cross-entropy loss is used for multiclass classification, i.e. $f(\mat{D},\mat{\Theta})$ is tasked to differentiate between a number of different classes or types of samples in $\mathcal{D}$. Commonly, classification tasks are represented using one-hot notation, i.e., a sample of class $c$ of $C$ classes can be labeled using a vector $\vect{t}$ of size $C$ with a $1$ at the $c$th element and zeros otherwise.
The categorical cross-entropy loss between the true one-hot labels $T(\mat{D}_i)=\vect{t}_i$ and the predicted distribution $f(\mat{D}_i,\mat{\Theta})=\vect{p}_i(\Theta)$ is given by
    \begin{equation}\label{CrossEntropy_eq}
    \mathcal{L}_{\text{CrossEntropy}}(\mat{\Theta}) = -\frac{1}{|\mathcal{B}|} \sum_{i =1}^{|\mathcal{B}|}  \vect{t}_{i}^{\top} \cdot\log \vect{p}_{i}(\mat{\Theta}).
    \end{equation}

For digital communication, the binary cross-entropy is often used for distinguishing between the binary states $0$ and $1$. This is a special case of the categorical cross-entropy with $C=2$.
It can be expressed as
    \begin{equation}\label{BCE}
        \mathcal{L}_{\text{BCE}}(\mat{\Theta}) = -\frac{1}{|\mathcal{B}|} \sum_{i =1}^{|\mathcal{B}|}  t_i \cdot\log p_i(\mat{\Theta}) + (1-t_i) \cdot\log (1-p_i(\mat{\Theta})).
    \end{equation}

After the loss function is calculated, the parameters $\mat{\Theta}$ are updated during training of $f(\mat{D},\mat{\Theta})$. In supervised learning, an iterative gradient-based update is usually performed through backpropagation. By choosing different optimizers and optimizer parameters, the amount of change of $\mat{\Theta}$ is controlled in each optimization iteration. Then the next batch $\mathcal{B}$ is generated, allowing to iteratively update $\mat{\Theta}$ until a predetermined number of iterations is reached or the performance measure reaches a target value.

There are multiple architectures in the literature that are used to implement $f(\mat{D},\mat{\Theta})$. They often consist of so-called neurons that calculate a weighted sum of all inputs and add a bias. To enable approximation of non-linear functions, non-linear activation functions $\sigma(\cdot)$ are often applied to neuron outputs. Multiple neurons, whose outputs can be calculated in parallel, are grouped in a layer.
In fully connected \acp{DNN}, every neuron of one layer is connected to every neuron in the adjacent layers.
Two other important \ac{NN} classes are \acp{CNN} and \acp{RNN}.
On the one hand, \acp{CNN} use convolutional layers that perform convolution of the input data with a trainable impulse response, denoted kernel in \ac{DL} jargon, with parameters $\mat{\Theta}$. They are well suited to detect structures inside an input frame and therefore often utilized in computer vision tasks~\cite{CNN_basics}, including processing of spectrograms, range-Doppler maps, or constellation diagrams, where their shift-invariance is utilized.
On the other hand, \acp{RNN} and transformer architectures are designed to capture (possibly long-term) temporal dependencies, and therefore suitable, e.g., for processing sequences of \ac{IQ} samples.
The recently introduced transformer architecture is based on a self-attention mechanism, which relates different positions within a single sequence to compute a representation of the sequence~\cite{Transformer}.
In summary, however, there are no strict rules, and \ac{NN} architectures allow a high degree of freedom to combine different layer types in application-specific architectures.

If the amount of training data is too limited or too specific, overfitting (or underfitting) can occur.
In many \ac{ML} domains like natural language processing or image processing, large text corpora or picture datasets are readily available.
In contrast to that, wireless communications and wireless sensing research often relies on channel models instead of measured data. Depending on the \ac{ML} task, obtaining models that approximate the real-world behavior well enough can be very challenging.
A selection of stochastic geometry-based channel models has been standardized \cite{5gchannelmodel} by generalizing measured behaviors into a statistical model.
Based on standardized models, channel simulators like QuaDRiGa \cite{burkhardt2014quadriga} produce channel realizations, so that the performance of various algorithms can be evaluated and compared on these channels.
Geometry-based stochastic channel models are less suitable for sensing applications, where deterministic quantities like target locations need to be estimated.
In these cases, electromagnetic ray tracers like Sionna RT \cite{Hoydis2023} or Remcom Wireless InSite \cite{Remcom} offer a reasonable compromise between the prohibitively high complexity of a \ac{FEM} simulation of the electromagnetic field equations and oversimplified geometric models.
Even with ray tracers, it is a good idea to specify and categorize particular environment scenarios to ensure comparability, like in the DeepMIMO datasets \cite{Alkhateeb2019}.
There are also efforts to collect and publish datasets for various frequency ranges, for example, the NextG Channel Model Alliance by the \ac{NIST}~\cite{nist_nextg_website}.

Raw measurement data can often be represented in various domains and formats, such as time domain, frequency domain, beamspace, and sample autocorrelation.  
Digitized data is usually acquired as a complex-valued signal, which comprises the in-phase (I) and quadrature (Q) components taken at specific time intervals.
These IQ samples are considered to be one of the richest representations of wireless signals for \acp{NN} that can be practically obtained.
Conventional \acp{NN} with real-valued inputs and outputs remain common in wireless communications and sensing research.
Usually, the real and imaginary parts of the complex-valued input vector are simply stacked and provided to the \ac{NN}.
Even though complex-valued neural networks (CVNNs) have been a research topic for a long time \cite{cvnns} and can natively process complex-valued signal representations, they have seen little adoption in the community, potentially due to limited support in common \ac{ML} frameworks like PyTorch or TensorFlow at the time of writing.

Although \acp{NN} can adapt to arbitrary input domains, providing data in a reasonable format can significantly improve performance and reduce training time.
Hence, feature engineering, i.e., converting the input into an appropriate format for processing by the \ac{NN}, can lead to efficient \ac{ML} solutions.
For many sensing tasks, a common preprocessing step is to convert raw signal measurements and a known transmit signal into so-called \ac{CSI} by performing channel estimation. Note that the dataset size can be reduced if only \ac{CSI} instead of \ac{TX} and \ac{RX} signals needs to be saved, but this framework is limited to sensing scenarios where pilot signals are utilized for sensing or where a known transmit sequence is assumed at the sensing \ac{RX}.
The \ac{CSI} is independent of the transmit signal and instead only dependent on the wireless channel, which could potentially include the hardware effects from transmitter and receiver chains, as long as the transmission system is sufficiently linear.
Since this is a useful signal representation for many sensing tasks, aforementioned channel datasets and channel simulators often directly contain or produce \ac{CSI} instead of raw \ac{IQ} samples.
For communication problems, \ac{CSI} can be used to simulate channel effects for particular transmit signals.
On the downside, preprocessing input data into features can also lead to loss of information.
Models with higher capacity, e.g., fed by multidimensional raw input data, can inherently learn the features of the wireless channel, including model deviations caused by hardware~\cite{OSheaHoydis2017}. 
This approach often yields better performance at the cost of higher inference time and complexity.

There are multiple challenges in \ac{ISAC} systems that seem uniquely suited to be addressed with \ac{ML}.
General models for channels and receivers exist, yet when integrating components developed on classical channel models into a measurement setup, some loss of performance might be observed. 
This performance degradation can be caused by model mismatches caused by hardware imperfections, distorted noise, quantization, or many other effects that have been ignored for the modeling of the channel.
By training an \ac{ML} system on real measurement data or even online using reinforcement learning, these effects can be addressed during optimization, even if only used for fine-tuning.
Another point making \ac{ISAC} specifically suited to approach using \ac{ML} is that joint optimization of communication and sensing algorithms is often non-convex. 
It can be very hard to (approximately) solve these optimization problems,e.g. optimal resource allocation in \ac{ISAC}, with classical approaches.
It is still possible for \ac{ML} solutions to get stuck in a local minimum and we cannot guarantee that the found solution is a global minimum. 
Nevertheless, e.g., comparison with theoretical bounds known from estimation and information theory can help verify the convergence of \ac{ML} systems.  

\subsection{Complexity Reduction}
Efforts to reduce complexity include reduction of computational complexity, simplification of data acquisition, increased flexibility of algorithms, and increased accessibility of training methods. As \ac{ISAC} research includes both spectrum and passive object sensing, we discuss it in a general section here.

The computational complexity of \ac{ISAC} algorithms is an important factor for associated power consumption, hardware requirements, and can play a role in the algorithm's latency. Consequently, achieving similar performance while reducing complexity remains a vital direction to explore. For \ac{ML} methods, we need to differentiate between the training and inference complexity. While training is usually performed once offline, being limited only by the time and hardware resources dedicated to training, the algorithm will be run many times with inference complexity on hardware with more limited resources. Research activities for complexity reduction tend to compare different \ac{ML} solutions based on complexity that achieve similar performance. For trainable algorithms, the complexity of training and inference has to be considered separately. Complexity reduction is not unique to \ac{ISAC}, but as \ac{ISAC} algorithms are to run on cost-effective hardware with low latency, it is of special interest.

\Ac{MBL}~\cite{shlezinger2023model} merges traditional model-based signal processing with \ac{DL} by parameterizing traditional model-based signal processing to overcome the limitations of conventional methods. 
Assuming $g(\mat{D})$ is a function with no learnable parameters (e.g., a conventional model-based algorithm), it can be enhanced with learnable parameters $\mat{\Theta}$ to $\tilde{g}(\mat{D},\mat{\Theta})$. How the classical and trainable components are entangled is up to the designer, often some tuning parameters of algorithms, such as weighting parameters, are made trainable.
The function $g$ and $\tilde{g}$ must be differentiable to enable backpropagation and learning of $\mat{\Theta}$. 
By incorporating domain knowledge, such as conventional receiver structures and algorithms, into the design of the \ac{NN} architecture, this approach improves interpretability while improving training and inference efficiency and reducing the amount of training data required.
Instead of training a \ac{DNN} to perform a specific task from scratch, \ac{ML} is employed to optimize the performance of existing algorithms, e.g., by tuning parameters to particular channel conditions and environments. A special type of \ac{MBL} is deep unfolding, where $g$ is an iterative algorithm, where each iteration is unrolled in order to finetune parameters $\mat{\Theta}$ for each iteration.

Methods like \ac{MBL} and deep unfolding limit the amount of trainable parameters, while being able to achieve the same performance as pure \ac{ML} solutions in some scenarios~\cite{shlezinger2023model}. Choosing the algorithm structure and which parts of a classical algorithm benefit from being enhanced with trainable features is vital to achieve sleek and powerful solutions for complex problems.
In~\cite{jiang2024isac}, deep unfolding is used for bistatic joint sensing. 
This work optimizes the hyper-parameters of an \ac{ISAC} algorithm for range-velocity estimation and communication demodulation, and then compares performance and complexity of this approach with other classical and trainable algorithms.

From a hardware-implementation point of view, complexity is a major indicator for cost of implementation and resource consumption.
As an alternative \ac{ML} paradigm, spiking \acp{NN} promise increased power efficiency in combination with neuromorphic hardware. 
In contrast to classical \acp{NN}, spiking \acp{NN} have emerged as a paradigm of neuromorphic computing promising higher energy efficiency~\cite{Eshraghian2023}. Spiking \acp{NN} are designed to emulate the function of biological neurons by processing discrete spike events. As biological neurons are known to be very energy efficient, emulating them promises complex \ac{AI} models with a fraction of the power consumption compared to current models. A neuron can have multiple inputs that contribute to the so-called membrane potential. When the membrane potential reaches a threshold, a spike is output by the neuron and the membrane potential is reset. Therefore, certain features such as clock-less computing, an inherent memory and a promise of high energy efficiency if implemented in hardware have attracted the attention of multiple research groups. In spiking \acp{NN}, different features of these neurons are made trainable.
We refer the reader to~\cite{Eshraghian2023} for a thorough introduction to spiking \acp{NN}.
In~\cite{chen2023neuromorphic}, spiking \acp{NN} are developed to perform object detection and communication demodulation in an \ac{ISAC} system. Although the paradigm shift to temporal spikes can be challenging, some applications such as impulse radio or pulse-based sensing can potentially be represented more easily in this signal space.

\section{Passive Object Sensing}\label{sec:passive_object_sensing}
One entirely new capability which \ac{ISAC} enables in future wireless communication networks will be the radar-like sensing of passive objects and the capability to get environmental properties of the propagation environment.
To this end, the wireless signal is exploited to identify reflections of radar sensing targets or to identify other environmental properties.
Different flavors of \ac{ISAC} radar sensing exist, ranging from transmitting dedicated sensing signals alongside communication signals to the full reuse of the available communication signals~\cite{JRC_Shatov, thomae2019}.

\begin{figure}
    \centering
    \def\svgwidth{0.82\columnwidth}
    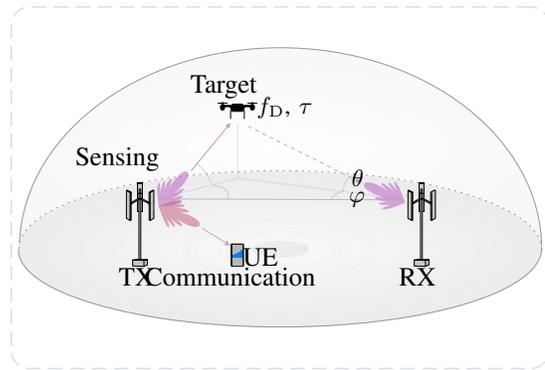
    \caption{A generalized ISAC setup for spatial sensing. Here TX sends a signal to sense the target at RX and simultaneously communicates with the \ac{UE}.}
    \label{fig:isac-spatial-general}
\end{figure}

The general sensing scenario for such an \ac{ISAC} network is shown in Fig.~\ref{fig:isac-spatial-general}.
After capturing the reflected signals at one or multiple sensing receivers, the wireless channel is estimated and utilized in further signal processing steps to estimate targets or environmental properties. 
The foreseen deployments of \ac{ISAC} in mono- and multistatic setups, the operation with standard communication hardware, and the constraint of economic feasibility create challenging tasks for the sensing signal processing.
Alongside solving the actual sensing task with often limited spectral resources compared to dedicated radar systems, the design choices of the communication system impose additional constraints. Compared to the high integration of \ac{ISAC}, conventional sensing solutions are employed in dedicated systems, where they are developed, tuned, and calibrated jointly with the dedicated measurement system to achieve their optimal operation points. 
Due to the seemingly uncountable and unforeseeable deployment scenarios of \ac{ISAC} sensing setups, a full site-specific development is infeasible, and therefore \ac{ISAC} mandates algorithmic solutions that are generalizable and computationally efficient.

In this section, we demonstrate how \ac{ML} methods can be applied to \ac{ISAC} systems to meet these challenges.
Specifically, we highlight the trade-offs between sensing and communication, detection, estimation, tracking, and classification in \ac{ISAC} radar sensing, as well as challenges in environmental and distributed sensing.
Each part provides a brief introduction to the specific challenges, conventional solutions, issues, and presents novel \ac{ML} solutions.
We specifically explore the role of \ac{ML} under the new paradigm shifts in \ac{ISAC} and highlight the associated computational complexity and flexibility. 
\subsection{Communication-Sensing Trade-Off}
The fundamental \ac{ISAC} challenges arise from the combination of two historically separate systems.
In this section, we consider the trade-offs unique to \ac{ISAC} which arise either from resource sharing or communication and sensing with the same \ac{TX} signal.
Fundamentally, \ac{ISAC} systems are subject to both the deterministic trade-off and the subspace or power trade-off~\cite{Xiong2023}.\par

The deterministic trade-off refers to the distribution of the modulations symbols.
On the one hand, communication systems are designed to maximize mutual information between two nodes. With sufficient \ac{SNR},  higher-order modulations are preferred, such as \ac{QAM}. The transmit sequence needs to be random in order to carry information. 
On the other hand, sensing systems perform optimally with symbols with equal power on each subcarrier, such as \ac{QPSK}. 
This results in conflicting optimization goals that can vary in importance, as the desired quality of service for both systems can change over time.  

The subspace or power trade-off refers to the transmit resources such as power or bandwidth being shared between sensing and communication tasks. In a general sense, it manifests as a resource allocation problem.
In systems where time or frequency resource blocks are allocated to either communication or sensing, the main indicator of sensing performance is the amount of resources reserved for sensing. 
However, there is a performance gap between time/frequency sharing methods and known performance bounds~\cite{Xiong2023}, which is difficult to close with rigid resource sharing methods. 
\Ac{ML} can be used to explore trade-off spaces and find good operating points that meet design criteria for both functionalities or provide adaptive operating points. 
Additionally, it is of interest to shed light on the synergies of both functionalities and optimize the transmit signal. 
Conventional\footnote{Please note we use \emph{conventional} solely to refer to existing non-\ac{ML} solutions for simplicity.} optimization methods can be computationally expensive and inflexible, whereas \ac{ML} can provide \textit{good-enough} approximations, often at lower computational cost.
The \ac{ISAC} resource allocation problem is addressed in~\cite{Yang2024a} and~\cite{Liu2025} using~\ac{RL}.

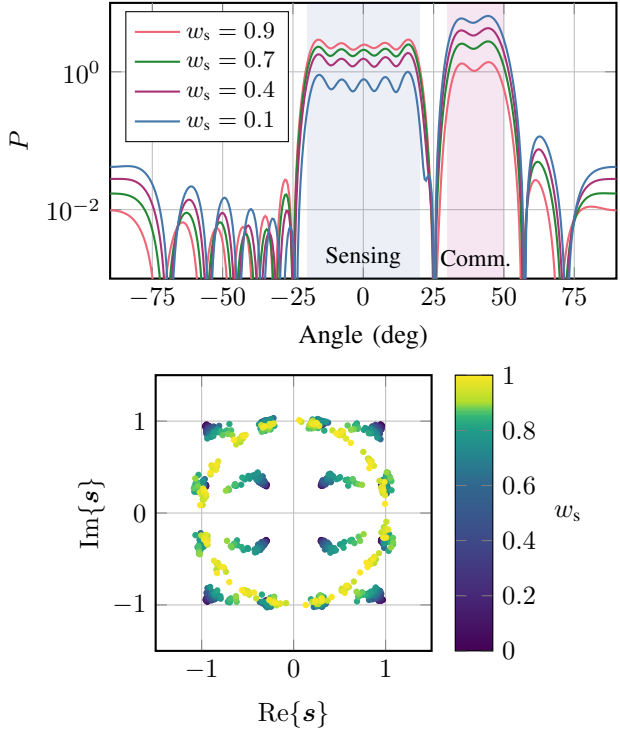
\begin{figure}
\pgfplotsset{
colormap={custom}{
samples of colormap={
10 of viridis,
target pos={0,200,400,600,700,800,850,875,900,1000},
}}}
\centering
\begin{subfigure}[c]{\columnwidth}
\vspace{4mm}
	\begin{tikzpicture}
		\begin{semilogyaxis}[
			xlabel=Angle (deg), ylabel=$P$,
			legend entries={Sensing,Communication, Sum}, 
			legend cell align={left},
			legend pos=north west,
			xmin=-90,xmax=90,
			ymin=0.001,ymax=10,
			axis line style=thick,
                width=0.95\columnwidth,
                height=0.6\columnwidth,
                extra x ticks={-25, 25, 75,-75},
                tick align=inside,
                grid=major,
                legend style={font=\small},
			]
            \fill[fill=KITblue, fill opacity=0.1] (axis cs: -20,0.001) rectangle (axis cs: 20,10);
            \fill[fill=KITpurple, fill opacity=0.1] (axis cs: 30,0.001) rectangle (axis cs: 50,10);
            \addlegendimage{thick, solid, cb-2} \addlegendentry{$w_{\text{s}}=0.9$}
		\addlegendimage{thick, solid, cb-3} \addlegendentry{$w_{\text{s}}=0.7$}
            \addlegendimage{thick, solid, cb-6} \addlegendentry{$w_{\text{s}}=0.4$}
            \addlegendimage{thick, solid, cb-1} \addlegendentry{$w_{\text{s}}=0.1$}
            \addplot [color=cb-2, thick] table [x expr= {\thisrow{angle}*180/pi}, y expr = {\thisrow{NN0.9}}]
		{\figures/ISAC/tradeoff/ws_beams.txt};
            \addplot [color=cb-3, thick] table [x expr= {\thisrow{angle}*180/pi}, y expr = {\thisrow{NN0.7}}]
		{\figures/ISAC/tradeoff/ws_beams.txt};
            \addplot [color=cb-6, thick] table [x expr= {\thisrow{angle}*180/pi}, y expr = {\thisrow{NN0.4}}]
		{\figures/ISAC/tradeoff/ws_beams.txt};
            \addplot [color=cb-1, thick] table [x expr= {\thisrow{angle}*180/pi}, y expr = {\thisrow{NN0.1}}]
		{\figures/ISAC/tradeoff/ws_beams.txt};
            \draw [black] (axis cs:0,0.002) node {\small Sensing};
            \draw [black] (axis cs:40,0.002) node {\small Comm.};
\end{semilogyaxis}
\end{tikzpicture}
\end{subfigure}

\begin{subfigure}[b]{1\columnwidth}
 \centering
	\begin{tikzpicture}
		\begin{axis}[
			xlabel=$\Re\{\vect{s}\}$, ylabel=$\Im\{\vect{s}\}$,
			grid=major,
            width=0.6\columnwidth,
            height=0.6\columnwidth,
			legend cell align={left},
			legend pos=north east,
			xmin=-1.5,xmax=1.5,
			ymin=-1.5,ymax=1.5,
			axis line style=thick,
                x label style={at={(axis description cs:0.5,-0.15)},anchor=north},
                y label style={at={(axis description cs:-0.15,.5)},anchor=south},
                colorbar,
                colormap name=custom,
                point meta max=1, 
                point meta min=0,
                cycle list={[samples of colormap={100 of custom}]},
                colorbar style={
                    ylabel=$w_{\text{s}}$,
                    ylabel style={rotate=-90},
                    },
			]
            \foreach \n in {1,2,...,100}{
            \addplot+ [
            only marks,
            mark size=1pt,
        ] table [x expr = \thisrow{1.0I}*cos(-14.8)+\thisrow{1.0Q}*sin(-14.8), y expr = -\thisrow{1.0I}*sin(-14.8)+\thisrow{1.0Q}*cos(-14.8)] {\figures/ISAC/tradeoff/ws_fine/\n.txt};
        };
		\end{axis}
	\end{tikzpicture}


\end{subfigure}
    \caption{Examples how power trade-off or deterministic tradeoff can manifest in an \ac{ISAC} system from~\cite{muth2024loss}.}
    \label{fig:isac-tradeoff}
\end{figure}

To illustrate the power trade-off, we use the example of beamforming.
Depending on the sensing task at hand, different requirements on the beam patterns can be imposed by defining an area of interest for sensing. This area should be illuminated with a certain amount of power to enable detection of targets, follow the movement of a target, or scan different sectors consecutively. For communication, we can also define an area of interest as the area where communication \acp{RX} can be present. Different approaches aim to optimize the \ac{SNR} at the communication \ac{RX}, directly minimize the \ac{BER} or optimize a different indicator for communication performance.
In this example, the beamforming weights $b_{\ell}$ are the output of an \ac{NN} and are optimized in an end-to-end \ac{ISAC} system with the loss term
\begin{align}
    L(\mat{\Theta}) = (1-w_{\text{s}})\mathcal{L}_{\text{BCE}}^{\text{commun}}(\mat{\Theta}) + w_{\text{s}}\left(\mathcal{L}_{\text{BCE}}^{\text{detect}}(\mat{\Theta}) + \mathcal{L}_{\text{MSE}}^{\text{angle}}(\mat{\Theta})\right),
    \label{eq:isac-tradeoffs}
\end{align}
with the \ac{NN} parameters $\mat{\Theta}$, trade-off weight $w_{\text{s}}$ and the loss terms referring to the definitions in \eqref{MSE_eq}-\eqref{BCE} applied to the tasks of communication, detection and angle estimation.
In Fig.~\ref{fig:isac-tradeoff}, the resulting beam patterns are shown. The \ac{ISAC} system trades off power between the area of interest for sensing and for communication. 
Depending on the importance of sensing indicated by $w_{\text{s}}$, power gets radiated either toward the sensing area or toward the communication \ac{RX}. 

Multiple research groups have addressed this trade-off using \ac{ML} methods.
In~\cite{MateosRamos2021, muth2023autoencoder, muth2024loss}, static beam patterns were obtained using a \ac{DNN} in an end-to-end fashion, with the \ac{MSE} of the \ac{AoA} estimation, the \ac{CE} for target detection and the \ac{MI} between \ac{TX} and \ac{RX} contributing to the loss term. 
In addition to common output normalization at the \ac{TX}, legal output constraints of a wireless transmission are considered in~\cite{fontanesi2024deep} for beamforming in \ac{UE}-centric sensing. 
The scenario in~\cite{fontanesi2024deep} consists of beamforming for sensing and communications with a choice of suitable base stations and generally leads to a non-convex optimization problem. 
A predictive beamforming approach is realized by a \ac{DNN}, using the absolute error between the targeted side lobe level and effective isotropic radiated power in the loss function.
As a last example, hybrid beamforming algorithms with multiple objectives are implemented in~\cite{Nguyen2023} using deep unfolding. 
A classical iterative optimizer with a fixed number of iterations is transformed into a \ac{DNN}, enabling fast convergence.

Concerning the deterministic trade-off, classical solutions to find good transmit signals have been hand-designed or limited to constant-modulus transmit signals to maximize sensing performance while enabling communication. 
Constellation shaping is being explored as a method to set flexible deterministic trade-offs, often being implemented using a \ac{DNN} mapping the input bit to constellation symbols. 
Probabilistic constellation shaping, which uses QAM constellations but attributes different probabilities to different constellation points, is used to jointly optimize communication and sensing performance in~\cite{Yang2024}. 
Geometric constellation shaping, where the position of the constellation points is optimized, is illustrated in Fig.~\ref{fig:isac-tradeoff} for $16$ constellation points. 
For high $w_{\text{s}}$, as defined in \eqref{eq:isac-tradeoffs}, the constellation tends towards a 16-PSK (sensing optimal) and with a trade-off tending toward communication with lower $w_{\text{s}}$ it resembles a 16-QAM. 
Constellation shaping has been analyzed in more detail in~\cite{geiger2025jointshaping} for \ac{OFDM}-\ac{ISAC} systems. 
The authors showed analytically that constellation kurtosis is the deciding factor in possible sensing performance, independent of the used shaping method. 
The deterministic trade-off can therefore be flexibly used to improve either communication or sensing performance compared to classical methods. 
Therefore, a loss function optimizing the communication performance with an added constraint for the kurtosis can be used instead of an end-to-end evaluation of the communication and sensing performance, respectively.

\subsection{Range-Doppler Detection}\label{sec:passive-sensing:range-doppler-detection}
\begin{figure}[t]
    \def\mathdefault#1{#1}
    \begin{center}
         \scalebox{0.75 }{
    \input{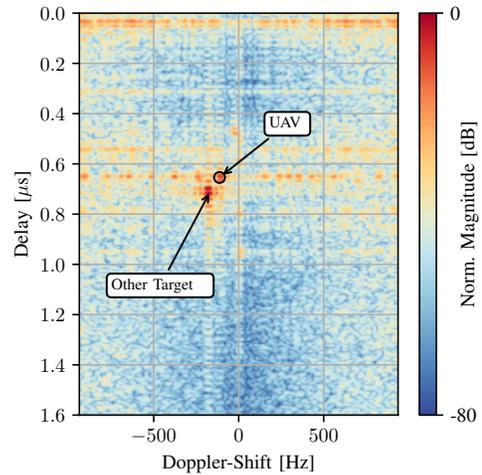}}
    \end{center}
    \caption{\ac{RDM} of an \ac{ISAC} outdoor measurement from~\cite{Beuster2023}. Strong clutter masks target visibility and limit target detection and estimation. A background subtraction removed clutter for better target visibility (UAV).}
    \label{fig:periodogram}
\end{figure}

Target detection refers to the decision if or how many targets are present in a given area. 
It is commonly used the \ac{RDM}, which describes the channels $2$D power profile in range (measured as delay) and Doppler shift, exemplified in Fig.~\ref{fig:periodogram}. 
The detection task is to identify which regions contain power reflected from a target, while keeping false alarms from clutter and noise to a minimum.

Generally, any waveform can yield an estimate of the \ac{RDM} which is commonly used for estimating target parameters. We show the calculation based on the \ac{OFDM} waveform which is a main contender of becoming the \ac{ISAC} waveform for 6G.
The \ac{RDM} is obtained from a series of $M$ transmit $\mat{X}\in \mathbb{C}^{K\times M}$ and received $\mat{R}\in \mathbb{C}^{K\times M}$ \ac{OFDM} symbols with $K$ subcarriers after cyclic prefix removal.
The $M$ channel transfer function estimates are obtained by element-wise division of $\mat{R}$ and $\mat{X}$. 
Then the \ac{RDM} is
\begin{align} \label{eq:rdm}
    \mathrm{RDM} =  \vert \bm{F}^*_K \left(\mat{R} \oslash \mat{X} \right)  \bm{F}_M^{\top} \vert^2, 
\end{align}
with $\bm{F}_K$ and $\bm{F}_M$ denoting the $K$- and $M$-point DFT unitary matrices, respectively. 
This is to say, we perform a column-wise IDFT to transform the frequency to delay domain, and row-wise DFT to transform the time to Doppler-shift domain. 

A statistically optimal detector in \ac{AWGN} is a Neyman-Pearson detector, which tests if a target is present at the channel estimate $\hat{h}$ at the discrete time delay $n$.
It can be formulated as~\cite[Chap. 10]{Trees2002}:
\begin{align}
    \frac{2}{\sigma_{\text{n}}^2} |\hat{h}_{n}|^2 \quad\mathop{\gtreqless}_{\hat{T}=0}^{\hat{T}=1}\quad \chi^2_{2}(1-P_{\text{f}}),
\end{align}
where $\chi^2_{2}(\cdot)$ denotes the chi-squared distribution with $2$ degrees of freedom, $P_{\text{f}}$ is the false alarm rate of detection and $\sigma_{\text{n}}^2$ denotes the noise power.
While the Neyman-Pearson detector is statistically optimal, it requires an explicit estimate of the noise power, which is difficult to obtain in practice. 
Hence, more sophisticated scenarios with extended targets and clutter often utilize more practical \ac{CFAR} approaches, such as cell-averaging- or ordered-statistics \ac{CFAR}, which also provide a constant false alarm rate under \ac{AWGN} noise.
With \ac{CFAR} the noise power for the decision threshold is estimated using training cells patterned around the cell under test, often accompanied by guard cells.

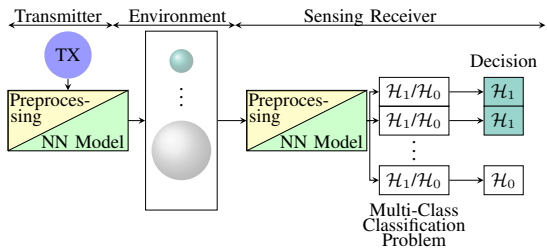
\begin{figure}
\centering
\begin{adjustbox}{width=0.4\textwidth}
\begin{tikzpicture}
\draw [shift={(0.775,5)},color=black] node[below] (tl) {$\text{ \small Transmitter}$};
\draw [shift={(2.775,5)},color=black] node[below] {$\text{ \small Environment}$};
\draw [shift={(6,5)},color=black] node[below] {$\text{ \small Sensing Receiver}$};

\draw[stealth-stealth] (0,4.6) -- (1.75,4.6);
\draw[stealth-stealth] (1.75,4.6) -- (3.8,4.6);
\draw[stealth-stealth] (3.8,4.6) -- (9,4.6);

\node[circle, fill=blue!40!white, minimum width=0.2cm, minimum height=0.2cm, align=center] (tx) at (1,4.1) {\small{TX}};

\node[rectangle, draw, minimum height=1cm, minimum width=2cm] (prnn1) at (1,3) {};
\draw [draw=black, fill=yellow, fill opacity=0.2]
       (prnn1.south west)  -- (prnn1.north west) -- (prnn1.north east);
\draw [draw=black, fill=green!20!white]
       (prnn1.south west) -- (prnn1.north east) -- (prnn1.south east)--(prnn1.south west);
       
\draw [shift={($(prnn1)+(-0.3,0.32)$)},color=black] node {$\text{\small Preproces-}$};
\draw [shift={($(prnn1)+(-0.7,0.05)$)},color=black] node {$\text{\small sing}$};
\draw [shift={($(prnn1)+(0.25,-0.35)$)},color=black] node {$\text{\small  NN Model}$};

\node[rectangle,minimum width=1.2cm, minimum height=3cm,draw] (env) at (2.9,3) {};
\shade[ball color = KITgreen, opacity = 0.4] (2.9,4) circle (0.2cm);
\node (d) at (2.9,3.5){\vdots};
\shade[ball color = gray!40, opacity = 0.4] (2.9,2.5) circle (0.5cm);

\draw [-stealth](tx) -- (prnn1);
\draw [-stealth](prnn1) -- (env);

\node[rectangle, draw, minimum height=1cm, minimum width=2cm] (prnn) at (5,3) {};
\draw [draw=black, fill=yellow, fill opacity=0.2]
       (prnn.south west)  -- (prnn.north west) -- (prnn.north east);
\draw [draw=black, fill=green!20!white]
       (prnn.south west) -- (prnn.north east) -- (prnn.south east)--(prnn.south west);

\draw [shift={($(prnn)+(-0.3,0.32)$)},color=black] node {$\text{\small Preproces-}$};
\draw [shift={($(prnn)+(-0.7,0.05)$)},color=black] node {$\text{\small sing}$};
\draw [shift={($(prnn)+(0.25,-0.35)$)},color=black] node {$\text{\small  NN Model}$};

\draw[-stealth] (env) -- (prnn);

\node[rectangle, minimum width=0.5cm, minimum height=0.2cm, align=center,draw] (h2) at (6.8,3){\small{$\mathcal{H}_1$}/{$\mathcal{H}_0$}};
\node[rectangle, minimum width=0.5cm, minimum height=0.2cm, anchor=south, draw] (h1) at ($(h2.north)-(0,0.5px)$){\small{$\mathcal{H}_1$}/{$\mathcal{H}_0$}};
\node (h) at ($(h2)-(0,0.4)$) {$\vdots$};
\node[rectangle, minimum width=0.5cm, minimum height=0.2cm, anchor=center,draw] (h3) at ($(h2)-(0,1)$){\small{$\mathcal{H}_1$}/{$\mathcal{H}_0$}};

\draw[-stealth] (prnn.east) -- ++(0.05,0) |- (h1.west);
\draw[-stealth] (prnn) -- (h2);
\draw[-stealth] (prnn.east) -- ++(0.05,0) |- (h3.west);

\node[minimum width=0.5cm, minimum height=0.15cm, align=center] at (6.8,1.5) {\text{\small Multi-Class}};
\node[minimum width=0.5cm, minimum height=0.15cm, align=center] at (6.8,1.25) {\text{\small Classification}};
\node[minimum width=0.5cm, minimum height=0.15cm, align=center] at (6.8,1) {\text{\small Problem}};
\node[minimum width=0.5cm, minimum height=0.15cm, align=center] at (8.3,4) {\text{\small Decision}};

\node[rectangle, draw, minimum width=0.5cm, minimum height=0.15cm, fill= KITgreen!40] (d1) at ($(h1)+(1.5,0)$) {\text{\small $\mathcal{H}_1$}};
\node[rectangle, draw, minimum width=0.5cm, minimum height=0.15cm, align=center, fill= KITgreen!40] (d2) at ($(h2)+(1.5,0)$) {\text{\small $\mathcal{H}_1$}};
\node[rectangle, draw, minimum width=0.5cm, minimum height=0.15cm, align=center] (d3) at ($(h3)+(1.5,0)$) {\text{\small $\mathcal{H}_0$}};

\draw [-stealth](h1) -- (d1);
\draw [-stealth](h2) -- (d2);
\draw [-stealth](h3) -- (d3);
\end{tikzpicture}
\end{adjustbox}
\caption{Generalized framework for learning-based detection. The different targets (spheres in the environment) are represented by classes encoded on a grid, which is e.g., aligned with the bins of the \ac{RDM}. Targets are detected by classifying each point in the grid as target or no-target. By extension, this approach is also suitable for multi-class classification of different target types.}
\label{fig:detection-learning}
\end{figure}

However, \ac{CFAR} methods also observe performance impacts, when their statistical assumptions, e.g., about the noise, are not matched.
For example, the estimate of the noise level is affected, when other targets are inside the training cells or the target extends over multiple cells.
These also require additional filtering techniques, such as windowing, to avoid suboptimal performance.
This not only causes increased false alarms but also prohibits the detection of closely-spaced targets.
Other issues stem from false-detections caused by distortions from mismatched transmit signal structure, imperfect hardware calibration, clutter, or ghost targets.
All of which are challenging to address and require additional specific filtering that is only applicable if the corresponding distortion source is known.
Lastly, the distinct extended target shapes or micro-Doppler can also cause multiple \ac{CFAR} detections of the same target, resulting in additional estimation and tracking efforts.
To combat those challenges, conventional algorithms require accurate modeling and subsequent, real-time applicable, filtering procedures to overcome these issues.
Comparably, \ac{ML} solutions may be suitable to overcome many of the presented challenges.

\ac{ML} methods for detection can be generally described using Fig.~\ref{fig:detection-learning}. Each output of the model outputs is interpreted as a detector which is compared to a certain threshold during inference to decide if a target is detected. This allows general multi-target detection or joint detection and estimation if each class is associated with a certain estimation bin. \ac{ML} components can be deployed both at the \ac{TX} and at the sensing \ac{RX}.
For example, in~\cite{jiang2019end} the waveform and detector are trained to approach a Neyman-Pearson detector, which is optimal in white noise. 
While no a priori information about the target and clutter model is assumed, their results show the trained solution achieves a more robust detection rate in clutter and colored noise compared to conventional approaches.
Furthermore, \cite{major2019vehicle} employs \ac{DL} to successfully differentiate targets, such as vehicles, from clutter based on the \ac{RDM}. 
It underlines how detection and object classification can be solved jointly, as a \ac{RDM} may contain distinct target features which are lost in later processing stages.
This is also underlined in~\cite{Dong2024}, where the features of space debris targets are exploited for detection, via a combination of support vector machines and \ac{CSI}-based sensing.
Notably, the algorithm is trained based on simulated scenarios and measurement data to achieve accurate detection and classification of the debris type.
In~\cite{Tosi2025}, the authors compare different \ac{CFAR} versions with the \ac{CNN}-method from~\cite{Schieler2025} trained on simulation data.
Their results highlight the \ac{ML} models ability to separate closely-spaced targets in scenarios with similar magnitude, while otherwise performing similar to the \ac{CFAR} methods. 
The discussion further emphasizes the additional need for further research to strengthen the understanding of the training procedures and simulation data models on the final model performance.
In~\cite{muth2023autoencoder} considering multiple targets and~\cite{muth2024loss} considering multiple snapshots, the comparison with a trainable detector and a classical statistical detector is performed. 
The simulations show it is possible to overcome the performance of generally optimal detectors without increasing false alarm rate, by using training data with scenario-specific features (e.g., derived from the area of interest).

\subsection{Estimation}
The estimation follows directly after a target is detected, with the goal to estimate the parameters of the propagation path caused by the target.  
Afterwards, the target can be localized by computing the possible locations with respect to the \ac{TX} and \ac{RX} positions.
Therefore, the estimations' goal is to find the propagation path parameters of the target reflection with higher accuracy compared to the grid-based detections, to improve the overall sensing accuracy.
Generally, the propagation path parameters relate a target's geometric position to its interaction with the impinging and reflected wave.
The geometry of the propagation, illustrated in Fig.~\ref{fig:isac-spatial-general}. The reflection from a target in the signal model is described by its delay $\tau_{\ell}$, Doppler-shift $f_{\mathrm{D},\ell}$, and \ac{DoA} in azimuth $\varphi$ and elevation $\theta$.
While the \ac{DoA} estimation typically requires antenna arrays at the \ac{RX} (unless \ac{SAR} is used), both $\tau_{\ell}$ and $f_{\mathrm{D},\ell}$ can be estimated from any \ac{SISO} link.
To model the wireless channel at the \ac{RX} it is commonly assumed, that each path impinging on the \ac{RX} is parametrized by its magnitude and phase, as well as $\tau_{\ell}$, $f_{\mathrm{D},\ell}$, $\varphi$ and $\theta$.
However, due to the linearity of the channel, the measurement represents the superposition of all planar waves, which presents a challenging inverse problem for the estimation.
E.g., the paths can be insufficiently separated, even with strongly different magnitudes (strong masks weak), as the propagation paths in the channel depend solely on the propagation scenario.

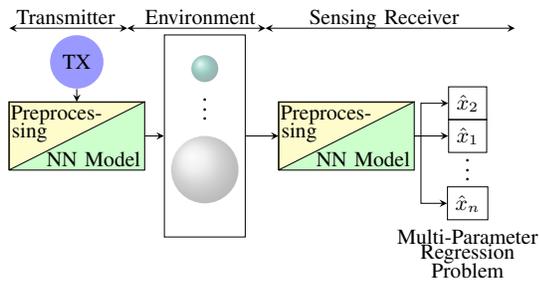
\begin{figure}
\centering
\begin{adjustbox}{width=0.4\textwidth}
\begin{tikzpicture}
\draw [shift={(0.775,5)},color=black] node[below] (tl) {$\text{ \small Transmitter}$};
\draw [shift={(2.775,5)},color=black] node[below] {$\text{ \small Environment}$};
\draw [shift={(5.5,5)},color=black] node[below] {$\text{ \small Sensing Receiver}$};

\draw[stealth-stealth] (0,4.6) -- (1.75,4.6);
\draw[stealth-stealth] (1.75,4.6) -- (3.8,4.6);
\draw[stealth-stealth] (3.8,4.6) -- (7.5,4.6);

\node[circle, fill=blue!40!white, minimum width=0.2cm, minimum height=0.2cm, align=center] (tx) at (1,4.1) {\small{TX}};

\node[rectangle, draw, minimum height=1cm, minimum width=2cm] (prnn1) at (1,3) {};
\draw [draw=black, fill=yellow, fill opacity=0.2]
       (prnn1.south west)  -- (prnn1.north west) -- (prnn1.north east);
\draw [draw=black, fill=green!20!white]
       (prnn1.south west) -- (prnn1.north east) -- (prnn1.south east)--(prnn1.south west);
       
\draw [shift={($(prnn1)+(-0.3,0.32)$)},color=black] node {$\text{\small Preproces-}$};
\draw [shift={($(prnn1)+(-0.7,0.05)$)},color=black] node {$\text{\small sing}$};
\draw [shift={($(prnn1)+(0.25,-0.35)$)},color=black] node {$\text{\small  NN Model}$};

\node[rectangle,minimum width=1.2cm, minimum height=3cm,draw] (env) at (2.9,3) {};
\shade[ball color = KITgreen, opacity = 0.4] (2.9,4) circle (0.2cm);
\node (d) at (2.9,3.5){\vdots};
\shade[ball color = gray!40, opacity = 0.4] (2.9,2.5) circle (0.5cm);

\draw [-stealth](tx) -- (prnn1);
\draw [-stealth](prnn1) -- (env);

\node[rectangle, draw, minimum height=1cm, minimum width=2cm] (prnn) at (5,3) {};
\draw [draw=black, fill=yellow, fill opacity=0.2]
       (prnn.south west)  -- (prnn.north west) -- (prnn.north east);
\draw [draw=black, fill=green!20!white]
       (prnn.south west) -- (prnn.north east) -- (prnn.south east)--(prnn.south west);

\draw [shift={($(prnn)+(-0.3,0.32)$)},color=black] node {$\text{\small Preproces-}$};
\draw [shift={($(prnn)+(-0.7,0.05)$)},color=black] node {$\text{\small sing}$};
\draw [shift={($(prnn)+(0.25,-0.35)$)},color=black] node {$\text{\small  NN Model}$};

\draw[-stealth] (env) -- (prnn);

\node[rectangle, minimum width=0.5cm, minimum height=0.2cm, align=center,draw] (h2) at (6.8,3){\small{$\hat{x}_1$}};
\node[rectangle, minimum width=0.5cm, minimum height=0.2cm, anchor=south, draw] (h1) at ($(h2.north)-(0,0.5px)$){\small{$\hat{x}_2$}};
\node (h) at ($(h2)-(0,0.4)$) {$\vdots$};
\node[rectangle, minimum width=0.5cm, minimum height=0.2cm, anchor=center,draw] (h3) at ($(h2)-(0,1)$){\small{$\hat{x}_n$}};

\draw[-stealth] (prnn.east) -- ++(0.1,0) |- (h1.west);
\draw[-stealth] (prnn) -- (h2);
\draw[-stealth] (prnn.east) -- ++(0.1,0) |- (h3.west);

\node[minimum width=0.5cm, minimum height=0.15cm, align=center] at (6.8,1.5) {\text{\small Multi-Parameter}};
\node[minimum width=0.5cm, minimum height=0.15cm, align=center] at (6.8,1.25) {\text{\small Regression}};
\node[minimum width=0.5cm, minimum height=0.15cm, align=center] at (6.8,1) {\text{\small Problem}};

\end{tikzpicture}
\end{adjustbox}
\caption{Example framework for learning-based estimation. Compared to the multi-class classification task in Fig.~\ref{fig:detection-learning}, the model is trained for a regression task to enable continuous parameter estimates.}
\label{fig:est-learning}
\end{figure}

Parameter estimation in \ac{ML} usually relies either on joint detection as shown in Fig.~\ref{fig:detection-learning} or on training a system on a regression problem as shown in Fig.~\ref{fig:est-learning}. 
In the first option, the targets are estimated on a grid, limiting the resolution of the estimation to the grid spacing. Comparably, regression enables grid-less estimation, but requires separate estimation of the model-order, e.g., the number of targets, and a learnable parameter permutation/order. Both approaches may also be combined, similar to well-known computer-vision models~\cite{redmon2016lookonceunifiedrealtime}, and implemented in \cite{schielerEstimation2025}.

Conventional algorithms include simple axis-reading of the peaks detected in the \ac{RDM}, and more advanced methods like \ac{MUSIC}, \ac{ESPRIT}, \ac{CS}, and \ac{MLE} approaches.
\ac{ISAC}-suitable solutions must be real-time capable, provide sufficient accuracy and target separability under the spectral resource constraints of mobile communication networks, and handle the plethora of different hardware measurement setups, for which synchronization and calibration are only possible within the limits of the communication system.
For example, iterative \ac{MLE} estimators like RIMAX \cite{richterEstimation2005} achieve the limits of the \ac{CRB}, but at a high runtime and significant computational cost, which is challenging to reduce~\cite{semper2024}. 
Other challenges for conventional methods are imperfections arising from non-ideal measurement hardware effects, such as non-linearities, quantization, and imperfect calibration.
Inevitably, such mismatches degrade the performance of algorithms reliant on the idealized signal models.
Extended targets present another issue, as they appear as a superposition of insufficiently separated paths (in terms of sampling resolution) to an estimator based on~\eqref{continuous_time_channel}. 
As a consequence, these paths create an additional overhead in the estimation, and possibly leveraging them for classification (as possible with \ac{ML}, see~\ref{sec:passive-sensing:range-doppler-detection}) requires additional hand-designed filtering. 

As noted before, many flavors of the estimation task share the same mathematical structure, and as a result, publications in the area stem from many different scientific communities and often contribute to domain specific solutions.
Therefore we decided to split the presentation of \ac{ML} approaches accordingly and present them in the context of the estimation domain they are presented in.
We pick \ac{AoA} estimation in a single dimension to introduce the mathematical background of harmonic retrieval problems, and then show the advances in different multidimensional estimation fields, focusing on the contributions of \ac{ML}.
We note that it is generally beneficial to perform a joint detection and estimation, as this always provides the optimal separation simply from more available dimensions for the separation, and hence detection, of targets.
E.g., a joint detection in range-Doppler-azimuth-elevation will yield better results compared to a range-only estimation. 
However, this is rarely possible under given hardware and real-time sensing constraints, as large chunks of data must be processed jointly, creating a need for lower-complexity algorithmic approximations.
We therefore also emphasize these \ac{ML} contributions where applicable.

\subsubsection{Angle of Arrival}
As described in \eqref{continuous_time_channel_node}, 
a waveform impinging on an antenna may yield a phase difference in the signal captured by different RXs. 
This phase difference is related to the angle $\theta$ from which the waveform comes. 
The problem of estimating $\theta$ from the received signal is commonly known as \ac{AoA} estimation.
Assume an array antenna 
estimating the \ac{AoA} from $M$ targets in the environment. 
Consider that the targets are in the far-field region of the antenna and that each target emits a signal $s_m(t)$ from an angle $\theta_m, m=0,\ldots,M-1$. 
Starting from the model in \eqref{continuous_time_channel_node}, consider (\textit{i}) that the targets have a \ac{LoS} path with the antenna array, i.e., $L(t)=1$ and (\textit{ii}) the delay from each target to the antenna array can be estimated and compensated for, i.e., $\tau_0(t) = 0$. Then, the channel model for each target is
\begin{align}
    h_i(t,\tau) = \alpha(t) \delta(t-\tau_{i,0}(t))e^{-\j 2\pi f_c \tau_{i,0}(t)}.
\end{align}
Here, $\tau_{i,0}(t)=id\sin(\theta(t))/c$, $i=0,\ldots,K-1$, where $d$ is the distance between RXs, $c$ is the speed of light, and $K$ is the number of sensors. 
In the case of a narrowband signal $s(t-\tau_{i,0})\approx s(t)$.\footnote{Extensions for wideband signals have been proposed, but we skip their description for simplicity.}
Following \eqref{ReceiverSignal} and considering the superposition of signals from different targets, the output signal at the $i$-th RX is
\begin{align} \label{eq:y_i(t)_angle}
    y_i(t) = \sum_{m=0}^{M-1} \alpha_m(t) s_m(t) e^{-\j 2\pi i\frac{d}{\lambda}\sin(\theta_m(t))} + n(t).
\end{align}
The continuous-time received signal in \eqref{eq:y_i(t)_angle} is sampled and  $T$ transmissions (often called \textit{snapshots}) can be collected for further processing. 
Aggregating the received samples across antenna elements and snapshots, the discrete-time received signal $\bm{Y}$ can be expressed as:
\begin{align} \label{eq:rx_signal_array_doa}
    \bm{Y} = \bm{A}(\theta)\bm{S} + \bm{N},
\end{align}
where $\bm{A} = [\bm{a}(\theta_1), \ldots, \bm{a}(\theta_M)]\in\complexset[K][M]$ is a matrix containing steering vectors of the form
\begin{align}
    \bm{a}(\theta) = [1, e^{-\j 2 \pi d \sin{(\theta)}/\lambda}, \ldots, e^{-\j 2 \pi d (K-1) \sin{(\theta)}/\lambda}]. 
\end{align}
The sampled received signal for each target and snapshot are aggregated in $\bm{S} = [\bm{s}(1), \ldots, \bm{s}(T)]$, where the complex channel gain $\alpha_m$ is absorbed in $s_m$ and $\bm{N}$ represents the additive noise.

There are different kinds of \ac{AoA} estimation methods: (\textit{i}) \ac{MLE} methods, (\textit{ii}) subspace methods, and (\textit{iii}) \ac{CS} methods. 
\ac{MLE} methods attempt to solve the following optimization problem:
\begin{align}
    \hat{\bm{\theta}} = \arg\max_{\bm{\theta}} \mathfrak{L}(\bm{\theta}|\bm{Y}),
\end{align}
where $\bm{\theta}=[\theta_1,\ldots,\theta_M]^\top$. 
The function $\mathfrak{L}(\bm{\theta}|\bm{Y})$ is the probability function of AoAs given the received signal, called the  \textit{likelihood function}. 
The likelihood function is in most cases multimodal, which hinders the implementation of MLE methods \cite{viberg91detection}. 
Several approaches have been proposed to solve this problem \cite{viberg91detection, stoica90maximum, sharman89genetic}, but they require high \ac{SNR}, large computational costs, 
or a small number of targets to estimate.

However, a common assumption of all \ac{AoA} estimation methods is that the model of \eqref{eq:rx_signal_array_doa} holds. 
Under hardware impairments, methods degrade in performance due to model mismatch.
In this context, \ac{ML} can be harnessed to reduce the complexity of \ac{CS} methods or to improve the estimation of subspace methods under few snapshots, coherent sources, low SNRs and modeling mismatches.
The first step to \ac{AoA} estimation with \ac{ML} is to feed the received signal $\bm{Y}$ as input to a \ac{NN} and obtain the \ac{AoA}, i.e., following the notation of Sec.~\ref{sec:ml-basics}, $\hat{\theta}~=~f(\bm{Y};\mat{\Theta})$.
To improve the angular resolution of the estimated \ac{AoA}, the field-of-view of the receive antenna is discretized in angular candidates $\{\theta_i\}_{i=1}^N$ and the output of the \ac{NN} is a probability vector that represents the probability that each angular candidate is the true angle, i.e., $\hat{\theta} = \sum_{i=1}^K \theta_i \hat{\bm{p}}_i$,
with $\hat{\bm{p}} = f(\bm{Y};\mat{\Theta})$. 
The \ac{AoA} estimation problem is regarded as a multi-class classification problem and the categorical \ac{CE} loss is used to train the \ac{NN} \cite{zheng2024deep}. 
An extension of this approach is to discretize a smaller angular sector instead of the entire field-of-view, which provides a more accurate angular estimation \cite{ma2022deep}. 
A similar way to consider a discrete angular grid as output to the NN is to compute the spatial spectrum, defined as $|\bm{a}^T(\theta)\hat{\bm{p}}|$ for different values of $\theta$. 
The estimated spatial spectrum computed from $\hat{\bm{p}}$ can be compared with the true spatial spectrum from the true \ac{AoA} \cite{chen2024sdoa}. 
Other ML methods are inspired by subspace methods, which use an  estimated covariance matrix of the received signal $\bmRY~=~\bm{Y}\bm{Y}^{\hermit}/T$ to perform AoA. By definition, $\bmRY$ is Hermitian. The upper triangular part of $\bmRY$ contains all the information about $\bmRY$ and it can be used as input to a NN instead. 
Another approach is to use a vectorized and normalized version of the covariance matrix \cite{zhu2020two}, i.e., 
\begin{align}
    \bm{r} = [&\mathcal{R}\{{\bm{R}_{1,2}}\}, \mathcal{I}\{{\bm{R}_{1,2}}\}, \ldots, \mathcal{R}\{{\bm{R}_{N,N}}\}, \mathcal{I}\{{\bm{R}_{N,N}}\}].
\end{align}
The input to a MLP can be $\bm{v} = \bm{r}/\lVert \bm{r} \rVert^2$ \cite{chen2022robust}. 
Moreover, the vector $\bm{v}$ can be reshaped back to a matrix and use a CNN to process the covariance matrix. 
The \ac{AoA} can be directly computed as the output of the CNN or estimated with a probability vector based on a grid of candidate \acp{AoA} \cite{chen2022robust, yu2023deep}. 

Most limitations of subspace methods result from the mismatch between the true and estimated covariance matrix $\bmRY$.
To improve the estimate of $\bmRY$, a CNN is trained to use an initial estimate of the covariance matrix (or its upper-triangular part) to produce a more accurate estimate, suitable root-MUSIC AoA estimation~\cite{shmuel2023subspacenet}. 
Another approach is to compute the MUSIC spectrum and then a MLP yields the estimated AoA~\cite{merkofer2024damusic}. 
Instead of training an estimation model, another direction in \ac{MBL} is to learn efficient calibration procedures instead.
In particular, for the case of hardware impairments, learning the parameters subject to impairments allows to calibrate the system. 
For instance, considering gain and position impairments in the ULA, the impaired steering vector becomes $[\bm{\tilde{a}}(\theta)]_k = g_k \exp(\jmath 2 \pi p_k \sin(\theta))$,
where $g_k$ and $p_k$ are the unknown gains and positions of the elements of the ULA. 
Considering as learnable parameters $\mat{\Theta} = \{g_k, p_k\}_{k=1}^K$ and using a subspace method that enables backpropagation, the performance of the impaired system can be improved \cite{chatelier2025physically}.
As a summary, in Table~\ref{tab:DoA_estimation}, we collect the most recent AoA estimation works, highlighting the problems from classical AoA estimation that they address. 

\begin{table*}[tb]
\centering
\scriptsize
\renewcommand{\arraystretch}{1.2} 
\caption{ML-based AoA estimation works.}
\label{tab:DoA_estimation}
\begin{tabular}{l|l|l|l|l|l|l|l|l}
\hline\hline
 & Ref. & ML method & \begin{tabular}[c]{@{}l@{}}Low \\ SNR\end{tabular} & \begin{tabular}[c]{@{}l@{}}Few\\ Snapshots\end{tabular} & \begin{tabular}[c]{@{}l@{}}Complexity \\ reduction\end{tabular} & \begin{tabular}[c]{@{}l@{}}Hardware \\ impairments\end{tabular} & \begin{tabular}[c]{@{}l@{}}Coherent \\ sources\end{tabular} & \begin{tabular}[c]{@{}l@{}}Broad-band\\ signals\end{tabular} \\ \hline
\multirow{8}{*}{DL} 
 & \cite{zheng2024deep} & CNN & \checkmark & \checkmark &  &  &  &  \\\cline{2-9}
 & \cite{ma2022deep} & CNN &  & \checkmark &  &  &  &  \\
 \cline{2-9}
 & \cite{chen2024sdoa} & CNN &  &  & \checkmark  & \checkmark &  &  \\
 \cline{2-9}
 & \cite{zhu2020two} & Ensemble DL &  &  & \checkmark &  &  &  \\\cline{2-9}
  & \cite{chen2022robust} & \begin{tabular}[c]{@{}l@{}}Denoising\\ autoencoder\end{tabular} &  &  &  & \checkmark &  &  \\ \cline{2-9}
   & \cite{yu2023deep} & CNN & \checkmark &  &  &  &  &  \\ 
\hline
\multirow{4}{*}{MBL} & \cite{shmuel2023subspacenet} & \begin{tabular}[c]{@{}l@{}}Data-augmented\\ root-MUSIC\end{tabular} & 
 \checkmark & \checkmark & \checkmark & \checkmark & \checkmark & \checkmark \\ \cline{2-9}
 & \cite{merkofer2024damusic} & \begin{tabular}[c]{@{}l@{}}Data-augmented\\ MUSIC\end{tabular} & \checkmark  & \checkmark & \checkmark & \checkmark & \checkmark & \checkmark \\ \cline{2-9}
 & \cite{chatelier2025physically} & \begin{tabular}[c]{@{}l@{}}Data-augmented\\ MUSIC\end{tabular} &   &  &  & \checkmark &  &  \\
 \hline\hline
\end{tabular}
\end{table*}

\subsubsection{Range-Doppler}
Estimating the range and Doppler-shift of the detected targets enables the determination of a target’s range and relative radial velocity. 
It is tightly coupled with the target detection (see~\ref{sec:passive-sensing:range-doppler-detection}), as both operate on the same data domain.
Typical algorithms include using peak positions from the \ac{RDM}, subspace methods like \ac{ESPRIT}, \ac{CS}, and \ac{MLE} approaches.
The computational complexity of \ac{MLE} methods is addressed in both \cite{Barthelme2021} and \cite{schielerGridFree2024}, which present approaches to alleviate it through a combination of conventional \ac{MLE} methods with \ac{ML}.
They illustrate that \acp{NN} can effectively initialize gradient-based \ac{MLE} methods, reducing runtime while maintaining accuracy. 
These hybrid approaches combine the efficiency of data-driven methods with the precision of model-based optimization, circumventing the need for exhaustive iterative searches.

Recent publications introduce \ac{ML}-based solutions, particulary such that are trained on model-based data \ac{MBL} instead of measurement data. 
Here the challenge is to design and train the model in a simulation environment such that it is suitable for the application in for actual real-world sensing applications. 
For example~\cite{schielerCNNMeasurement} demonstrates it is possible to train a \ac{CNN} for estimation on synthetic data and transfer it to measurement data, alleviating the need for costly measurement data collection.
It is noted, that models like the presented CNN could also be re-used for target classification based on the \ac{RDM} features similar to their use in computer vision applications. 
Traditional target classification typically involves a small set of classes~\cite{zyweck96radar, sorowka15pedestrian}, and hence, computer vision models receives interest due to their successful classification with many classes. 
However, it must be noted that sensing measurements as the \ac{RDM} provide less features for classification compared with camera recordings.
Yet, similar models are successfully applied to classify targets from radar signals, involving three classes in~\cite{gupta2021target}, four classes in~\cite{ulrich2021deep}, and six classes in \cite{patel2019deep}.

\subsubsection{Angle-Range}
\begin{figure}[tb]
    \centering
    \begin{tikzpicture}[font=\scriptsize, spy using outlines={circle, magnification=2.5, connect spies}]

\usetikzlibrary{spy}

\begin{axis}[
width=8cm,
height=3.5cm,
at={(0,0)},      
log basis x={10},
tick align=outside,
tick pos=left,
legend columns = 1,
legend style={
  fill opacity=1,
  draw opacity=1,
  text opacity=1,
  at={(0.37,1.1)},
  anchor=center,
  align=left,
  legend cell align={left}
},
x grid style={white!69.0196078431373!black},
xlabel={False alarm rate},
xmajorgrids,
xmin=1e-4, xmax=1,
xtick style={color=black},
y grid style={white!69.0196078431373!black},
ylabel={Position error [m]},
ymajorgrids,
ymin=0, ymax=2,
xmode=log,
ytick style={color=black},
]
\addplot[color=black, mark=*, dashed, mark options={scale=0.8}] table {figures/ISAC/angle-range/data_results/baseline_ideal_pfa_pos_error.txt};\addlegendentry{Baseline, known impairments};

\addplot[color=red!80!black, mark=diamond] table {figures/ISAC/angle-range/data_results/impairment_pfa_pos_error.txt}; \addlegendentry{Model-based learning};

\addplot[color=blue!80!black, dashdotted, mark=star, mark options={scale=0.8, solid}] table {figures/ISAC/angle-range/data_results/baseline_learn_pfa_pos_error.txt}; \addlegendentry{Model-based calibration};

\end{axis}

\end{tikzpicture}
     
    \caption{Position error as a function of the false alarm rate. The results are computed from synthesized data considering a MIMO-OFDM radar with 64 antenna elements and 256 subcarriers. A maximum number of 5 targets in the scene are assumed, considering a random number in each snapshot. Inter-antenna spacing impairments were considered in the ISAC transceiver.} 
    \label{fig:results_angle_range}
\end{figure}
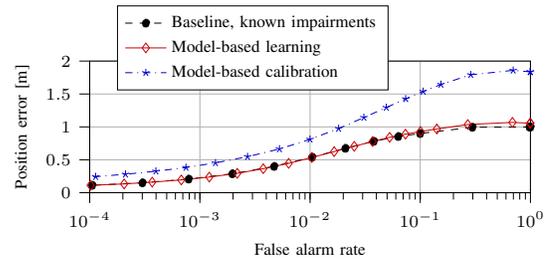

The position of a target can be estimated transmitting a signal that contains different frequency components and using the same signal for different antenna elements. 
At the \ac{RX} side, an \textit{angle-range} matrix can be computed from the received signal across antenna elements and frequencies similarly to the RDM of \eqref{eq:rdm}.
\Ac{ML} has been used in angle-range estimation to compensate for hardware impairments. 
Particularly, \ac{MBL} was harnessed in \cite{mateos2024model} to compensate for hardware impairments while performing angle-range estimation and target detection. 
The considered impairments are inter-antenna spacing impairments, similarly to 
$\bm{\tilde{a}}(\theta)$ and the learnable parameters are the positions of the antenna elements in the \ac{ULA}. 
In Fig.~\ref{fig:results_angle_range}, sensing results are represented as a function of the false alarm rate for the multi-target scenario of \cite{mateos2024model}. 
The MBL approach is compared with the baseline with perfect knowledge of the impairments and with a model-based calibration technique that requires no gradient backpropagation. 
Fig.~\ref{fig:results_angle_range} suggests that with MBL the performance of the system without impairments can be closely approached and with MBL the performance of model-based calibration can be outperformed.  
This work was extended in \cite{mateos2024semi} to reduce the amount of labeled data by designing an unsupervised loss function.
\subsubsection{Angle-Range-Doppler}
Angle-Range-Doppler estimation is a fundamental part of advanced driving assistance systems (ADAS). 
Light detection and ranging (LIDAR) is commonly used in ADAS since it is lightweight, cost-effective, and precise. 
However, the operation of LIDAR is severely affected by adverse weather conditions. 
Conversely, mmWave radar is more robust under harsh conditions at the expense of sparser clouds of points to detect targets. 
To improve the performance of mmWave radar for radar estimation, in \cite{hu2023radnet} the accuracy of existing standard processing methods was improved with \ac{DL}. 
Another use case of \ac{ML} in radar estimation is to improve the resilience of the estimation against jamming, demonstrated in~\cite{Benyahia2022squeeze}, where \ac{ML} is utilized to improve radar estimation against clutter and jamming.
A scheme for a joint 4D detection and estimation of targets is presented in~\cite{Schieler2023}, based on \ac{MBL} with data simulated for a \ac{URA}.
There the \ac{CNN} approach from~\cite{Schieler2022} is extended to jointly estimate the delay, Doppler-shift, azimuth, and elevation parameters of an unknown number of paths using simulation data. 
Their results show, that joint processing of 4D radar sensing data 
can enable significant runtime advantages and high accuracy when combined with \ac{MLE} methods.

\subsection{Tracking}
So far, we have considered that we obtain an estimate of the target position and velocity every time that one or multiple snapshots were transmitted. 
These estimates over time form the trajectory of each target in the scene. 
If target trajectories are considered instead of independent estimates, the estimate of the target position and velocity in the next time step can be improved. 
Tracking refers to the estimation of these trajectories.
The major challenge of tracking is to associate new measurements with object trajectories including: (\textit{i}) creating and removing trajectories
and (\textit{ii}) distinguishing clutter from targets. 
There have been several algorithms proposed for multi-object tracking, such as the probability hypothesis density~\cite{garcia19trajectory}, the generalized labeled multi-Bernoulli~\cite{vo19multi}, and the multi-scan trajectory Poisson multi-Bernoulli mixture~\cite{xia2019multi}. 
However, the unknown measurement-target association makes these approaches have a super-exponential complexity in the number of time steps to process~\cite{mahler2007statistical}.
\Ac{ML} can reduce the computational complexity of conventional tracking algorithms.
At the \ac{TX} side, tracking is used to predict the user position and perform beamforming, which can help in the design of narrower beams with larger antenna arrays to increase the data rate of users. 
Predictive and adaptive beamforming was studied in \cite{liu2022learning}, where an unsupervised approach with an \ac{LSTM} network is proposed. 
The work in \cite{wang2023deep} also uses an \ac{LSTM} network with the raw data instead of a two-step approach where the positions of targets are estimated and then beamforming is performed. 
Strategies to reduce the training overhead have also been proposed \cite{demirhan2022radar}.
At the \ac{RX} side, multi-object tracking was addressed with a tailored transformer in \cite{pinto2023deep}. 
\Ac{MBL} solutions have also been developed based on the Kalman filter (KF). 
The KF provides the \ac{MMSE} estimator for a time-varying discrete system under \ac{AWGN}. 
It is based on linear equations that describe how the system evolves over time, referred to as the state-space model.
However, many problems involve non-linear state-space models. 
Several solutions have been proposed, such as the extended KF \cite{gruber1967approach} and the unscented KF \cite{julier1997new}. 
The family of particle filters
\cite{gordon1993novel} was also introduced for non-linear, non-Gaussian state-space models, which obtains better results than the KF with higher computational complexity. 
The main weakness of the aforementioned filters is that they rely on the state-space model and filters exhibit performance degradation if the state-space model is not accurate. Several works deal with some level of uncertainty in the state-space model, but they cannot achieve the same performance as the filters with full knowledge.
\ac{MBL} methods can enhance conventional filters to adapt to unexpected deviations from the assumed state-space models. 
Regarding the KF, the work in \cite{revach2022kalmannet} augments the KF to deal with non-linearities and model mismatches. 
This approach was extended to make the \ac{NN} robust against changes in the model \cite{ni2024adaptive} or to quantify the uncertainty in the estimations by extracting the error covariance matrix \cite{dahan2024uncertainty}. 
The particle filter was also augmented with \ac{DL} to mitigate the model mismatch~\cite{nuri2024learning}.

\subsection{Gesture Recognition}
Apart from the propagation parameters discussed in previous sections, targets also possess other useful features, depending on the sensing application. 
For example, we already discussed applications for target classification, e.g. pedestrian, bike, car, from the radar sensing data. 
Here we focus on gesture recognition, particularly with multi-sensor fusion techniques. 

Gesture recognition is one of the most important aspects of human-computer interaction, with applications such as the development of hearing aids, monitoring of a patient's emotional state, or lie detection~\cite{mitra07gesture}. 
At first, gesture recognition was based on visual aids~\cite{na19deep} or wearable sensors \cite{mukhopadhyay15wearable}. 
However, these sensors exhibit issues under inappropriate light conditions and usability issues. 
Comparably, the use of radar signals is robust against light conditions and users do not need to carry sensors. 
While model-based algorithms rely on wave propagation theory and statistical models, requiring hand-crafted feature extraction~\cite{qi2023resource}, \ac{ML} can combine feature extraction and classification and be trained to jointly optimize both tasks. 
Gesture recognition under the lens of \ac{AI} was first considered using synthetic data. 
The work in \cite{qi2023resource} performs simulations with publicly available datasets and in \cite{du2020three}, the radar echoes are first transformed into a range-Doppler time points, which are further processed by \ac{DL}. 
But recently, \ac{ML} has also been successfully deployed in practical implementations. 
In \cite{bhat2023gesture}, mmWave access points were used to perform gesture recognition and in \cite{zhao2022angle}, the authors use a 4D imaging radar ---a type of radar that can measure elevation, azimuth, range, and Doppler of targets--- to classify between different human postures.

\subsection{Environmental Sensing}
Environmental sensing refers to sensing applications, where the material parameters or a particle distribution in the air is measured instead of estimating geometrical features. 
An early example is radar for weather forecasting which has been in commercial deployment since the 1950s. 
With different available bandwidths and \acp{DT} becoming more accessible, additional environmental features become available for sensing. 
\begin{figure}
    \centering
    \includegraphics[width=0.75\columnwidth]{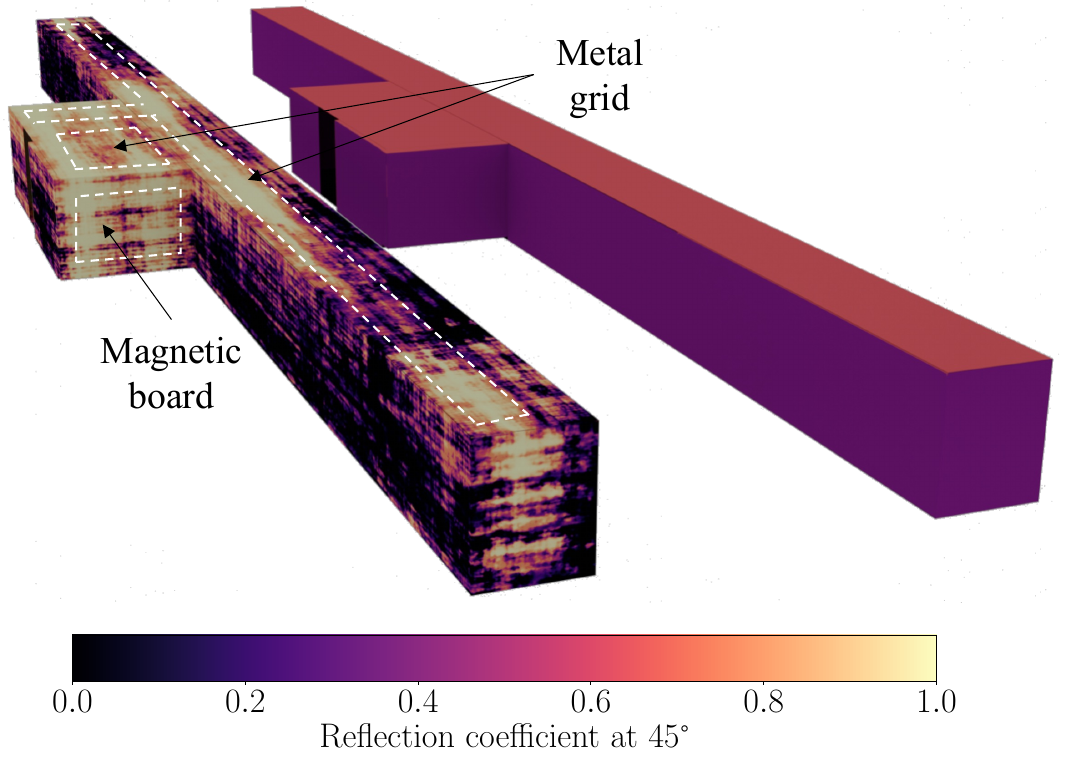}
    \caption{Learned materials from~\cite{Hoydis2023a} to obtain fine our rough representation of the environment. This graphic is under \faCreativeCommons\ \faCreativeCommonsBy CC-BY 4.0 International license. }
    \label{fig:material-sensing}
\end{figure}

A more recent example for weather prediction is given in~\cite{bouget2021fusion}, which uses fused radar images and wind forecasts to classify between different weather classes with an \ac{NN}. 
In this work, the open MeteoNet dataset was used for training with a \ac{CE} loss function to achieve more accurate predictions compared to conventional optical flow benchmarks. 
In~\cite{aidin2021deep}, the rain fade for satellite communications is predicted using \ac{DL} from satellite radar imagery, as well as link power measurements. 
The authors utilize a \ac{CNN} to predict the probability of rain fade from the satellite radar images. 
In the community of  weather prediction data fusion is generally well established as multiple different sensor types are available, including radars.
In the future, this set of sensors can be complemented by the \ac{ISAC} system to enhance weather prediction services.

In the field of material parameter sensing,~\cite{Helal2022} gives an overview of \ac{ML} algorithms for THz sensing to measure the parameters of solid and gaseous materials. 
Lastly, in~\cite{Hoydis2023a} parameter sensing based on a known geometry is proposed as a way to improve the accuracy of virtual twins. 
Through ray tracing, channels accurately representing the actual environment are simulated. It is still unclear how much detail is needed to build the environment and materials. 
In Fig.~\ref{fig:material-sensing}, a hallway is shown with the measured reflectivity of the walls, floor, and ceiling. 
Depending on the complexity of the targeted material model, the radiation environment can be approximated with few materials or on a fine grid. 

\subsection{Distributed Sensing}
Distributed sensing is generally of interest, as 6G networks are expected to provide more sensing nodes, including mobile nodes, such as \ac{UE}s. 
Different concepts for distributed sensing are discussed, such as a network composed of multistatic links~\cite{JRC_Shatov} and more advanced multi-sensor MIMO systems with multi-sensor coordination~\cite{ThomaeDallmann2023}.
Regardless, preprocessing the initial measurements of the wireless channel at the nodes, e.g., via target detection and estimation, is essential to avoid significant signaling overhead.
Therefore, we consider distributed sensing also as a fusion task, where estimates from multiple nodes must be jointly processed. 
The problem of distributed sensing is therefore twofold.
Firstly, sensing data representation must be communicated without too much overhead, and secondly, a fusion algorithm must be capable of estimating the localization results.
Such an algorithm must handle fusion of different links, frequency bands, and possibly even additional sensor types, such as cameras or LIDARs.
For such concepts, multimodal fusion based on \ac{ML} is proposed in~\cite{AI_ISAC_2024}.
However, to the best of our knowledge, distributed sensing for the fusion of different \ac{ISAC} nodes has not been demonstrated to benefit from \ac{ML} methods in the current literature. 
Many studies use classical fusion methods, but some apply \ac{ML} to more difficult scenarios~\cite{Zhuang2023}.
Nevertheless, \ac{ML} is a common tool in the data fusion literature~\cite{MENG2020115} and might show successful applications when distributed sensing tasks receive more attention.

\section{Spectrum and Emitter Analysis}\label{sec:spectral_sens}

This section addresses sensing and localization tasks performed by sensors without an illumination source.
More specifically, we focus on \ac{ML}-powered methods for dealing with non-cooperative \acp{TX} and aim to answer the question of who is transmitting from which location, which is crucial, for example, for spectrum management, resource allocation, and jammer detection.
Often overlooked in the context of \ac{ISAC}, these tasks are essential in \ac{CR} and cognitive radar, and their broader implementation could greatly improve the intelligence and resilience of future wireless networks.

\subsection{Spectrum Sensing}

Spectrum sensing aims to detect underutilized bands in the monitored portion of the spectrum.
The concepts of spectrum sensing and spectrum sharing underpin ideas of \ac{CR}~\cite{Wideband_SS_CR_2008} and cognitive radar~\cite{CognitiveRadarOverview_2019}.
Specifically, Listen-Before-Talk (LBT) function is a part of many \ac{MAC} protocols in wireless communications.
LBT is a key mechanism for the sharing of spectrum in an uncoordinated way, detecting primary signals and spectrum opportunities~\cite{LBT_2016}.
The current trend toward the coexistence of various wireless technologies in the same spectrum imposes even stricter requirements for their coordination~\cite{Martone2021}. 
Traditional detectors like energy detectors, matched filters, and cyclic feature detectors achieve good performance under ideal conditions where the assumed models of channels and signals are accurate and known in advance. 
However, this assumption is often unrealistic in practical \ac{CR} systems, resulting in problems caused by model mismatches that can degrade performance.
To overcome these issues, the potential of AI-enhanced methods for spectrum sensing in the context of \ac{CR} became evident early on~\cite{EarlySS2008}, followed by reviews at the beginning of the 2010s~\cite{CR_AI_Survey2010},\cite{CR_AI_Survey2013}.

We begin by formulating the problem of wideband spectrum sensing, which is crucial for both current and future wireless networks as it enables multi-band opportunistic channel access. 
Single-band spectrum sensing can be viewed as a special case of this.
In wideband sensing, an \ac{RX} monitors the spectrum that consists of $N_f$ frequency sub-bands.
The signal $\mathbf{s}_{\text{RX}}$ received at the \ac{RX} front-end is expressed as shown in (\ref{ReceiverSignal}). 
Subsequently, to extract the spectrum features of the received signal, power spectrum density (PSD) is computed by applying the Fourier transform to the autocorrelation function, i.e., 

\begin{equation}
\mathbf{S_{\text{RX}_{n}}} = \mathrm{FFT}[\mathrm{Corr}(\mathbf{s}_{\text{RX}_{n}})], 
\label{eq:SpectrumSensingPSD}
\end{equation}
where $\mathrm{FFT}[\cdot]$ denotes the fast Fourier transform, and $\mathrm{Corr}(\cdot)$ is the signal autocorrelation.
Note that since the time domain signal from (\ref{ReceiverSignal}) contains noise, the spectrum $\mathbf{S_{\text{RX}_{n}}}$ is also noisy.
The sub-band measurements are then concatenated to represent the full band spectrum $\mathbf{S_{\text{RX}}}$ as

$\mathbf{S_{\text{RX}}} = [\mathbf{S}_{\text{RX}_{1}}, \mathbf{S}_{\text{RX}_{2}},...,\mathbf{S}_{\text{RX}_{n}},...,\mathbf{S}_{\text{RX}_{N_f}}]$.

The detector's task is to decide on each sub-band's $n$ occupancy based on noisy measurement vector $\mathbf{S_{\text{RX}}}$. 
The decision represents a binary hypothesis testing problem formulated as

\begin{equation}
f(\mathbf{S}_{\text{RX}_{n}})\underset{\mathcal{H}_0}{\overset{\mathcal{H}_1}{\gtreqless}} \gamma,
\label{eq:SpectrumSensingHypothesisTest}
\end{equation}
where $f(\cdot)$ is some function of measurement $\mathbf{S}_{\text{RX}_{n}}$, $\mathcal{H}_0$ and $\mathcal{H}_1$ denote the absence and presence of (primary) signals, respectively, and $\gamma$ is the decision threshold. 
The value of $\gamma$ is associated with the detection probability, with the default choice being 0.5. 
In practice, it has to be tuned to balance the importance of missed detection and false alarm events for a particular application.
Fig.~\ref{fig:SpectrumSensing} shows the wideband spectrum sensing scenario, where only a few sub-bands are occupied. 

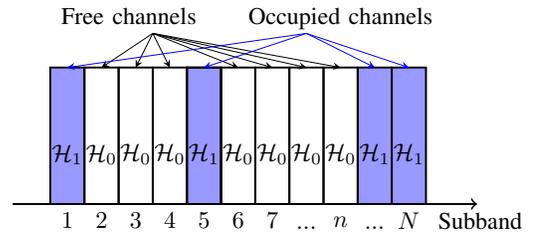
\begin{figure}
\centering
\begin{adjustbox}{width=0.38\textwidth}
\begin{tikzpicture}
\draw[thick,->] (0.45,1) -- (7.25,1) node[anchor=north west,xshift=-2em] {Subband};
\fill[blue!40!white, draw=black, thick] (1,1) rectangle (1.5,3);
\fill[blue!40!white, draw=black, thick] (3,1) rectangle (3.5,3);
\fill[blue!40!white, draw=black, thick] (5.5,1) rectangle (6,3);
\fill[blue!40!white, draw=black, thick] (6,1) rectangle (6.5,3);
\draw[thick] (1.5,1) rectangle (2,3);
\draw[thick] (2,1) rectangle (2.5,3);
\draw[thick] (2.5,1) rectangle (3,3);
\draw[thick] (3.5,1) rectangle (4,3);
\draw[thick] (3,1) rectangle (3.5,3);
\draw[thick] (4,1) rectangle (4.5,3);
\draw[thick] (4.5,1) rectangle (5,3);
\draw[thick] (5,1) rectangle (5.5,3);
\draw [shift={(1.25,1)},color=black] node[below] {$1$};
\draw [shift={(1.75,1)},color=black] node[below] {$2$};
\draw [shift={(2.25,1)},color=black] node[below] {$3$};
\draw [shift={(2.75,1)},color=black] node[below] {$4$};
\draw [shift={(3.25,1)},color=black] node[below] {$5$};
\draw [shift={(3.75,1)},color=black] node[below] {$6$};
\draw [shift={(4.25,1)},color=black] node[below] {$7$};
\draw [shift={(4.75,0.8)},color=black] node[below] {$...$};
\draw [shift={(5.25,0.94)},color=black] node[below] {$n$};
\draw [shift={(5.75,0.8)},color=black] node[below] {$...$};
\draw [shift={(6.25,1)},color=black] node[below] {$N$};
\draw [shift={(1.25,2)},color=black] node[below, thick] {$\mathcal{H}_1$};
\draw [shift={(1.75,2)},color=black] node[below, thick] {$\mathcal{H}_0$};
\draw [shift={(2.25,2)},color=black] node[below] {$\mathcal{H}_0$};
\draw [shift={(2.75,2)},color=black] node[below] {$\mathcal{H}_0$};
\draw [shift={(3.25,2)},color=black] node[below] {$\mathcal{H}_1$};
\draw [shift={(3.75,2)},color=black] node[below] {$\mathcal{H}_0$};
\draw [shift={(4.25,2)},color=black] node[below] {$\mathcal{H}_0$};
\draw [shift={(4.75,2)},color=black] node[below] {$\mathcal{H}_0$};
\draw [shift={(5.25,2)},color=black] node[below] {$\mathcal{H}_0$};
\draw [shift={(5.75,2)},color=black] node[below] {$\mathcal{H}_1$};
\draw [shift={(6.25,2)},color=black] node[below] {$\mathcal{H}_1$};
\draw [shift={(2.15,4)},color=black] node[below] {$\text{Free channels}$};
\draw[-stealth] (2.5,3.5) -- (1.75,3);
\draw[-stealth] (2.5,3.5) -- (2.25,3);
\draw[-stealth] (2.5,3.5) -- (2.75,3);
\draw[-stealth] (2.5,3.5) -- (3.75,3);
\draw[-stealth] (2.5,3.5) -- (4.25,3);
\draw[-stealth] (2.5,3.5) -- (4.75,3);
\draw[-stealth] (2.5,3.5) -- (5.25,3);
\draw [shift={(5.25,4)},color=black] node[below] {$\text{Occupied channels}$};
\draw[-stealth,blue] (4.75,3.5) -- (1.25,3);
\draw[-stealth,blue] (4.75,3.5) -- (3.25,3);
\draw[-stealth,blue] (4.75,3.5) -- (5.75,3);
\draw[-stealth,blue] (4.75,3.5) -- (6.25,3);
\end{tikzpicture}
\end{adjustbox}
\caption{An illustration of a multi-band spectrum sensing. Free and occupied sub-channels are marked with '$\mathcal{H}_0$' and '$\mathcal{H}_1$' labels, respectively.}
\label{fig:SpectrumSensing}
\end{figure}

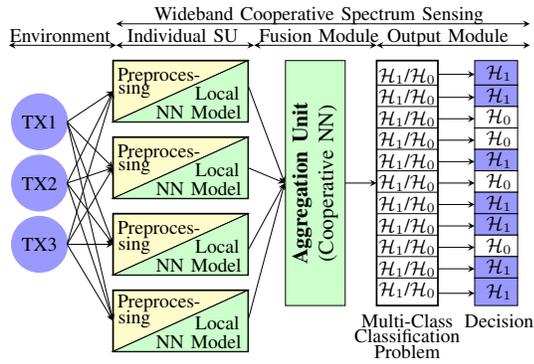
\begin{figure}
\centering
\begin{adjustbox}{width=0.4\textwidth}
\begin{tikzpicture}
\draw [shift={(5.,6)},color=black] node[below] {$\text{ \small Wideband Cooperative Spectrum Sensing}$};
\draw [shift={(0.775,5.65)},color=black] node[below] {$\text{ \small Environment}$};
\draw [shift={(2.775,5.65)},color=black] node[below] {$\text{ \small Individual SU}$};
\draw [shift={(5,5.65)},color=black] node[below] {$\text{ \small Fusion Module}$};
\draw [shift={(7.1,5.65)},color=black] node[below] {$\text{ \small Output Module}$};
\draw[stealth-stealth] (1.75,5.6) -- (8.5,5.6);
\draw[stealth-stealth] (0,5.25) -- (1.75,5.25);
\draw[stealth-stealth] (1.75,5.25) -- (4.,5.25);
\draw[stealth-stealth] (4.,5.25) -- (6.,5.25);
\draw[stealth-stealth] (6,5.25) -- (8.5,5.25);
\node[circle, fill=blue!40!white, minimum width=0.2cm, minimum height=0.2cm, align=center] at (0.5,4) {\small{TX1}};
\node[circle, fill=blue!40!white, minimum width=0.2cm, minimum height=0.2cm, align=center] at (0.5,3) {\small{TX2}};
\node[circle, fill=blue!40!white, minimum width=0.2cm, minimum height=0.2cm, align=center] at (0.5,2) {\small{TX3}};

\draw [-stealth](0.95,2) -- (1.7,0.75);
\draw [-stealth](0.95,2) -- (1.7,2);
\draw [-stealth](0.95,2) -- (1.7,3.25);
\draw [-stealth](0.95,2) -- (1.7,4.5);

\draw [-stealth](0.95,3) -- (1.7,0.75);
\draw [-stealth](0.95,3) -- (1.7,2);
\draw [-stealth](0.95,3) -- (1.7,3.25);
\draw [-stealth](0.95,3) -- (1.7,4.5);

\draw [-stealth](0.95,4) -- (1.7,0.75);
\draw [-stealth](0.95,4) -- (1.7,2);
\draw [-stealth](0.95,4) -- (1.7,3.25);
\draw [-stealth](0.95,4) -- (1.7,4.5);

\draw [-stealth](3.9,0.75) -- (4.5,3);
\draw [-stealth](3.9,1.9) -- (4.5,3);
\draw [-stealth](3.9,3.25) -- (4.5,3);
\draw [-stealth](3.9,4.6) -- (4.5,3);

\draw [-stealth](5,3) -- (6,3);

\draw[thick] (1.7,0.25) rectangle (3.9,1.25);
\draw[thick] (1.7,1.5) rectangle (3.9,2.5);
\draw[thick] (1.7,2.75) rectangle (3.9,3.75);
\draw[thick] (1.7,4) rectangle (3.9,5);
\draw [draw=black, fill=yellow, fill opacity=0.2]
       (1.7,0.25) -- (1.7,1.25) -- (3.9,1.25) -- cycle;
\draw [draw=black, fill=green!20!white]
       (1.7,0.25) -- (3.9,0.25) -- (3.9,1.25) -- cycle;
\draw [draw=black, fill=yellow, fill opacity=0.2]
       (1.7,1.5) -- (1.7,2.5) -- (3.9,2.5) -- cycle;
\draw [draw=black, fill=green!20!white]
       (1.7,1.5) -- (3.9,1.5) -- (3.9,2.5) -- cycle;
\draw [draw=black, fill=yellow, fill opacity=0.2]
       (1.7,2.75) -- (1.7,3.75) -- (3.9,3.75) -- cycle;
\draw [draw=black, fill=green!20!white]
       (1.7,2.75) -- (3.9,2.75) -- (3.9,3.75) -- cycle;
\draw [draw=black, fill=yellow, fill opacity=0.2]
       (1.7,4) -- (1.7,5) -- (3.9,5) -- cycle;
\draw [draw=black, fill=green!20!white]
       (1.7,4) -- (3.9,4) -- (3.9,5) -- cycle;
\draw [shift={(2.45,1.3)},color=black] node[below] {$\text{\small Preproces-}$};
\draw [shift={(2.05,1.05)},color=black] node[below] {$\text{\small sing}$};
\draw [shift={(3.4,0.95)},color=black] node[below] {$\text{\small Local}$};
\draw [shift={(3.1,0.65)},color=black] node[below] {$\text{\small  NN Model}$};

\draw [shift={(2.45,2.5)},color=black] node[below] {$\text{\small Preproces-}$};
\draw [shift={(2.05,2.25)},color=black] node[below] {$\text{\small sing}$};
\draw [shift={(3.4,2.2)},color=black] node[below] {$\text{\small Local}$};
\draw [shift={(3.1,1.9)},color=black] node[below] {$\text{\small  NN Model}$};

\draw [shift={(2.45,3.75)},color=black] node[below] {$\text{\small Preproces-}$};
\draw [shift={(2.05,3.55)},color=black] node[below] {$\text{\small sing}$};
\draw [shift={(3.4,3.45)},color=black] node[below] {$\text{\small Local}$};
\draw [shift={(3.1,3.15)},color=black] node[below] {$\text{\small  NN Model}$};

\draw [shift={(2.45,5)},color=black] node[below] {$\text{\small Preproces-}$};
\draw [shift={(2.05,4.8)},color=black] node[below] {$\text{\small sing}$};
\draw [shift={(3.4,4.7)},color=black] node[below] {$\text{\small Local}$};
\draw [shift={(3.1,4.4)},color=black] node[below] {$\text{\small  NN Model}$};

\node[rectangle, draw=black, fill=green!20!white, rotate=90, minimum width=4cm, minimum height=0.75cm, align=center] at (5,3) {\textbf{Aggregation Unit}\\ (Cooperative NN)};

\draw[thick] (6,1) rectangle (7,5);

\node[minimum width=0.5cm, minimum height=0.15cm, align=center] at (6.5,0.75) {\text{\small Multi-Class}};
\node[minimum width=0.5cm, minimum height=0.15cm, align=center] at (6.5,0.5) {\text{\small Classification}};
\node[minimum width=0.5cm, minimum height=0.15cm, align=center] at (6.5,0.25) {\text{\small Problem}};
\node[minimum width=0.5cm, minimum height=0.15cm, align=center] at (8.,0.75) {\text{\small Decision}};

\node[rectangle, minimum width=0.5cm, minimum height=0.25cm, align=center] at (6.5,1.25){\small{$\mathcal{H}_1$}/{$\mathcal{H}_0$}};
\node[rectangle, minimum width=0.5cm, minimum height=0.25cm, align=center] at (6.5,1.6){\small{$\mathcal{H}_1$}/{$\mathcal{H}_0$}};
\node[rectangle, minimum width=0.5cm, minimum height=0.25cm, align=center] at (6.5,1.95){\small{$\mathcal{H}_1$}/{$\mathcal{H}_0$}};
\node[rectangle, minimum width=0.5cm, minimum height=0.25cm, align=center] at (6.5,2.3){\small{$\mathcal{H}_1$}/{$\mathcal{H}_0$}};
\node[rectangle, minimum width=0.5cm, minimum height=0.25cm, align=center] at (6.5,2.65){\small{$\mathcal{H}_1$}/{$\mathcal{H}_0$}};
\node[rectangle, minimum width=0.5cm, minimum height=0.25cm, align=center] at (6.5,3){\small{$\mathcal{H}_1$}/{$\mathcal{H}_0$}};
\node[rectangle, minimum width=0.5cm, minimum height=0.25cm, align=center] at (6.5,3.35){\small{$\mathcal{H}_1$}/{$\mathcal{H}_0$}};
\node[rectangle, minimum width=0.5cm, minimum height=0.25cm, align=center] at (6.5,3.7){\small{$\mathcal{H}_1$}/{$\mathcal{H}_0$}};
\node[rectangle, minimum width=0.5cm, minimum height=0.25cm, align=center] at (6.5,4.05){\small{$\mathcal{H}_1$}/{$\mathcal{H}_0$}};
\node[rectangle, minimum width=0.5cm, minimum height=0.25cm, align=center] at (6.5,4.4){\small{$\mathcal{H}_1$}/{$\mathcal{H}_0$}};
\node[rectangle, minimum width=0.5cm, minimum height=0.25cm, align=center] at (6.5,4.75){\small{$\mathcal{H}_1$}/{$\mathcal{H}_0$}};

\draw (6,1.45) -- (7,1.45);
\draw (6,1.8) -- (7,1.8);
\draw (6,2.15) -- (7,2.15);
\draw (6,2.5) -- (7,2.5);
\draw (6,2.85) -- (7,2.85);
\draw (6,3.2) -- (7,3.2);
\draw (6,3.55) -- (7,3.55);
\draw (6,3.9) -- (7,3.9);
\draw (6,4.25) -- (7,4.25);
\draw (6,4.6) -- (7,4.6);

\draw [draw=black, fill=blue!40!white]
       (7.6,4.6) -- (8.3,4.6) -- (8.3,5) -- (7.6,5) -- cycle;
\draw [draw=black, fill=blue!40!white]
       (7.6,4.25) -- (8.3,4.25) -- (8.3,4.6) -- (7.6,4.6) -- cycle;
\draw [draw=black, fill=blue!0!white]
       (7.6,3.9) -- (8.3,3.9) -- (8.3,4.25) -- (7.6,4.25) -- cycle;
\draw [draw=black, fill=blue!0!white]
       (7.6,3.55) -- (8.3,3.55) -- (8.3,3.9) -- (7.6,3.9) -- cycle;
\draw [draw=black, fill=blue!40!white]
       (7.6,3.2) -- (8.3,3.2) -- (8.3,3.55) -- (7.6,3.55) -- cycle;
\draw [draw=black, fill=blue!0!white]
       (7.6,2.85) -- (8.3,2.85) -- (8.3,3.2) -- (7.6,3.2) -- cycle;
\draw [draw=black, fill=blue!40!white]
       (7.6,2.5) -- (8.3,2.5) -- (8.3,2.85) -- (7.6,2.85) -- cycle;
\draw [draw=black, fill=blue!40!white]
       (7.6,2.15) -- (8.3,2.15) -- (8.3,2.5) -- (7.6,2.5) -- cycle;
\draw [draw=black, fill=blue!0!white]
       (7.6,1.8) -- (8.3,1.8) -- (8.3,2.15) -- (7.6,2.15) -- cycle;
\draw [draw=black, fill=blue!40!white]
       (7.6,1.45) -- (8.3,1.45) -- (8.3,1.8) -- (7.6,1.8) -- cycle;
\draw [draw=black, fill=blue!40!white]
       (7.6,1.0) -- (8.3,1.0) -- (8.3,1.45) -- (7.6,1.45) -- cycle;

\node[minimum width=0.5cm, minimum height=0.15cm, align=center] at (8,4.8) {\text{\small $\mathcal{H}_1$}};
\node[minimum width=0.5cm, minimum height=0.15cm, align=center] at (8,4.425) {\text{\small $\mathcal{H}_1$}};
\node[minimum width=0.5cm, minimum height=0.15cm, align=center] at (8,4.075) {\text{\small $\mathcal{H}_0$}};
\node[minimum width=0.5cm, minimum height=0.15cm, align=center] at (8,3.725) {\text{\small $\mathcal{H}_0$}};
\node[minimum width=0.5cm, minimum height=0.15cm, align=center] at (8,3.375) {\text{\small $\mathcal{H}_1$}};
\node[minimum width=0.5cm, minimum height=0.15cm, align=center] at (8,3.025) {\text{\small $\mathcal{H}_0$}};
\node[minimum width=0.5cm, minimum height=0.15cm, align=center] at (8,2.675) {\text{\small $\mathcal{H}_1$}};
\node[minimum width=0.5cm, minimum height=0.15cm, align=center] at (8,2.325) {\text{\small $\mathcal{H}_1$}};
\node[minimum width=0.5cm, minimum height=0.15cm, align=center] at (8,1.975) {\text{\small $\mathcal{H}_0$}};
\node[minimum width=0.5cm, minimum height=0.15cm, align=center] at (8,1.625) {\text{\small $\mathcal{H}_1$}};
\node[minimum width=0.5cm, minimum height=0.15cm, align=center] at (8,1.225) {\text{\small $\mathcal{H}_1$}};

\draw [-stealth](7,4.775) -- (7.6,4.775);
\draw [-stealth](7,4.4) -- (7.6,4.4);
\draw [-stealth](7,4.05) -- (7.6,4.05);
\draw [-stealth](7,3.7) -- (7.6,3.7);
\draw [-stealth](7,3.35) -- (7.6,3.35);
\draw [-stealth](7,3) -- (7.6,3);
\draw [-stealth](7,2.65) -- (7.6,2.65);
\draw [-stealth](7,2.3) -- (7.6,2.3);
\draw [-stealth](7,1.95) -- (7.6,1.95);
\draw [-stealth](7,1.6) -- (7.6,1.6);
\draw [-stealth](7,1.2) -- (7.6,1.2);

\end{tikzpicture}
\end{adjustbox}
\caption{Generalized architecture for learning-based wideband cooperative spectrum sensing.}
\label{fig:CooperativeSpectrumSensing}
\end{figure}

Over the recent years, a huge number of research articles applying \ac{ML}-based methods to the problem of spectrum sensing were published. 
Next, we overview methods based on a single- and multiple-node spectrum observation.

\subsubsection{Single-User Spectrum Sensing}
Single-node spectrum sensing is based on independent judgment of a single user.
It allows for an easier system that does not require data fusion at the cost of limited performance.
In~\cite{SS_CNN_China_2020}, authors presented a \ac{CNN}-based method capable of adapting to various noise types and detecting the presence of various signals, including those unseen.
They used the signal power spectrum as \ac{CNN} input and the maximum-minimum eigenvalue ratio-based method as a baseline~\cite{maxminSpecrumSensing}.
The results achieved under various false alarm probabilities show that the NN-based approach performed better than the baseline, especially for low SNR.
The~\textit{DeepSweep}, another \ac{CNN}-based model that exploits real-world recorded IQ samples of Wi-Fi signals as input, was presented in~\cite{Robinson_ICMLCN2024}.
Focusing on narrowband interference detection in the 2.4\,GHz IMS band via subsequent processing of the spectrum parts, the authors report a real-time inference capability of less than 1\,ms with up to 98\% accuracy.
In \cite{SpectrumTransformer2024}, a spectrum detector is implemented based on a Transformer architecture. 
The main idea of this work is to exploit the self-attention mechanism to learn not only the inner-band but also the inter-band spectral features in a wideband scenario.
To benchmark their model, the authors use an earlier CNN-based framework called \textit{DeepSense} from~\cite{SS_DeepSense_Melodia2021}, reporting a significant complexity reduction and higher detection accuracy.
In practical wireless systems, obtaining a sufficient amount of training data is often challenging and not realistic.
This issue was addressed in~\cite{GAN_SS_2018}, where the authors generated additional labeled data using a GAN framework.
The authors of more recent~\cite{CNN_Spectrum_Sensing_2022} followed this concept of data augmentation with a GAN to train their \ac{CNN}-based sensing algorithm with improved generalizability and robustness.

\subsubsection{Cooperative Spectrum Sensing}
Cooperative spectrum sensing refers to an approach in which a decision on spectrum occupancy is based on multiple observations of the area of interest. 
Exploiting cooperative sensing nodes distributed in the area of interest promises increased coverage, performance, and robustness~\cite{Wymeersch2009_IEEEProc}.
For instance, collaborative spectrum sensing mitigates the hidden node problem in \ac{CR} networks based on carrier-sense multiple access with collision avoidance~\cite{HiddenNode2008}.
Furthermore, distributed observations enable the design of more complex and flexible data fusion pipelines and learning paradigms, resulting in diverse research challenges.

In~\cite{LSTM_SS_2023}, a hierarchical method that integrates convolutional layers with \acp{LSTM} was proposed. 
This study models \acp{SU} equipped with multiple antennas that receive signals over various time slots, presenting spatial features as correlation matrices that are processed through the \ac{CNN} and \ac{LSTM} layers. 
Subsequently, features extracted separately for each RX are concatenated and passed through additional \ac{LSTM} and fully connected layers, followed by a Softmax function, which estimates the probabilities of $\mathcal{H}_0$ and $\mathcal{H}_1$ and decides on channel occupancy. 
The authors also provided calculations of $\gamma$ based on a targeted false alarm rate. 
The reported inference time is 1.3\,ms with simulations performed on NVIDIA Quadro P5000 GPU 16GB RAM, thus indicating the real-time feasibility of the algorithm.
The authors of~\cite{Coop_Unsupervised_SS_2022} presented an unsupervised collaborative \ac{DL} method based on PCA and dimensionality reduction, obtaining performance comparable to supervised learning.
A recent work~\cite{SpectrumSensing2024} focuses on wideband cooperative spectrum sensing under partial observations, i.e., a scenario where each of \acp{RX} monitors only a few consecutive sub-channels $N_f$, whereas the \ac{TX} can occupy any sub-channel. 
The original \ac{CNN} for a multi-task problem is decoupled into sub-band relevant sub-networks, reducing the computational cost without performance impairment.

\subsection{Signal Classification}

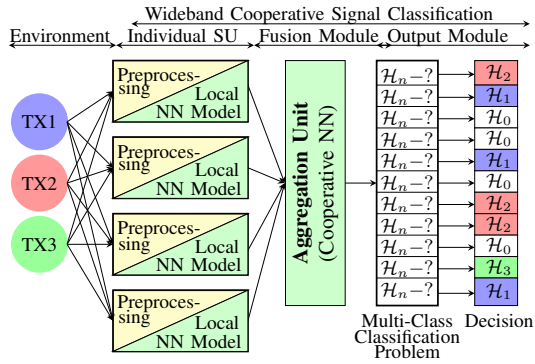
\begin{figure}
\centering
\begin{adjustbox}{width=0.4\textwidth}
\begin{tikzpicture}
\draw [shift={(5.,6)},color=black] node[below] {$\text{ \small Wideband Cooperative Signal Classification}$};
\draw [shift={(0.775,5.65)},color=black] node[below] {$\text{ \small Environment}$};
\draw [shift={(2.775,5.65)},color=black] node[below] {$\text{ \small Individual SU}$};
\draw [shift={(5,5.65)},color=black] node[below] {$\text{ \small Fusion Module}$};
\draw [shift={(7.1,5.65)},color=black] node[below] {$\text{ \small Output Module}$};
\draw[stealth-stealth] (2,5.6) -- (8.5,5.6);
\draw[stealth-stealth] (0,5.25) -- (1.75,5.25);
\draw[stealth-stealth] (1.75,5.25) -- (4.,5.25);
\draw[stealth-stealth] (4.,5.25) -- (6.25,5.25);
\draw[stealth-stealth] (6,5.25) -- (8.5,5.25);
\node[circle, fill=blue!40!white, minimum width=0.2cm, minimum height=0.2cm, align=center] at (0.5,4) {\small{TX1}};
\node[circle, fill=red!40!white, minimum width=0.2cm, minimum height=0.2cm, align=center] at (0.5,3) {\small{TX2}};
\node[circle, fill=green!40!white, minimum width=0.2cm, minimum height=0.2cm, align=center] at (0.5,2) {\small{TX3}};

\draw [-stealth](0.95,2) -- (1.7,0.75);
\draw [-stealth](0.95,2) -- (1.7,2);
\draw [-stealth](0.95,2) -- (1.7,3.25);
\draw [-stealth](0.95,2) -- (1.7,4.5);

\draw [-stealth](0.95,3) -- (1.7,0.75);
\draw [-stealth](0.95,3) -- (1.7,2);
\draw [-stealth](0.95,3) -- (1.7,3.25);
\draw [-stealth](0.95,3) -- (1.7,4.5);

\draw [-stealth](0.95,4) -- (1.7,0.75);
\draw [-stealth](0.95,4) -- (1.7,2);
\draw [-stealth](0.95,4) -- (1.7,3.25);
\draw [-stealth](0.95,4) -- (1.7,4.5);

\draw [-stealth](3.9,0.75) -- (4.5,3);
\draw [-stealth](3.9,1.9) -- (4.5,3);
\draw [-stealth](3.9,3.25) -- (4.5,3);
\draw [-stealth](3.9,4.6) -- (4.5,3);

\draw [-stealth](5,3) -- (6,3);

\draw[thick] (1.7,0.25) rectangle (3.9,1.25);
\draw[thick] (1.7,1.5) rectangle (3.9,2.5);
\draw[thick] (1.7,2.75) rectangle (3.9,3.75);
\draw[thick] (1.7,4) rectangle (3.9,5);
\draw [draw=black, fill=yellow, fill opacity=0.2]
       (1.7,0.25) -- (1.7,1.25) -- (3.9,1.25) -- cycle;
\draw [draw=black, fill=green!20!white]
       (1.7,0.25) -- (3.9,0.25) -- (3.9,1.25) -- cycle;
\draw [draw=black, fill=yellow, fill opacity=0.2]
       (1.7,1.5) -- (1.7,2.5) -- (3.9,2.5) -- cycle;
\draw [draw=black, fill=green!20!white]
       (1.7,1.5) -- (3.9,1.5) -- (3.9,2.5) -- cycle;
\draw [draw=black, fill=yellow, fill opacity=0.2]
       (1.7,2.75) -- (1.7,3.75) -- (3.9,3.75) -- cycle;
\draw [draw=black, fill=green!20!white]
       (1.7,2.75) -- (3.9,2.75) -- (3.9,3.75) -- cycle;
\draw [draw=black, fill=yellow, fill opacity=0.2]
       (1.7,4) -- (1.7,5) -- (3.9,5) -- cycle;
\draw [draw=black, fill=green!20!white]
       (1.7,4) -- (3.9,4) -- (3.9,5) -- cycle;
\draw [shift={(2.45,1.3)},color=black] node[below] {$\text{\small Preproces-}$};
\draw [shift={(2.05,1.05)},color=black] node[below] {$\text{\small sing}$};
\draw [shift={(3.4,0.95)},color=black] node[below] {$\text{\small Local}$};
\draw [shift={(3.1,0.65)},color=black] node[below] {$\text{\small  NN Model}$};

\draw [shift={(2.45,2.5)},color=black] node[below] {$\text{\small Preproces-}$};
\draw [shift={(2.05,2.25)},color=black] node[below] {$\text{\small sing}$};
\draw [shift={(3.4,2.2)},color=black] node[below] {$\text{\small Local}$};
\draw [shift={(3.1,1.9)},color=black] node[below] {$\text{\small  NN Model}$};

\draw [shift={(2.45,3.75)},color=black] node[below] {$\text{\small Preproces-}$};
\draw [shift={(2.05,3.55)},color=black] node[below] {$\text{\small sing}$};
\draw [shift={(3.4,3.45)},color=black] node[below] {$\text{\small Local}$};
\draw [shift={(3.1,3.15)},color=black] node[below] {$\text{\small  NN Model}$};

\draw [shift={(2.45,5)},color=black] node[below] {$\text{\small Preproces-}$};
\draw [shift={(2.05,4.8)},color=black] node[below] {$\text{\small sing}$};
\draw [shift={(3.4,4.7)},color=black] node[below] {$\text{\small Local}$};
\draw [shift={(3.1,4.4)},color=black] node[below] {$\text{\small  NN Model}$};

\node[rectangle, draw=black, fill=green!20!white, rotate=90, minimum width=4cm, minimum height=0.75cm, align=center] at (5,3) {\textbf{Aggregation Unit}\\ (Cooperative NN)};

\draw[thick] (6,1) rectangle (7,5);

\node[minimum width=0.5cm, minimum height=0.15cm, align=center] at (6.5,0.75) {\text{\small Multi-Class}};
\node[minimum width=0.5cm, minimum height=0.15cm, align=center] at (6.5,0.5) {\text{\small Classification}};
\node[minimum width=0.5cm, minimum height=0.15cm, align=center] at (6.5,0.25) {\text{\small Problem}};
\node[minimum width=0.5cm, minimum height=0.15cm, align=center] at (8.,0.75) {\text{\small Decision}};

\node[rectangle, minimum width=0.5cm, minimum height=0.25cm, align=center] at (6.5,1.25){\small{$\mathcal{H}_n - ?$}};
\node[rectangle, minimum width=0.5cm, minimum height=0.25cm, align=center] at (6.5,1.6){\small{$\mathcal{H}_n - ?$}};
\node[rectangle, minimum width=0.5cm, minimum height=0.25cm, align=center] at (6.5,1.95){\small{$\mathcal{H}_n - ?$}};
\node[rectangle, minimum width=0.5cm, minimum height=0.25cm, align=center] at (6.5,2.3){\small{$\mathcal{H}_n - ?$}};
\node[rectangle, minimum width=0.5cm, minimum height=0.25cm, align=center] at (6.5,2.65){\small{$\mathcal{H}_n - ?$}};
\node[rectangle, minimum width=0.5cm, minimum height=0.25cm, align=center] at (6.5,3){\small{$\mathcal{H}_n - ?$}};
\node[rectangle, minimum width=0.5cm, minimum height=0.25cm, align=center] at (6.5,3.35){\small{$\mathcal{H}_n - ?$}};
\node[rectangle, minimum width=0.5cm, minimum height=0.25cm, align=center] at (6.5,3.7){\small{$\mathcal{H}_n - ?$}};
\node[rectangle, minimum width=0.5cm, minimum height=0.25cm, align=center] at (6.5,4.05){\small{$\mathcal{H}_n - ?$}};
\node[rectangle, minimum width=0.5cm, minimum height=0.25cm, align=center] at (6.5,4.4){\small{$\mathcal{H}_n - ?$}};
\node[rectangle, minimum width=0.5cm, minimum height=0.25cm, align=center] at (6.5,4.75){\small{$\mathcal{H}_n - ?$}};

\draw (6,1.45) -- (7,1.45);
\draw (6,1.8) -- (7,1.8);
\draw (6,2.15) -- (7,2.15);
\draw (6,2.5) -- (7,2.5);
\draw (6,2.85) -- (7,2.85);
\draw (6,3.2) -- (7,3.2);
\draw (6,3.55) -- (7,3.55);
\draw (6,3.9) -- (7,3.9);
\draw (6,4.25) -- (7,4.25);
\draw (6,4.6) -- (7,4.6);

\draw [draw=black, fill=red!40!white]
       (7.6,4.6) -- (8.3,4.6) -- (8.3,5) -- (7.6,5) -- cycle;
\draw [draw=black, fill=blue!40!white]
       (7.6,4.25) -- (8.3,4.25) -- (8.3,4.6) -- (7.6,4.6) -- cycle;
\draw [draw=black, fill=blue!0!white]
       (7.6,3.9) -- (8.3,3.9) -- (8.3,4.25) -- (7.6,4.25) -- cycle;
\draw [draw=black, fill=blue!0!white]
       (7.6,3.55) -- (8.3,3.55) -- (8.3,3.9) -- (7.6,3.9) -- cycle;
\draw [draw=black, fill=blue!40!white]
       (7.6,3.2) -- (8.3,3.2) -- (8.3,3.55) -- (7.6,3.55) -- cycle;
\draw [draw=black, fill=blue!0!white]
       (7.6,2.85) -- (8.3,2.85) -- (8.3,3.2) -- (7.6,3.2) -- cycle;
\draw [draw=black, fill=red!40!white]
       (7.6,2.5) -- (8.3,2.5) -- (8.3,2.85) -- (7.6,2.85) -- cycle;
\draw [draw=black, fill=red!40!white]
       (7.6,2.15) -- (8.3,2.15) -- (8.3,2.5) -- (7.6,2.5) -- cycle;
\draw [draw=black, fill=blue!0!white]
       (7.6,1.8) -- (8.3,1.8) -- (8.3,2.15) -- (7.6,2.15) -- cycle;
\draw [draw=black, fill=green!40!white]
       (7.6,1.45) -- (8.3,1.45) -- (8.3,1.8) -- (7.6,1.8) -- cycle;
\draw [draw=black, fill=blue!40!white]
       (7.6,1.0) -- (8.3,1.0) -- (8.3,1.45) -- (7.6,1.45) -- cycle;

\node[minimum width=0.5cm, minimum height=0.15cm, align=center] at (8,4.8) {\text{\small $\mathcal{H}_2$}};
\node[minimum width=0.5cm, minimum height=0.15cm, align=center] at (8,4.425) {\text{\small $\mathcal{H}_1$}};
\node[minimum width=0.5cm, minimum height=0.15cm, align=center] at (8,4.075) {\text{\small $\mathcal{H}_0$}};
\node[minimum width=0.5cm, minimum height=0.15cm, align=center] at (8,3.725) {\text{\small $\mathcal{H}_0$}};
\node[minimum width=0.5cm, minimum height=0.15cm, align=center] at (8,3.375) {\text{\small $\mathcal{H}_1$}};
\node[minimum width=0.5cm, minimum height=0.15cm, align=center] at (8,3.025) {\text{\small $\mathcal{H}_0$}};
\node[minimum width=0.5cm, minimum height=0.15cm, align=center] at (8,2.675) {\text{\small $\mathcal{H}_2$}};
\node[minimum width=0.5cm, minimum height=0.15cm, align=center] at (8,2.325) {\text{\small $\mathcal{H}_2$}};
\node[minimum width=0.5cm, minimum height=0.15cm, align=center] at (8,1.975) {\text{\small $\mathcal{H}_0$}};
\node[minimum width=0.5cm, minimum height=0.15cm, align=center] at (8,1.625) {\text{\small $\mathcal{H}_3$}};
\node[minimum width=0.5cm, minimum height=0.15cm, align=center] at (8,1.225) {\text{\small $\mathcal{H}_1$}};

\draw [-stealth](7,4.775) -- (7.6,4.775);
\draw [-stealth](7,4.4) -- (7.6,4.4);
\draw [-stealth](7,4.05) -- (7.6,4.05);
\draw [-stealth](7,3.7) -- (7.6,3.7);
\draw [-stealth](7,3.35) -- (7.6,3.35);
\draw [-stealth](7,3) -- (7.6,3);
\draw [-stealth](7,2.65) -- (7.6,2.65);
\draw [-stealth](7,2.3) -- (7.6,2.3);
\draw [-stealth](7,1.95) -- (7.6,1.95);
\draw [-stealth](7,1.6) -- (7.6,1.6);
\draw [-stealth](7,1.2) -- (7.6,1.2);

\end{tikzpicture}
\end{adjustbox}
\caption{Generalized architecture for learning-based wideband cooperative signal classification for 3 TX signal types.}
\label{fig:SignalClassification}
\end{figure}

Non-cooperative signal classification can be seen as an advanced version of standard spectrum sensing and is sometimes referred to as intelligent spectrum sensing~\cite{CognitiveRadio_ML_Survey2024}.
The main difference is that it requires not only to decide if the subband is free or not but also to recognize the signal type.
Fig.~\ref{fig:SignalClassification} illustrates the generic signal classification problem.

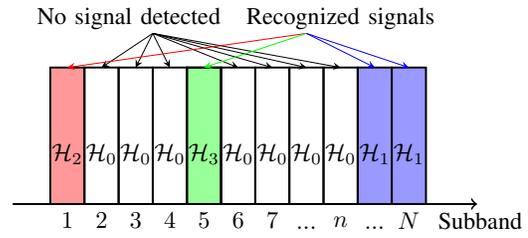
\begin{figure}
\centering
\begin{adjustbox}{width=0.38\textwidth}
\begin{tikzpicture}
\draw[thick,->] (0.45,1) -- (7.25,1) node[anchor=north west,xshift=-2em] {Subband};
\fill[red!40!white, draw=black, thick] (1,1) rectangle (1.5,3);
\fill[green!40!white, draw=black, thick] (3,1) rectangle (3.5,3);
\fill[blue!40!white, draw=black, thick] (5.5,1) rectangle (6,3);
\fill[blue!40!white, draw=black, thick] (6,1) rectangle (6.5,3);
\draw[thick] (1.5,1) rectangle (2,3);
\draw[thick] (2,1) rectangle (2.5,3);
\draw[thick] (2.5,1) rectangle (3,3);
\draw[thick] (3.5,1) rectangle (4,3);
\draw[thick] (3,1) rectangle (3.5,3);
\draw[thick] (4,1) rectangle (4.5,3);
\draw[thick] (4.5,1) rectangle (5,3);
\draw[thick] (5,1) rectangle (5.5,3);
\draw [shift={(1.25,1)},color=black] node[below] {$1$};
\draw [shift={(1.75,1)},color=black] node[below] {$2$};
\draw [shift={(2.25,1)},color=black] node[below] {$3$};
\draw [shift={(2.75,1)},color=black] node[below] {$4$};
\draw [shift={(3.25,1)},color=black] node[below] {$5$};
\draw [shift={(3.75,1)},color=black] node[below] {$6$};
\draw [shift={(4.25,1)},color=black] node[below] {$7$};
\draw [shift={(4.75,0.8)},color=black] node[below] {$...$};
\draw [shift={(5.25,0.94)},color=black] node[below] {$n$};
\draw [shift={(5.75,0.8)},color=black] node[below] {$...$};
\draw [shift={(6.25,1)},color=black] node[below] {$N$};
\draw [shift={(1.25,2)},color=black] node[below] {$\mathcal{H}_2$};
\draw [shift={(1.75,2)},color=black] node[below] {$\mathcal{H}_0$};
\draw [shift={(2.25,2)},color=black] node[below] {$\mathcal{H}_0$};
\draw [shift={(2.75,2)},color=black] node[below] {$\mathcal{H}_0$};
\draw [shift={(3.25,2)},color=black] node[below] {$\mathcal{H}_3$};
\draw [shift={(3.75,2)},color=black] node[below] {$\mathcal{H}_0$};
\draw [shift={(4.25,2)},color=black] node[below] {$\mathcal{H}_0$};
\draw [shift={(4.75,2)},color=black] node[below] {$\mathcal{H}_0$};
\draw [shift={(5.25,2)},color=black] node[below] {$\mathcal{H}_0$};
\draw [shift={(5.75,2)},color=black] node[below] {$\mathcal{H}_1$};
\draw [shift={(6.25,2)},color=black] node[below] {$\mathcal{H}_1$};
\draw [shift={(2.15,4)},color=black] node[below] {$\text{No signal detected}$};
\draw[-stealth] (2.5,3.5) -- (1.75,3);
\draw[-stealth] (2.5,3.5) -- (2.25,3);
\draw[-stealth] (2.5,3.5) -- (2.75,3);
\draw[-stealth] (2.5,3.5) -- (3.75,3);
\draw[-stealth] (2.5,3.5) -- (4.25,3);
\draw[-stealth] (2.5,3.5) -- (4.75,3);
\draw[-stealth] (2.5,3.5) -- (5.25,3);
\draw [shift={(5.25,4)},color=black] node[below] {$\text{Recognized signals}$};
\draw[-stealth,red] (4.75,3.5) -- (1.25,3);
\draw[-stealth,green] (4.75,3.5) -- (3.25,3);
\draw[-stealth,blue] (4.75,3.5) -- (5.75,3);
\draw[-stealth,blue] (4.75,3.5) -- (6.25,3);
\end{tikzpicture}
\end{adjustbox}
\caption{An illustration of a multi-band multi-signal detection and type classification. $\mathcal{H}_1$, $\mathcal{H}_2$, and $\mathcal{H}_3$ denote 3 signal types, while $\mathcal{H}_0$ denotes the signal absence.}
\label{fig:SpectrumSensing}
\end{figure}

For the input $\text{x}$, which can be raw data or feature vector, signal classification can be represented as the multiple hypothesis testing problem:

\begin{equation}
    \hat{\mathcal{H}} = \argmax_{\text{n}\in{[1,|\mathcal{S}|]}} f(\mathcal{H_\text{n}}|\text{x},\Theta),
\end{equation}
where $\mathcal{H}$ and $\hat{\mathcal{H}}$ are the ground-truth signal type and the predicted signal type, respectively; $\mathcal{S}$ is the pool of possible signal types with $|\mathcal{S}|$ being the number of signal types in $\mathcal{S}$; $f(\Theta)$ is the mapping function from samples to modulation types, where $\Theta$ denotes the model parameters.

Signal classification can be roughly arranged into two categories: (\textit{i}) a broader task of waveform classification (or \ac{RAT} recognition), and (\textit{ii}) \ac{AMC}. 
By classifying waveforms, networks can verify the legitimacy of signals. 
For instance, detecting anomalous or unauthorized waveforms could signal potential security threats like jamming or spoofing~\cite{JammerDetection2024}. 
This helps improve the security and integrity of communication systems. 
\ac{AMC} is a more specific task that can enable real-time adjustments to \ac{RX} settings and accelerate the demodulation procedure.
The interest in \ac{AMC} can be traced back to the turn of the 1990s, with \cite{AMC_AISBETT_1987} and \cite{ACM_old_features_1992} exemplifying early works on analog and digitally modulated communication signals classification, respectively.
Nowadays, wireless networks use numerous modulation schemes, protocols, and waveforms, making signal classification way more complex. 
Moreover, wireless signals can overlap or undergo distortion due to interference or fading, challenging the recognition task.

The methods developed over the next decade can be categorized into two categories which are based on the likelihood function and statistical features of the signal, respectively~\cite{Dobre2007}.
Since feature-based approaches are more practical and easier to implement, they attracted significant interest.
The list of popular statistical features includes a ratio of in-phase and quadrature power components, the standard deviation of the direct instantaneous phase, the standard deviation of the absolute value of the normalized instantaneous signal amplitude, the mean value of the signal magnitude, etc~\cite{DNN_ACM_features_2016}.
A popular set of statistical features can be illustrated by higher order moments (HOMs), defined for a complex-valued input sequence $\text{x}$ of any length as

\begin{equation}\label{HOM}
M_{pq} = \mathds{E}[\text{x}^{(p-q)}\cdot{(\text{x}^{*})}^q].
\end{equation}

Furthermore, combinations of HOMs can be leveraged to obtain higher order cumulants (HOCs) denoted as $C$; for a full list of formulas, refer to~\cite{DNN_ACM_features_2016}).

Classification based on a single feature is hardly possible due to the abundance of signals from various wireless standards that exhibit similar statistical parameters.
In principle, all extracted features can be taken into account for signal classification problems, serving as input to the ML-based classifier.
However, the number of features used to train the algorithm affects the computational complexity and inference time. 
Hence, for practical applications, the proper feature selection is of great importance to choose the most distinguishable and discard those that are redundant.
To this end, methods from statistics were adopted to quantify the dissimilarity between features.
For instance, \ac{BD} is often used in feature selection tasks to compare feature probability distributions based on their overlap and shared information~\cite{BD1943}.
For two discrete probability distributions $P$ and $Q$ in the same domain $\mathcal{X}$, \ac{BD} is computed as

\begin{equation}
D_{B}(P,Q)=-\ln \left(\sum_{x\in {\mathcal {X}}}{\sqrt {P(x)Q(x)}})\right).
\end{equation}

The \ac{BD} is always non-negative: a value of 0 means that the two distributions are identical, while higher values indicates a greater difference between them.
The task of effective feature selection based on \ac{BD} in the context of AMC was studied and evaluated on three neural network-based classifiers for AWGN and frequency-selective fading channels in~\cite{BD_AMC}.

The rest of the subsection provides an overview of popular ML-based signal classification approaches, datasets, and the evolution of the signal classification task with advancements in wireless technology.

\subsubsection{Knowledge-Driven Signal Classification}
\begin{figure}
\centering
  \begin{subfigure}{0.36\textwidth}
    \includegraphics[width=\linewidth]{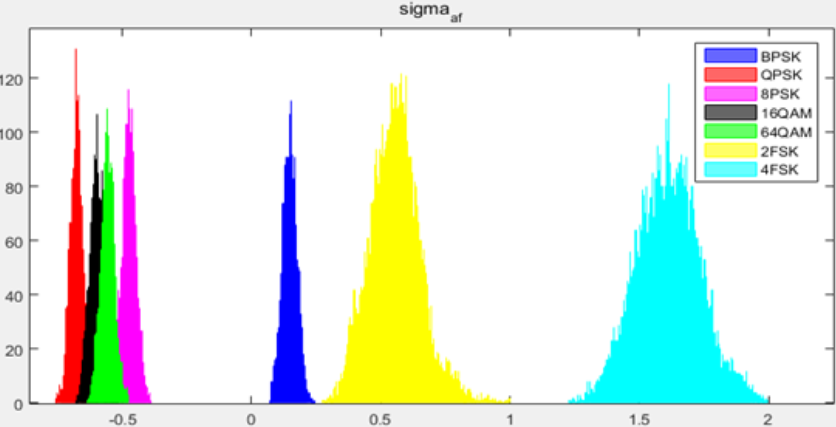}
    \caption{Standard deviation of the absolute normalized centered instantaneous frequency for the signal segment for \ac{AMC}~\cite{DNN_ACM_features_2016}(\copyright[2016]IEEE)} \label{fig:feat_a}
  \end{subfigure}%
  
  \begin{subfigure}{0.36\textwidth}
    \includegraphics[width=\linewidth]{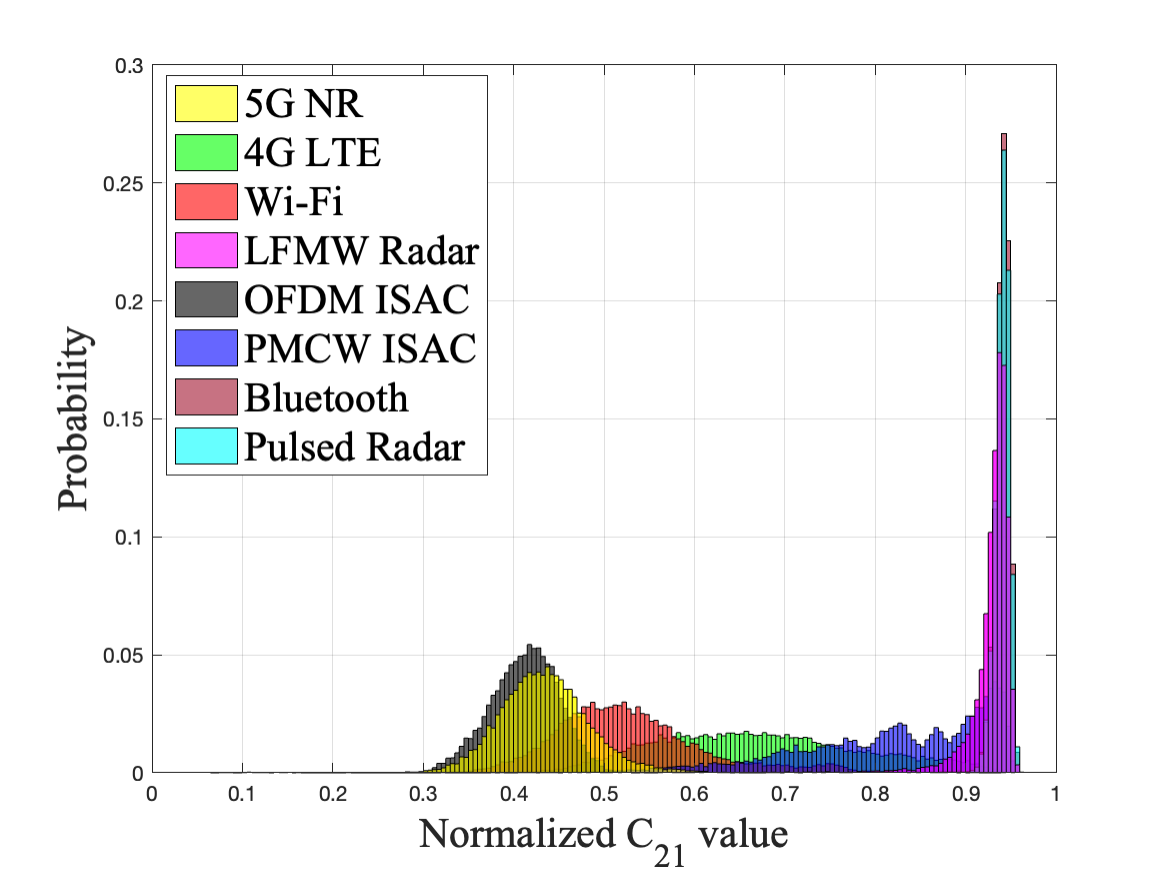}
    \caption{Normalized histogram illustrating the probability distribution of statistical feature $C_{21}$ of waveforms in the dataset from \cite{JRC_classification2025}.} \label{fig:feat_b}
  \end{subfigure}

\caption{Histograms that illustrate the probability distribution of statistical features of signals in the datasets.} \label{Distr}
\end{figure}

Vectors of statistical features, such as moments, cumulants, and cyclic cumulants of the signal itself, were used as input in early \ac{ML}-based signal classification applications~\cite{EarlyAMC2005}.
These features can be fed into simple decision trees~\cite{DecisionTree2000}, random forests~\cite{RandomForest2017}, or SVMs~\cite{SVM2004}.
This approach remains popular nowadays since it results in lightweight \ac{ML} models and promises robust performance under poor \ac{SNR} conditions.
For instance, in 2016, the authors of~\cite{DNN_ACM_features_2016} mapped 20000 signal samples into 21 statistical features as an input to a small DNN to classify 7 modulation types, namely BPSK, QPSK, 8-PSK, 16-QAM, 64-QAM, 2-FSK, and 4-FSK.
Fig.~\ref{fig:feat_a} shows the standard deviation of the absolute normalized centered instantaneous frequency for the signal segment.
Based on this feature alone, it is already possible to uniquely separate 3 classes.
Furthermore, the 4 rest classes have a rather small overlap.
In contrast, \cite{JRC_classification2025} studies waveform classification for potential B5G scenarios, considering a synthetic dataset comprised of various realizations of 8 waveforms, including 3GPP-compliant 4G, 5G, Wi-Fi, and Bluetooth waveforms for communications, \ac{LFMW} and pulsed waveforms for radar application, and two fully integrated waveform candidates based on \ac{OFDM} and \ac{PMCW}.
Fig.~\ref{fig:feat_b} shows normalized histograms which approximate pdfs of the underlying distribution of a cumulant $C_{21} = \mathds{E}[{|\text{x}|}^2]$, where $\text{x}$ is an input time series.
Due to significant feature overlaps, it is difficult to expect an accurate signal classification in this scenario.
This example shows that methods driven solely by expert knowledge, which operate on a feature space of reduced dimension, have limited potential for certain tasks.

\subsubsection{Data-Driven Classification with Raw Input Data}

To circumvent the aforementioned limitations of the feature-based approach, T. O'Shea and J. Hoydis, in their seminal article from 2017, proposed utilizing raw time-series data as NN input~\cite{OSheaHoydis2017}. 
Using raw data, namely IQ samples or their amplitude-phase (AP) representation, increases computational complexity but promises enhanced classification accuracy.
The classifier architecture in~\cite{OSheaHoydis2017} itself is a straightforward \ac{CNN} composed of narrowing convolutional layers followed by FC layers and a dense Softmax output layer.
Following that work, many researchers studied various aspects of improving this purely data-driven learning approach.
For example, the authors of~\cite{Beyond5G_AMC_2020} employed a similar CNN-based method and concentrated on inference time reduction, directly targeting real-time capability.
In~\cite{WaveformClass2021}, the signals were classified into in-network users, out-network users, and jammers coexisting in the same wireless network. 
Similarly, this paper used IQ data as input to a \ac{CNN}.
A \ac{CNN}-based \ac{AMC} that accounts for phase imperfections is performed in~\cite{Karagiannidis2024}, where inputs are represented by images of modulation constellations with dimensions of $256\times256$, each containing 1000 points.
However, since the original raw time-domain IQ samples represent a 1D time series, architectures based on standard \acp{RNN}, \acp{LSTM}, and Transformers are of special interest and have gained popularity.
Another study on \ac{AMC} for \ac{OFDM} systems leveraged various signal representations, specifically IQ and AP, as bi-stream input, thus utilizing the synergy between distinct signal modalities~\cite{AMC_Kumar2024}. 
In this paper, the input sequence length equals 2048.
Additionally, an attention-based \ac{CNN}-\ac{LSTM} model has been used to design the classifier.

Although data-driven methods are currently far more popular due to their superior classification accuracy, they lack explainability.
Combining raw data with the expert-extracted features and prior knowledge not only makes a step towards better explainability but also aims to reap the benefits of both approaches~\cite{Dobre2024}.
Finally, when exploiting raw data as an input for classification, the performance-complexity trade-off should be wisely considered.
The signal should be adequately represented by a sufficient number of samples to enhance classification accuracy while avoiding excessive computational burden to maintain low inference time.

\subsubsection{Signal Classification with Spectrograms as Input}

Spectrograms are a popular alternative signal representation for classification tasks.
A spectrogram visually represents the spectrum of frequencies in a signal as it changes over time. 
They are generated by dividing the original time-domain signal into overlapping segments and using the short-time Fourier transform (STFT) to determine the magnitude of the frequency spectrum for each segment~\cite{STFT}.

The conversion of wireless signals into spectrograms allows the application of image processing techniques for signal classification.
Specifically, an adapted version of You Only Look Once (YOLO), a supervised \ac{CNN}-based algorithm to learn how to find and classify the objects in an image, was successfully leveraged for wideband signal recognition task in~\cite{Vagollari2023}.
This work introduced a two-step lightweight \ac{AMC} framework: In the first stage, YOLO was exploited to detect, localize, and classify narrowband signals in wideband spectrograms, and in the second stage, a custom \ac{CNN} refined the classification results.

Since constructing spectrograms requires computationally heavy processing stages that allow some degree of freedom, it is vital to be aware of several involved parameters on the algorithm performance.
In this context, \cite{FFT_AMC_Comparison2022} studied the influence of STFT window size and type on \ac{AMC} accuracy for their designed CNN-based algorithm.
The reported result, performed over 26 signal types affected by multipath fading channels, shows the marginal dependency of \ac{AMC} accuracy on STFT with 128, 256, and 512 samples for window overlap of 25\%, 50\%, and 75\%, and for six window functions.
The study shows some advantage of using shorter window lengths and bigger overlaps. 
However, it lacks inference times for different methods, which is important for fair trade-off estimation.

Overall, despite their apparent merits, spectrograms may be impractical for real-time wireless signal classification applications. 
First, they require additional signal processing steps, which can be computationally expensive. 
Therefore, it is recommendable to take into account preprocessing time while evaluating the inference of the \ac{ML}-based algorithms that rely on spectrograms as input.
Second, although spectrograms ensure a rich signal representation, they lack information regarding the signal phase~\cite{Zhang_infocom_2021}.

\subsubsection{Jamming and Anomaly Detection}

Discovering jammers, intentional interferers, and other types of spectrum anomalies can be seen as a subproblem of signal classification, dealing exclusively with the unwanted \acp{TX}.
Jamming is often studied for wireless applications in the industrial domain, where it can be a serious issue, as equipment downtime can cause multimillion-dollar losses.
\ac{ML} techniques are expected to play a crucial role in detecting jammers in wireless networks by identifying anomalies in environments caused by malicious \acp{TX}.
For example,~\cite{Jam_factory_floor_2023} studied an autoencoder-based jammer detection and mitigation for automated guided vehicles in mmWave networks for manufacturing environments.
Further,~\cite{DT_spectrum2024} presents an \ac{ML}-based framework for anomaly and jammer detection in the context of \acp{DT} for industrial environments.
To detect spectrum anomalies, the authors employed unsupervised learning based on distributed \ac{RSS} measurements.
Specifically, the algorithm compares legitimate users’ expected radio environment map estimates with actual distributed \ac{RSS} measurements, thus detecting the unauthorized \ac{TX} presence and roughly localizing it.

In summary, since jammers often target critical infrastructure, AI-enhanced detection algorithms should be able to continuously sense the environment and detect attacks as soon as possible.
Hence, the learning algorithms’ real-time capability and adaptability are essential.
For this purpose, unsupervised and \ac{RL} techniques that allow the model to update with new data and maintain responsibility for new types of jamming attacks require detailed investigation.
We refer readers to~\cite{JammerDetection2024} for a comprehensive taxonomy of anomaly and jammer attacks along with state-of-the-art learning techniques to detect and mitigate them.

\subsubsection{Cooperative Signal Classification}

Similarly to cooperative spectrum sensing, cooperative signal classification by multiple \acp{RX} observing the same environment can extend the monitored coverage area, offer spatial diversity, and remove the dependency on a single node and channel which might be under poor conditions.
Standard cooperative algorithms can be exemplified by~\cite{Co_AMC_legacy_1_2010}, which performs likelihood ratio-based distributed fusion using multiple radios, where each \ac{RX} makes a local classification decision and then these decisions are merged to make a global decision.
An alternative approach is to fuse the data in a centralized way, where the data can be raw~\cite{Co_AMC_legacy_2_2013} or be represented as a set of statistical features, e.g., cumulants~\cite{Co_AMC_legacy_cumulants_2017}.
Note that when done not properly, the fusion of data from multiple \acp{RX} can even result in degraded performance compared to a single \ac{RX} solution.

A prominent work~\cite{Rajendran2018} studies \ac{DL} models for signal classification task in a distributed sensor network only using averaged magnitude FFT data.
This approach resulted in a practical model which comprises $i$ low-cost sensors which scan narrow bands at frequencies $f_i$ from 50\,MHz to 6\,GHz, and transfer small amount of data to the cloud, where \ac{LSTM}-based models perform classification on the over-the-air dataset which consists of 6 waveforms.
The reported accuracy varies depending on a specific structure, and yields more than 80\% for the largest model.
Further, the \ac{CNN} was used in~\cite{DL_CoAMC_2020} in the context of \ac{MIMO} systems, where the RX, equipped with multiple antennas, classifies the modulation types. 
Signals received at each RX antenna are fed to \ac{CNN}, which obtains sub-results, which are subsequently merged to make the final decision.
Four cooperative decision rules, namely direct voting,
weighty voting, weighty averaging, and direct averaging are applied to infer the final modulation type, with weighty averaging reported to be superior.
The performance of centralized and distributed CNN-based \ac{AMC} were compared in~\cite{Collab_AMC_2020}.
The centralized approach performs slightly better, but this comes at the cost of higher training time and increased requirements for the edge computer’s computational power.
A recently proposed collaborative \ac{AMC} framework for \ac{OFDM} signals~\cite{OFDM_class_features_collab2024} focuses on signal-level fusion.
As a result of multi-stage signal processing, a signal distribution property vector is constructed, which is finally fed to a simple lightweight \ac{NN} which comprises 2 convolutional and 2 fully connected layers with multi-class entropy loss function.

Currently, there are relatively few papers that utilize distributed observations for various signal classification tasks using \ac{ML}. However, this topic has significant potential, with various problems to tackle.
Table~\ref{Tab:SpecrtumSensingOverview} summarizes the literature on spectrum sensing and signal classification.

\begin{table*}[]
\caption{An overview of the \ac{ML}-based approaches for spectrum sensing and signal classification from literature.}
\centering
\scalebox{0.85}{
\begin{tabular}{|c|c|c|c|c|l|l|}
\hline\hline
               Task   & Ref. (Year) & Input& ML method & Distr. & \phantom{to center}Reported contributions & \phantom{shift to center}Limitations \\ \hline
\multirow{18}{*}{\textbf{\rotatebox{90}{Spectrum Sensing}}} 

& \cite{LSTM_SS_2023} (2023) & \makecell[l]{Covariance matrices\\of energy correlations} & Hybrid CNN-LSTM & \checkmark & \makecell[l]{High accuracy and robustness\\under imperfect CSI} & \makecell[l]{High computational complexity} \\ \cline{2-7} 
 
& \cite{Robinson_ICMLCN2024} (2024) & IQ samples & CNN-based &  & \makecell[l]{Real-time capability ($<$1\,ms inference\\ time), 98\% accuracy} &  \makecell[l]{Limited functionality,\\ only narrowband operation}\\ \cline{2-7} 
 
& \cite{SpectrumTransformer2024} (2024)& Rearranged PSD & Transformer & & \makecell[l]{Operates in wide band, reduced\\ complexity, higher detection accuracy} &\makecell[l]{Requires validation with\\over-the-air measurements}  \\ \cline{2-7} 
 
& \cite{SpectrumSensing2024} (2024)& Rearranged PSD & CNN-based  & \checkmark & \makecell[l]{Wideband cooperative sensing solution\\ under only partial observations} &\makecell[l]{High inference time, not suitable\\ for real-time 6G applications}  \\ \cline{2-7} 
 
& \cite{SS_DeepSense_Melodia2021} (2021)& IQ samples & CNN-based & & \makecell[l]{Over-the-air setup, real-time capability \\(0.61\,ms inference time), 98\% accuracy} & \makecell[l]{Relatively low accuracy and high \\inference time with a large-scale dataset} \\ \cline{2-7} 
 
& \cite{GAN_SS_2018} (2018)& \makecell[c]{Samples in\\time domain\\(not detailed)} & \makecell[c]{SVM and \\Random Forest} & &\makecell[l]{Demonstrates advantages of GANs in\\ training data augmentation for sensing} &\makecell[l]{Results are obtained in\\ a basic narrowband scenario}  \\ \cline{2-7} 
 
& \cite{Coop_Unsupervised_SS_2022} (2022) & \makecell[c]{Energy vectors} & \makecell[c]{Unsupervised learn.\\using Gaussian mixture\\ model and PCA} &\checkmark & \makecell[l]{Useful for scenario with little labeled\\ data, robust in low SNR} & \makecell[l]{Noticeable performance gap \\compared to supervised learning} \\ \cline{2-7} 
 
& \cite{Reinf_CSS_2024} (2024) & Energy vectors & \makecell[c]{Attention-enhanced\\ RL}  & \checkmark &  \makecell[l]{Reliable sensor scheduling,\\reduced communication overhead} &\makecell[l]{Unclear real-time capability and\\ practical feasibility, limited\\ performance accuracy comparison}\\ 

\hline
          
\multirow{24}{*}{\textbf{\rotatebox{90}{Signal Classification}}}

& \cite{DNN_ACM_features_2016} (2016)&\makecell[c]{Statistical\\feature vectors} & \makecell[c]{Small DNN with\\5 FC layers} &  & High accuracy AMC &\makecell[l]{Limited and relatively simple set\\of low-order modulated signals\\with distinct statistical properties}  \\ \cline{2-7} 

& \cite{OSheaHoydis2017} (2017) & Raw IQ samples & CNN-based & & \makecell[l]{AMC with higher accuracy compared\\ to feature-based AMC}& \makecell[l]{Brute force solution with high\\computational complexity} \\ \cline{2-7} 
                  
& \cite{AMC_Kumar2024} (2024) & \makecell[c]{Raw IQ samples\\and their AP form} & \makecell[c]{Attention-based\\hybrid CNN-LSTM} &  & \makecell[l]{High-accuracy AMC for OFDM systems\\with over-the-air captured data} &\makecell[l]{Limited to OFDM waveform, too high\\inference time ($>$1\,s)}  \\ \cline{2-7}
                  
& \cite{FFT_AMC_Comparison2022} (2022) & Spectrograms & CNN-based & &\makecell[l]{Studies dependency of AMC accuracy\\ on STFT with 128, 256, and 512 samples} &\makecell[l]{No information regarding inference\\time}  \\ \cline{2-7} 

& \cite{Rajendran2018} (2018) & \makecell[c]{Averaged FFT\\ magnitudes} & LSTM-based & \checkmark & \makecell[l]{Experimentally proven low-cost\\efficient AMC approach} & \makecell[l]{Limited accuracy (around 80\%) even \\with a small number of classes}  \\ \cline{2-7}

& \cite{DL_CoAMC_2020} (2022) & Raw IQ samples & CNN-based & \checkmark & \makecell[l]{Presents a discussion on four cooperative\\decision rules in AMC}&  \makecell[l]{Small number of candidate classes, low-\\order modulations (BPSK, QPSK, 8PSK,\\ 16QAM), no information on complexity} \\ \cline{2-7} 

& \cite{OFDM_class_features_collab2024} (2024) & \makecell[c]{Signal distribution\\ property vector} & \makecell[c]{Lightweight NN\\ based on CNN} & \checkmark & \makecell[l]{Effective AMC under unbalanced SNR \\and varying channel conditions} & \makecell[l]{Only 6 low-order modulation\\formats, requires real-world validation,\\ no discussion on complexity} \\ \cline{2-7} 

& \cite{Fontaine2020} (2020) & \makecell[c]{(i) IQ and (ii) FFT} & CNN & &\makecell[l]{RAT classification in sub-GHz \\implemented with low-cost SDRs\\ and 2 NN input types} & \makecell[l]{Only ultra narrowband signals,\\ relatively high inference time}\\ \cline{2-7}
                  
& \cite{Vagollari2021} (2021) & Spectrograms & YOLO-based & &\makecell[l]{Adoption of a popular ML object detector\\for efficient detection and localization\\ of signals in wideband RF spectrum} & \makecell[l]{Limited recognition accuracy of \\some signals, required real-\\data validation}\\ \cline{2-7}

& \cite{Rad_Com_Coex_Classif_2020} (2020) & \makecell[c]{Fourier synchro-\\squeezing trans-\\formation (FSST)} & CNN-based & &\makecell[l]{Studies waveform classification in\\ radar-communications coexistence\\ scenarios} &\makecell[l]{Show good performance only\\ with high SNR; no information \\regarding inference time}\\ 
\hline\hline
\end{tabular}
}
\label{Tab:SpecrtumSensingOverview}
\end{table*}

\subsubsection{Scenarios and Types of Signals to Recognize}

For decades, the standard tasks were to classify modulation formats, waveforms or other parameters of pure communications signals.
Recently, due to spectrum congestion and cognitive radar and radio evolution, the inclusion of radar for classification signals has become more common~\cite{Rad_Com_Coex_Classif_2020}.
The advancement of radar-communications convergence will impose more difficult challenges on wireless \acp{RX} and networks in general to ensure correct operation.

Signal classification was recognized as an important task that could be tackled in the future \ac{ISAC} \ac{RX} design~\cite{ISAC_vision_Liu_2022}.
Learning of modulation schemes and/or radar sensing and communications signals waveforms can be used instead of, e.g., successive interference cancellation (SIC), thus reducing \ac{RX} complexity and facilitating faster acquisition.
A step in this direction is~\cite{RadComClass2024Latest}, where the authors presented a \ac{CNN}-based waveform classifier for \ac{ISAC} \ac{RX} design which operates on time–frequency images.
However, this work only considers a co-existence radar-communications scenario.
In \ac{6G} and beyond, the inclusion of potential fully integrated \ac{ISAC} signal waveforms is under investigation.
For instance, in such a case, radar-centric \ac{PMCW} or communications-centric \ac{OFDM}can be added to the standard set of waveforms.
Since the \ac{ISAC} waveforms are the tweaked versions of the existing dedicated radar and communications waveforms, they retain a strong resemblance, thus hindering the classification.
It can be envisioned that in such a scenario, only processing raw IQ data by a relatively large \ac{DNN} might obtain acceptable classification accuracy.

In general, waveform and \ac{RAT} classification are more difficult compared to \ac{AMC} because they require analyzing the signal in a broader scope and considering a wider range of features: Protocols, modulation schemes, access techniques, and combinations of these parameters.
For example, 4G LTE, 5G NR, and Wi-Fi waveforms are all \ac{OFDM}-based and use similar sets of modulation schemes, therefore, achieving high classification accuracy can be very challenging.
In contrast, \ac{AMC} is more focused on identifying a particular characteristic (modulation scheme) of the signal, making it a less complex task overall, although still challenging under low \ac{SNR} or with high-order modulations.
Fig.~\ref{Distr} illustrates the difference in complexity of tasks by examples of statistical features distributions.
In literature, \ac{AMC} dominates the field; however, it can be anticipated that blind waveform and access technology recognition will play a significant role in future wireless networks, contributing to tasks such as interference monitoring and management, network security, and anomaly detection. 
In this context, \ac{ML} techniques will be essential to improve automation and intelligent aspects.
It can be safely envisioned that with the growth of modulation orders used in wireless systems, types of \acp{RAT}, and waveforms, intelligent spectrum sensing in 6G networks will require novel learning approaches capable of balancing high performance with acceptable complexity for real-time operation.

\subsubsection{Datasets}
Common benchmark datasets are of high importance to compare various \ac{ML} models for radio signal analysis.
RadioML 2018.01A, a canonical dataset for emitter types classification used in hundreds of academic research articles since 2018, was generated in~\cite{RML16_OShea} and improved in~\cite{RadioML_Dataset_2018}\footnote{https://www.deepsig.ai/datasets/}. 
It contains signals of 24 radio signal types generated and recorded under wireless propagation environment using \ac{SDR} platforms.
These works laid foundation for a recent RML22~\cite{AMC_Dataset2023}, where the authors claim a more realistic and correct methodology for signal generation.
Apart from a dataset, its main contribution is the open source Python-based framework that allows for generation of wireless signals impaired by a variety of propagation effects, which can be used to evaluate \ac{ML} models for modulation classification.

A wideband dataset that enables detection and localization in time and frequency domains is presented in \cite{WidebandDataset2021}.
This dataset is represented by synthetically generated data which emulates wideband signals through a combination of diverse layouts of narrowband emissions.
The authors of~\cite{dataset_fraunhofer2022} provide another wideband spectrogram images dataset containing 20000 labeled samples of Wi-Fi and Bluetooth signals along with tools to create new data sets for specific requirements.

While the aforementioned datasets consist exclusively of communications signals, modern wireless environments allow radar users to use the same spectrum. 
Given this,~\cite{SpectrumSensingDataset2023} provides a collection of pulse radar and 4G LTE signals recorded over the air in the shared Citizen Broadband Radio Service (CBRS) band (3.55-3.7\,GHz), where signals coexist in overlapping and non-overlapping manner in a number of varying \ac{SINR} conditions. 
The authors also describe the data collection setup, consisting of three Ettus X310 USRP \ac{SDR} systems, one each for the radar \ac{TX} and 4G LTE signals in the CBRS band and a third USRP for receiving IQ samples.

Overall, pioneering RadioML 2018.01A from 2018 remains the most popular dataset for testing wireless signal classification approaches. 
While it is still useful to examine the performance of the designed algorithms, this dataset does not represent the realms of wireless standards of 2025. 
Therefore, the research community apparently needs to adopt a new benchmark; however, it is difficult to dictate or predict when and what it will be.
The vast majority of articles on \ac{ML}-based signal classification evaluate models whose input is a signal captured by a single \ac{RX}.
A distributed spectrum analysis approach, where networked sensors monitor the same environment, could be of interest for signal classification as well as jamming and anomaly detection.
Further, its potential advantages and trade-offs (classification/detection accuracy vs. complexity) can be investigated. 
In this context, novel datasets emulating distributed sensing scenarios are welcome. 
In principle, modern software tools allow fast standard-compliant signal generation using MATLAB Communication Toolbox, while the realistic propagation environment can be modeled leveraging ray tracing software like Altair Winprop or Sionna RT.
Further, over-the-air real signals can be recorded for various scenarios using \acp{SDR}.

\subsection{Machine Learning-based Remote Transmitter Positioning}

Obtaining low-complexity and good-performing positioning algorithms is often challenging due to the nonlinear relationship between the signal source coordinates and positioning parameters, i.e., as \ac{TDoA}, \ac{AoA} and \ac{RSS}. 
Indoor environments likely present \ac{NLoS} propagation conditions that significantly damage the performance of model-based position estimators.
Moreover, traditional positioning parameter modeling neglects hardware-related impairments, such as power amplifier nonlinearity. Dealing with such issues usually means designing computationally intensive approaches.
In such a context, \ac{ML}-based techniques are able to circumvent the nonlinear characteristics and provide a lower complexity solution.

Related works on \ac{ML}-based localization seek to represent the positioning parameters, such as \ac{RSS}, \ac{CSI}, \ac{ToA} or \ac{AoA} in image-like formats to exploit the proven capabilities of \acp{CNN} architectures used in image processing applications \cite{deepmtl,deeptxfinder,ml_fingerprinting_survey}.

\subsubsection{Blind Transmitter Positioning}

Remote positioning assumes that within an area of interest, multiple \acp{SU} are connected to a \ac{CU}.
The \acp{SU} receive the incoming signals from the signal source(s) and send them to the \ac{CU}, where its coordinates are estimated. 
The source coordinates can be communicated to the source, or be only employed for network management. 
Positioning can be further classified as cooperative or non-cooperative depending on whether the devices aid in their position estimation, e.g., by transmitting positioning-specific signaling. 
Private wireless networks that are threatened by jamming transmissions can benefit from non-cooperative remote positioning for reliability.

Within an area of interest, a total of $N$ \acp{SU} are placed in fixed locations and communicate their measurements to a \ac{CU}.
Blind \ac{TX} positioning refers to a remote positioning approach where the sources that need to be localized do not aid in the process.
In this case, position estimation can be obtained through \ac{AoA} or \ac{RSS} measurements. 
In the \ac{AoA} case, \acp{SU} have to be equipped with an antenna array for estimating \ac{AoA}, consequently increasing the \acp{SU} size and cost.
On the other hand, \ac{RSS} measurements can be obtained through simpler and low-cost \acp{SU}.

Large-scale propagation models are used to predict the \ac{RSS} based on the distance between the source and \acp{SU}. 
These models typically combine deterministic and random components and are applicable for distances beyond the Fresnel distance, corresponding to far-field propagation conditions. 
The deterministic component follows an exponential decay, where the decay exponent varies depending on the environment. 
Some models also incorporate an additional attenuation factor to account for fixed obstructions, such as walls, glass windows, and floors. 
Meanwhile, the statistical component captures power fluctuations caused by multipath propagation and random blockages.

The \ac{RSS} at the  $j$-th \ac{SU} can be expressed by the lognormal \ac{PLM} given by 
\begin{equation} \label{received_power_linear}
    P_j =  P_0 - 10\beta\log_{10}\left(d_j/d_0\right) + n_j,
\end{equation}
where $\beta$ represents the decay exponent, $n_j$ is Gaussian distributed shadowing noise, $d_j$ denotes the Euclidean distance between the source and the $j$-th \ac{SU}, and $P_{0}$ is the power received at the reference distance $d_0$, which should be larger than the Fresnel distance. 

The lognormal \ac{PLM} aligns relatively well with empirical data, but is mainly used for coverage prediction. Thus, it is not a positioning-specific propagation model.
Unfortunately, this model does not lead to a low-complexity closed-form position estimator, which means that approximations must be applied to obtain real-time solutions.
Hence, \ac{ML} is an attractive solution to \ac{RSS}-based positioning.
Recent investigations have shown that a relatively small fully-connected \ac{NN} outperforms model-based positioning techniques~\cite{experimental_dnnrss}.

\subsubsection{Coherent Cooperative Transmitter Positioning}

In positioning systems, all relevant parameters are estimated from the received signals. 
Theoretically, utilizing these raw signals as input to an \ac{NN} should enable positioning accuracy comparable to classical methods, as they inherently contain all the necessary information about the source’s location.
However, a significant challenge arises when training such a model as the number of required training examples increases exponentially with the input dimension, thus rendering the learning process computationally expensive and difficult to generalize effectively.
To address this issue, one common approach is to reduce the dimensionality of the received signals by extracting specific features, such as the \ac{RSS}. 
By summarizing the signal information into a more compact representation, the computational complexity is reduced, allowing for more efficient model training. 
However, this simplification comes at a cost, as valuable information about the characteristics of the signal is inevitably lost in the process.

In this context, feature engineering plays a critical role in ensuring that \ac{ML} algorithms perform optimally. 
The success of a model is directly influenced by how well the input data is represented, as different representations can significantly impact learning efficiency and accuracy. 
The primary objective of feature engineering is to derive representations that capture the underlying structure of the data in a way that facilitates learning and generalization. 
By carefully selecting and transforming relevant features, it is possible to enhance model performance while maintaining computational efficiency.
Striking the right balance between dimensionality reduction and information retention is key to developing robust and scalable positioning solutions.

If the \acp{SU} are clock synchronized, the \ac{CU} can coherently exploit the position-related information in the received signals, and \ac{ML} offers us the flexibility for engineering an enhanced position information parameter.
In \cite{picm}, the authors define a \ac{PICM} where the main diagonal matrix contains the \ac{RSS} values at each \ac{SU}, and the off-diagonal elements of the matrix indicate the correlation levels between sensing units, which depend on their relative positions with respect to the source. By analyzing both the diagonal and off-diagonal elements, it is possible to gain insights into the spatial distribution of received signals, improving the accuracy of \ac{ML} models.

\subsection{Multi-Task Learning for Spectrum Analysis}

The information extracted from wireless signals via spectrum monitoring can be used for various tasks.
It is, however, possible to go further and leverage the same input to solve multiple tasks jointly, thus yielding extra functionality.
The main argument behind the \ac{MTL} concept is that such an extra functionality can often be obtained at a reasonable \ac{DNN} complexity cost and, if the tasks are related, it can even improve the performance of one or a few tasks~\cite{MultiTaskLearning2022}.

For instance, detection of the unused or underused portions of a radio frequency spectrum can be done jointly or subsequently with signal classification.
This straightforward \ac{MTL} application is explored in~\cite{Miloshevski2024}, where the \ac{DNN} is trained without prior knowledge of the signals, thus following an unsupervised learning paradigm.
To this end, the captured radio signals represented as 1D FFT are fed to a \ac{CNN}-based network, which learns how to cluster the signals. 
Before the model can be used for the inference, the clusters, whose number is not predefined, are manually labeled by a human expert into categories, e.g., 4G LTE, 5G NR, Wi-Fi, Noise, etc.
After testing on 3 datasets and evaluating against \ac{AE}-based state-of-the-art architectures, the authors report superior performance and reduced computational complexity of their proposed approach.

Furthermore, analysis of the same signal can be used to extract the information related to the \ac{TX} position.
Based on this premise, the authors of~\cite{Shatov_ISWCS2024} proposed a framework to simultaneously classify wireless signal waveforms and locate the \ac{TX}.
To this end, they model an indoor scenario with a single TX, which broadcasts a signal captured by several distributed sensors in the form of IQ samples.
Subsequently, portions of these IQ samples are used as input for a dual-task \ac{DNN}, which jointly estimates the wireless signal waveform classes and the \ac{TX} coordinates.
The framework is refined, extended, and further analyzed in~\cite{DualTaskLearning2025}. 
The loss function comprises weighted cross-entropy loss and \ac{MSE} for waveform classification and TX localization, defined by (\ref{MSE_eq}) and (\ref{CrossEntropy_eq}), respectively.
While the values of $\mathcal{L}_{\text{CrossEntropy}}$ and $\mathcal{L}_\text{MSE}$ characterize the learning of individual tasks, the \ac{DNN} parameters are optimized to minimize the composite loss
\begin{equation}\label{TotalLoss}
    \mathcal{L}_\text{total} = \mathcal{L}_{\text{MSE}} + \lambda\cdot\mathcal{L}_{\text{CrossEntropy}},
\end{equation}
where the weight $\lambda$ balances individual classification and regression tasks loss contributions since they might have considerably different values~\cite{weighted_multitask}.
The reported results show that dual functionality can be implemented at a low additional computational cost.

\begin{figure}
\centering
    \includegraphics[width=0.85\linewidth]{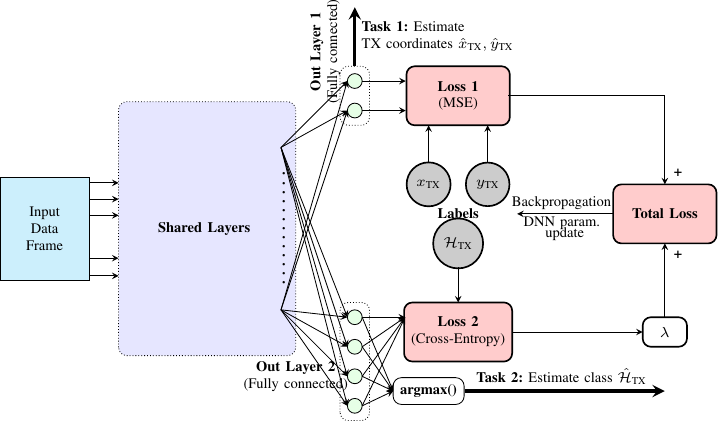}
\caption{Overview of the \ac{ML}-based framework for dual-task learning for 4 waveform classes proposed in~\cite{DualTaskLearning2025}.} \label{fig:dualtask}
\end{figure}

Fig.~\ref{fig:dualtask} illustrates a generic dual-task learning framework to jointly localize the \ac{TX} and classify its signal.
The supervised learning requires labeled training data where each input frame $s_{\text{RX}}$ has attached labels of waveform class and 2D coordinates $\{s_{\text{RX}}, \mathcal{H}_{\text{TX}}, x_{\text{TX}}, y_{\text{TX}}\}$.
Input data $s_{\text{RX}}$ in the original work~\cite{DualTaskLearning2025} consists of raw IQ samples, however, reduced dimensionality representations could be an option to improve the algorithm complexity and inference time.
The shared layers are shown as a black box since their configuration can vary significantly and their precise architecture is chosen empirically\footnote{In~\cite{DualTaskLearning2025}, the input data frame comprises the stacked vectors of 256 I and Q samples recorded by sensing units. 
The shared layers are represented by 4 \ac{LSTM} and 2 fully connected layers.}, the \ac{DNN} outputs the estimated TX coordinates in 2D space $\hat{x}_{\text{TX}}, \hat{y}_{\text{TX}}$ (Output Layer 1) and the \ac{TX} waveform class $\hat{\mathcal{H}}_{\text{TX}}$ (Output Layer 2), where four neurons correspond to the number of waveform classes in the dataset.
The introduced architecture assumes the presence of a \ac{TX}. 
Note that it can be further extended and trained to identify pure Noise input by simply adding another node to the Output Layer 2. 
Since this potential extension can be interpreted as spectrum sensing functionality, the \ac{NN} becomes a triple-task system.

\subsection{Channel Charting}

Model-based \ac{TX} positioning techniques operate on the premise of an assumed channel model, such as a \ac{LoS} channel or a channel with limited multipath components.
In contrast, data-based techniques explicitly exploit the unique characteristics of the wireless channel, often bypassing the need for a predefined model.
Examples include fingerprinting, which leverages location-specific channel signatures, and channel charting \cite{studer_cc}, a novel method that maps high-dimensional \ac{CSI} into a low-dimensional space while preserving geometric relationships.

Fingerprinting requires a reference positioning system to acquire labels for supervised training.
Channel charting, on the other hand, is based on dimensionality reduction techniques to extract meaningful representations from high-dimensional \ac{CSI} data.
This process facilitates the creation of a "chart" where the relative positions of users in the environment are captured.
Unlike fingerprinting, channel charting is self-supervised and does not require absolute positioning information.
The data-based techniques only rely on the assumption that there exists an injective mapping between the physical location of a \ac{TX} and its associated channel, which is reasonable for sufficiently static environments.

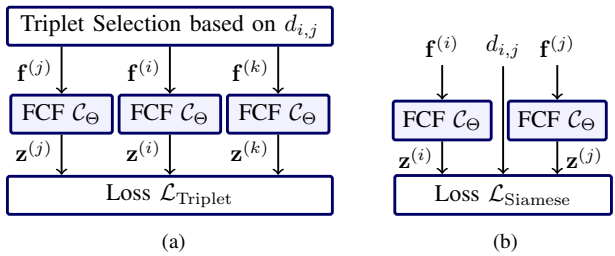
\begin{figure}
    \centering
    \begin{subfigure}[b]{0.59\columnwidth}
        \centering
        \scalebox{0.9}{
            \begin{tikzpicture}
    \tikzstyle{box1} = [rectangle, draw = blue!40!black, very thick, rounded corners=1pt,inner sep = 2pt, align = center, fill = blue!5!white, minimum height=16pt,minimum width=1.5cm]
    \tikzstyle{box2} = [rectangle, draw = blue!40!black, very thick, rounded corners=1pt,inner sep = 2pt, align = center, minimum height=16pt,minimum width=4.8cm]
    
    \node[box2] (tripletsel) at (4.0, 0.2) {Triplet Selection based on $d_{i,j}$};
    
    \node[box1] (dnn_1) at (2.4,-1.1) {FCF $\mathcal{C}_\Theta$};
    \node[box1] (dnn_2) at (4,-1.1) {FCF $\mathcal{C}_\Theta$};
    \node[box1] (dnn_3) at (5.6,-1.1) {FCF $\mathcal{C}_\Theta$};
    
    
    \node[box2] (triplet_loss) at (4.0,-2.3) {Loss $\mathcal{L}_\mathrm{Triplet}$};
    
    \draw [->, thick] ($(tripletsel.south) + (-1.6, 0)$) -- (dnn_1.north) node[midway,anchor=east] {$\mathbf{f}^{(j)}$};
    \draw [->, thick] ($(tripletsel.south) + (0, 0)$) -- (dnn_2.north) node[midway,anchor=east] {$\mathbf{f}^{(i)}$};
    \draw [->, thick] ($(tripletsel.south) + (1.6, 0)$) -- (dnn_3.north) node[midway,anchor=east] {$\mathbf{f}^{(k)}$};
    
    
    \draw [->, thick]  (dnn_1.south) -- (dnn_1.south|-triplet_loss.north)
    node[pos=0.4,anchor=east]{$\mathbf{z}^{(j)}$};
    \draw [->, thick]  (dnn_2.south) -- (dnn_2.south|-triplet_loss.north)
    node[pos=0.4,anchor=east]{$\mathbf{z}^{(i)}$};
    \draw [->, thick]  (dnn_3.south) -- (dnn_3.south|-triplet_loss.north)
    node[pos=0.4,anchor=east]{$\mathbf{z}^{(k)}$};
\end{tikzpicture}
        }
        \caption{}
        \label{fig:triplet_nn_structure}
    \end{subfigure}
    \begin{subfigure}[b]{0.39\columnwidth}
        \centering
        \scalebox{0.9}{
            \begin{tikzpicture}
    \tikzstyle{box1} = [rectangle, draw = blue!40!black, very thick, rounded corners=1pt,inner sep = 4pt, align = center, fill = blue!5!white, minimum height=16pt,minimum width=1.4cm]
    \tikzstyle{box2} = [rectangle, draw = blue!40!black, very thick, rounded corners=1pt,inner sep = 4pt, align = center, minimum height=16pt,minimum width=3.2cm]

    \node (in_1) at (2.1,0) {$\mathbf{f}^{(i)}$};
    \node (in_2) at (3.8,0) {$\mathbf{f}^{(j)}$};
    
    \node (in_3) at (3,0) {$d_{i,j}$};
    
    \node[box1] (dnn_1) at (2.1,-1.1) {FCF $\mathcal{C}_\Theta$};
    \node[box1] (dnn_2) at (3.8,-1.1) {FCF $\mathcal{C}_\Theta$};
    
    
    \node[box2] (contrastive_loss) at (3,-2.2) {Loss $\mathcal{L}_\mathrm{Siamese}$};
    
    \draw [->, thick]  (in_1.south) -- (dnn_1.north)
    node[midway,anchor=east]{};
    \draw [->, thick]  (in_2.south) -- (dnn_2.north)
    node[midway,anchor=east]{};
    
    \draw [->, thick]  (in_3.south) -- (contrastive_loss)
    node[midway,anchor=east]{};
    
    
    \draw [->, thick]  (dnn_1.south) -- (dnn_1.south|-contrastive_loss.north)
    node[midway,anchor=east]{$\mathbf{z}^{(i)}$};
    \draw [->, thick]  (dnn_2.south) -- (dnn_2.south|-contrastive_loss.north)
    node[midway,anchor=west]{$\mathbf{z}^{(j)}$};
\end{tikzpicture}
        }
        \caption{}
        \label{fig:siamese_nn_structure}
    \end{subfigure}
    \label{fig:cc_nn_architectures}
    \caption{Structure of Neural Networks commonly used for channel charting: (a) Triplet Neural Network and (b) Siamese Neural Network.}
\end{figure}

\begin{figure}
    \centering
    \begin{subfigure}[t]{0.49\columnwidth}
        \begin{tikzpicture}
            \tikzstyle{every node}=[font=\small]
            \begin{axis}[
                width=0.65\textwidth,
                height=0.65\textwidth,
                scale only axis,
                xmin=-14,
                xmax=4.2,
                ymin=-15.8,
                ymax=-0.2,
                xlabel = {Coordinate $\tilde z_1 ~ [\mathrm{m}]$},
                ylabel = {Coordinate $\tilde z_2 ~ [\mathrm{m}]$},
                ylabel shift = -8 pt,
                xlabel shift = -4 pt,
                xtick={-10, -6, -2, 2}
            ]
                \addplot[thick,blue] graphics[xmin=-12.5,ymin=-14.5,xmax=2.5,ymax=-1.5] {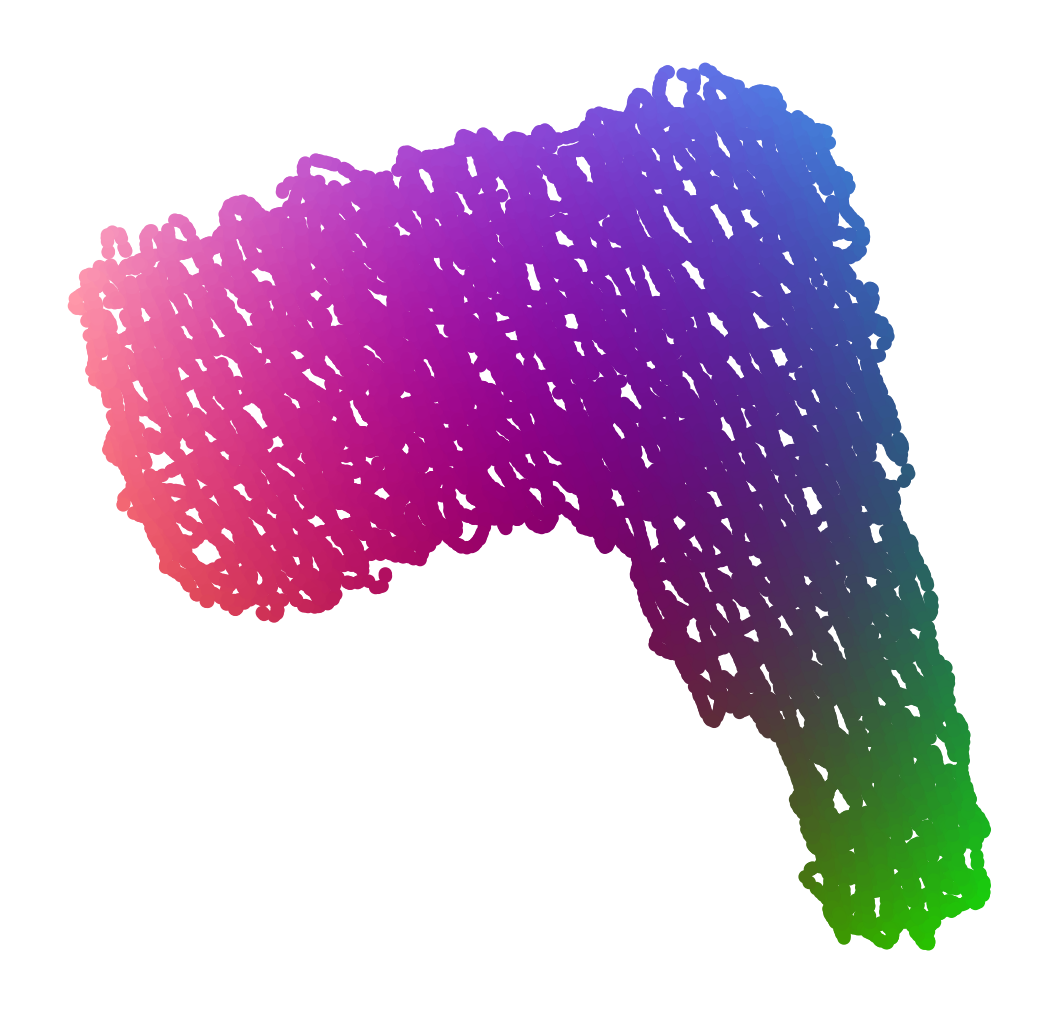};
            \end{axis}
        \end{tikzpicture}
        \caption{Reference positions}
    \end{subfigure}
    \begin{subfigure}[t]{0.49\columnwidth}
        \begin{tikzpicture}
            \tikzstyle{every node}=[font=\small]
            \begin{axis}[
                width=0.65\textwidth,
                height=0.65\textwidth,
                scale only axis,
                xmin=-14,
                xmax=4.2,
                ymin=-15.8,
                ymax=-0.2,
                xlabel = {Coordinate $\tilde z_1 ~ [\mathrm{m}]$},
                ylabel = {Coordinate $\tilde z_2 ~ [\mathrm{m}]$},
                ylabel shift = -8 pt,
                xlabel shift = -4 pt,
                xtick={-10, -6, -2, 2}
            ]
                \addplot[thick,blue] graphics[xmin=-13,ymin=-14.8,xmax=3.2,ymax=-1.2] {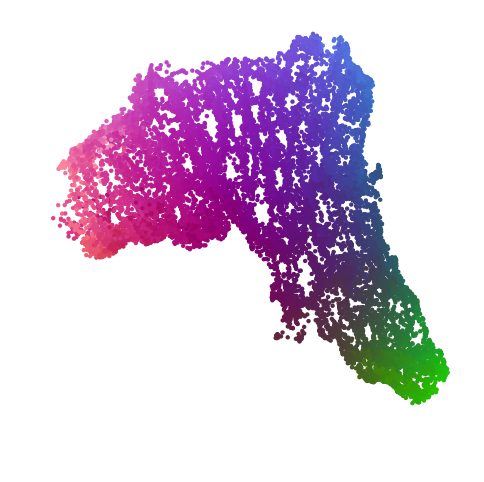};
                \node (arrayc) at (axis cs: -1.531375, -15.0595) [fill = green!50!black, minimum width = 1.5, minimum height = 6, inner sep = 0pt, rotate = 77.8, anchor = center] {}; 
            \end{axis}
        \end{tikzpicture}
        \caption{Learned channel chart and antenna array (green rectangle)}
    \end{subfigure}
    \caption{Reference positions and learned channel chart in a top view chart. Measurement points are colorized and colors are preserved in the channel chart.}
    \label{fig:channelchart}
\end{figure}

The charting process typically involves training the so-called \ac{FCF} $\mathcal C_\Theta$ to identify and exploit intrinsic structures within the \ac{CSI} data, making it more robust in complex environments.
Often, channel charting relies on the notion of a \emph{dissimilarity} between channels: Channels (each corresponding to a location in physical space) that are similar with respect to some dissimilarity metric (e.g., angle delay profile, timestamps), are said to have a low dissimilarity.
If the \ac{FCF} that performs the dimensionality reduction is realized as a \ac{NN}, contrastive learning based on the dissimilarity can be applied.
Two common \ac{NN} structures for contrastive learning are the triplet \ac{NN} shown in Fig. \ref{fig:triplet_nn_structure} and the Siamese \ac{NN} shown in Fig. \ref{fig:siamese_nn_structure}.
In the case of the triplet \ac{NN}, a triplet selection step presents three channel feature vectors to a weight-sharing \ac{NN}: Two features vectors $\mathbf f^{(i)}$ and $\mathbf f^{(j)}$, which are considered to be similar to each other, and a third feature vector $\mathbf f^{(k)}$, which is considered to be dissimilar to $\mathbf f^{(i)}$ according to the dissimilarity metric.
The triplet loss function is then designed to be minimal for channel charts where the low-dimensional latent space representations $\mathbf z^{(i)}$ and $\mathbf z^{(j)}$ are close to each other over all considered triplets $(i,j,k)$, while also ensuring that $\mathbf z^{(k)}$ is far away from $\mathbf z^{(i)}$.
For the Siamese \ac{NN}, on the other hand, the loss function is designed to ensure that the Euclidean distance between any two latent space representations $\mathbf z^{(i)}$ and $\mathbf z^{(j)}$ in the Channel Chart matches the dissimilarity $d_{i,j}$ between those points as closely as possible.
In general, the Triplet \ac{NN}-based approach is more flexible and is compatible with various dissimilarity metrics.
However, training the \ac{FCF} in Siamese \ac{NN} can lead to better performance, if a good dissimilarity metric is chosen~\cite{stephan2024angle}.

An exemplary channel chart, taken from \cite{euchner2024uncertainty}, is shown in Fig. \ref{fig:channelchart} next to a chart of reference positions.
Every point in the channel chart and reference position chart corresponds to one \ac{CSI} measurement between a \ac{UE} and a single antenna array with $2\times4$ antennas, operating at a carrier frequency of $1.272\,\mathrm{GHz}$ and a bandwidth of $50\,\mathrm{MHz}$.
The \ac{CSI} dataset is called \textit{dichasus-cf0x} and taken from the DICHASUS dataset collection \cite{dichasus2021}.
Even though large parts of the measurement are in a \ac{NLoS} region for the antenna array or at least severely affected by multipath propagation, channel charting is able to reconstruct the global L-shaped topology of the environment so that the learned \ac{FCF} can now be used for localization in both \ac{LoS} and \ac{NLoS} areas.

Channel charting can not only enable accurate user tracking and navigation, but also provides relative position estimates to the wireless network, which offers opportunities for improving network management.
For example, pilots or spectral resources may be allocated based on user mobility patterns inferred from the channel chart.
\section{Exploring \ac{AI}-Driven Synergies Between Different Radio Sensing Modalities: Opportunities and Challenges}\label{sec:integr}

\begin{figure*}
\centering
    \includegraphics[width=0.875\linewidth]{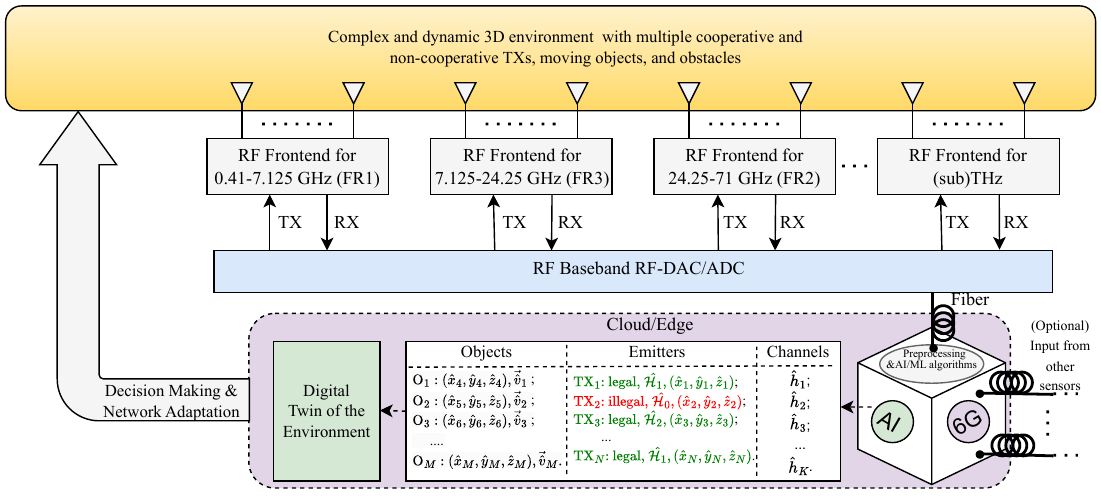}
\caption{Illustration of the envisioned \ac{AI}-enhanced multi-functional radio sensing integrated into the wireless network.
We consider the dynamic environment with a complex propagation structure where multiple \ac{TX} and passive objects are present.
The sensing system is built on high-performance hardware, starting with the \ac{RF} frontends, which are capable of scanning multiple channels in various FRs with an instantaneous bandwidth of a few GHz.
\ac{RF} baseband module is an \ac{FPGA} which serves, among other things, as a DAC/ADC. In the \ac{RX} mode, it outputs the complex baseband signals with high sampling frequency, which are subsequently fed via the fiber (or, alternatively, high-speed wireless backhaul links) to the cloud/edge computer, which performs the \ac{AI}-based signal processing.
Diverse information regarding the location and properties of the physical objects, \acp{TX}, and propagation structure, obtained through \ac{AI}-based algorithms, is then used to build an ultimate \ac{DT} of the environment.
Such a \ac{DT} facilitates the network's self-management functions, including interference management, beamforming, threat detection, etc.
Incoming from the right side connections denote a system extension opportunity to implement distributed configurations, allowing data fusion algorithms.} \label{fig:Integrated_Vision}
\end{figure*}

This section outlines the vision and potential strategies for integrating sensing modalities discussed in sections~\ref{sec:passive_object_sensing} and~\ref{sec:spectral_sens} into future wireless networks.
We reinforce the significance of both the physical and electromagnetic environments' awareness, while also considering the major challenges and potential synergies between these domains.
Fig.~\ref{fig:Integrated_Vision} illustrates several aspects of the \ac{AI}-powered \ac{ISAC} system presented in this section, and shows how system components and functionalities could be integrated in \ac{6G}.

\subsection{Potential Benefits}

It has been envisioned that in the future, there will be more integration between various sensing tasks~\cite{MultiModalSensing_Cheng_2024}.
By engaging different yet logically adjacent sensing modalities, the system can develop a holistic understanding of complex physical and electromagnetic environments and execute various tasks, thus fulfilling high expectations for future radio.
Denoted as \ac{MM-MTL}~\cite{MMMTL_2021}, this paradigm is capable of simultaneously tackling tasks in a variety of domains and promising improved generalization and robustness.
Below are a few possible examples of how \ac{MM-MTL} can benefit wireless networks.

\subsubsection{Improved Security}

Sensing is a cornerstone of network security.
In a critical infrastructure setting (e.g., an airport surveillance system), radar could be used to detect the presence of unauthorized drones~\cite{JRC_Shatov}, while spectrum sensing identifies if those drones interfere by transmitting on the same frequencies the wireless network uses.
The system could then adapt by switching the frequency or adjusting the transmit power to minimize the negative impact.
Spectrum sensing alone can detect an interferer, but signal classification can provide additional insights about interference type and source.
Furthermore, after detection, radar and non-cooperative \ac{TX} localization techniques can complement each other to pinpoint the drone.
Consequently, the wireless system might initialize counter-measures (e.g., removal) against the detected object.
This synergetic approach can be applied to other scenarios, e.g., industrial monitoring systemsdetecting unauthorized sources.

\subsubsection{Optimized Waveform Representation for Sensing}
AI can not only help to perform the sensing task itself, but also identify the optimal waveform representation for specific sensing problems under certain conditions.
For instance, it can help choose the minimum acceptable number of IQ samples for signal representation for raw data-driven \ac{AMC} task, given the \ac{SINR} conditions and the required accuracy.
Furthermore, in some cases (e.g., low \ac{SINR}), \ac{AI} might decide to switch to statistical feature-based signal representation due to its potentially higher robustness, aiming to balance computational burden and traffic load according to accuracy requirements.

\subsubsection{\ac{AI}-Empowered Multi-Modal Radio Sensing as a Self-Sustaining Network Enabler}

Early vision papers predicted that \ac{6G} will be a \ac{SSN} capable of continuously maintaining its \acp{KPI} under dynamic and complex environments~\cite{6Gvision_2020}.
Realization of the \ac{SSN} concept can only rely on a high volume of environmental data, which can be provided via multi-modal sensing, while \ac{AI} algorithms ensure its optimal utilization.

The network's awareness of both the physical environment and electromagnetic situation, potentially empowered by \ac{MM-MTL}, enables \acp{SSN} as follows:

\begin{itemize}
    \item Radar can be used to detect blockers to adapt the beamforming.
    \item Spectrum sensing can help to identify idle spectrum, optimize channel assignment, and transmit power.
    \item Signal classification can leverage interference management, providing detailed information about \acp{TX} in the monitored area.
    \item Non-cooperative \ac{TX} localization and channel charts can be used to improve the accuracy of positioning-based services as well as refine beamforming.
\end{itemize}

\ac{AI} models can automatically optimize the network parameters, significantly increasing the degree of network autonomy.
The concept of \ac{SSN} aligns with \acp{DT}, where building a \ac{DT} containing full knowledge of current \ac{EM} and the physical environments can be seen as an intermediate step.

\subsection{Implementation Aspects}

Integrating the multi-band, multi-modal sensing functionality into future wireless networks will require tackling many practical issues.
We assume a sensing-centric system, where communication is only used to exchange information regarding captured environmental data and those necessary for inference (e.g., parameters for \ac{FL}-based model).
The sensing can be realized as a dedicated multifunctional sensor module.
Multifunctional sensors capable of spectrum sensing in the context of cognitive radar coexisting with communications networks were proposed in~\cite{Martone2021}.
For the considered \ac{MM-MTL} system, we take a step forward and imply a multifunctional sensor module as a platform integrating several sensing functionalities $\text\textendash$ for instance, radar, spectrum sensing, signal classification, non-cooperative TX localization, into an AI-powered dedicated sensing module. 
This module(s) could then be integrated into the radio access network, the facility surveillance system, etc., similarly to~\cite{FNWF2023}.

In the following, we outline hardware and real-time learning aspects of such a multi-task sensor module. 
In addition, we discuss architectural considerations and advocate for a distributed sensing approach to amplify the performance and reliability of this ambitious and complex hypothetic system.

\subsubsection{Hardware}
Bringing the multi-task \ac{ISAC} to reality in a chaotic wireless environment will require advanced hardware components with a high degree of adaptivity.
The hardware-related topics and challenges to be overcome for joint radar-communications are scrutinized in~\cite{JRC_Shatov}.
To also enable intelligent spectrum analysis functionality, high-performance \ac{RF} frontends are essential since they provide algorithms with input raw data.
This includes the need for wideband ADCs, \ac{RF} amplifiers, and antenna arrays to capture a wide range of signal types.
In particular, they should be capable of simultaneously tracking multiple frequencies and signal acquisition in multiple frequency bands, i.e., conversion of analog waveforms into digital form (baseband IQ samples) for storage and further processing by \ac{ML} algorithms, as shown in Fig.~\ref{fig:Integrated_Vision}.
For example, an \ac{RF} frontend based on direct sampling architecture from~\cite{Schuster2023} is designed specifically for diverse sensing applications.
It enables simultaneous sampling of all frequency bands below 6\,GHz used in present-day mobile communications systems in Germany\footnote{Although most standard radar frequency ranges are above 6\,GHz, it is widely used, e.g., for long-range air traffic control and surveillance, long-range weather, and marine radar applications.}.

In 6G networks, antennas will be reconfigurable, driving increased interest towards dynamic metasurface antennas (DMAs).
DMAs can precisely and simultaneously control all key properties of electromagnetic waves, including amplitude, phase, frequency, polarization, etc~\cite{antennas2023}. 
This capability enables operation across wide frequency ranges, making them ideal for wideband spectrum sensing and diverse signal classification.
Metasurfaces can act as selective filters, allowing specific signal types to pass through, simplifying signal classification based on spectral, temporal, or polarization characteristics.
The tunable receiving patterns of DMAs are particularly valuable for passive sensing applications, while tunable radiation patterns are of interest for radar or channel sounding.
A recently developed DMA prototype, operating in the 60 GHz mmWave band, is presented in~\cite{60GHz_metasurface}. 
Combined in large numbers, such DMAs form a foundation for emerging holographic \ac{MIMO} (HMIMO) receivers, spectrally agile structures capable of recording and manipulating electromagnetic fields with extreme flexibility and accuracy~\cite{Hol_MIMO_6G}.
Enhanced with \ac{AI}, HMIMO technology shows great potential for advancing 6G wireless applications, including futuristic intelligent spectrum monitoring and \ac{ISAC} systems.

The bottleneck of many \ac{DL} algorithms reviewed in Sec.~\ref{sec:passive_object_sensing} and~\ref{sec:spectral_sens} is that they often impose a heavy computational burden, especially those operating on high-dimensional raw data. 
It might cause problems like a slow training process and inference time, which then fail to meet real-time constraints.
Leveraging \acp{GPU} and \acp{TPU} can help to meet real-time requirements of present and future wireless networks. 
For large capacity \acp{DNN}, \acp{GPU} and \acp{TPU} can accelerate the process by orders of magnitude.
High-performance \acp{GPU} and \acp{TPU} are relatively expensive, leaving the question of practicality and affordability to a specific use case and budget constraints. 
Alternatives are the use of lightweight models or reduced dimensionality data representation. However, it usually results in performance loss.
It is up to the particular application whether the accuracy is acceptable.

Besides high-performance components for wireless systems, an effort is underway to develop holistic testbeds for \ac{6G} and beyond research.
For this purpose, some solutions aim not only to fulfill the communications requirements, but also specifically focus on enabling extra sensing features, such as distributed \ac{MIMO} propagation measurements, radar target characterization, and \ac{ISAC}~\cite{Open6GHubTestbed2024},\cite{TUI_Testbed2024}.
In particular, the broadband \ac{6G} \ac{MIMO} testbed of the Open6GHub Project~\cite{Open6GHubTestbed2024} offers a versatile hardware platform that supports multiple frequency bands for different use cases and scenarios, focusing on radar-based sensing and spectrum sensing.
This testbed comprises high-performance hardware capable of wideband sampling and a backend server system that allows running scripts in MATLAB or Python in real time to test algorithms, including those \ac{AI}-driven.

\subsubsection{Architectural Aspects}

Many algorithms described in sections~\ref{sec:passive_object_sensing} and~\ref{sec:spectral_sens} either explicitly use distributed observations or could potentially benefit from them.
Being integrated into a single node, sensing performance might suffer from blockage and other detrimental channel effects or the disability to monitor the whole area of interest due to lack of sensitivity, not to mention their potential breakdown.
Moreover, some applications, like blind \ac{TX} localization, are infeasible through a single observation.
We therefore advocate for distributed and collaborative sensing since it helps to overcome these issues, increasing the accuracy, coverage, scalability, and robustness~\cite{DISAC}.
Concerning \ac{AI}-powered networks, distributed sensing will naturally ensure more data for data-driven algorithm design.

The nodes of such a distributed sensing network might (\textit{i}) comprise ubiquitous heterogeneous \ac{IoT} devices with diverse sensing capabilities sharing the sensed data with an edge computer, (\textit{ii}) be a network of high-performance dedicated modules at predefined locations (possibly, but not necessarily co-located with base stations or access points, or edge computers) explicitly designed to obtain spatial and spectrum awareness, or (\textit{iii}) combine and balance contributions of both approaches, leveraging device diversity of (\textit{i}) and high performance of individual nodes of (\textit{ii}).

In addition to the nodes' functionality, a distributed architecture must also address how to balance the computational load and efficiently share information across the network to ensure real-time decision-making. 
With this in mind, the role of \ac{AI} is not limited to the inference of some sensing parameters.
Addressing multiple access issues is essential to enable efficient communication and ensure reliable data transmission in dense, resource-limited networks~\cite{MultipleAccess6G_2024}.
\ac{AI} algorithms can optimize computational load across the network by analyzing the current load on network nodes and communication links.
Thus, the concepts of \ac{FL} and \ac{RL} are highly important to ensure real-time adaptability in dynamic and unknown environments.

\ac{RL} is a natural choice for practical wireless applications since it is capable of learning while interacting with the environment in real-time.
\ac{RL} is widely used for radio resource allocation for real-time dynamic configurations in multi-device networks~\cite{RL_MAC_2012}.
In the context of sensing applications, addressing privacy concerns is particularly important.
This issue is explicitly tackled via \ac{FL}, where the local data recorded through sensing are not shared over the air.
Furthermore, since \ac{FL} assumes only sharing the model parameters, it reduces the stress over wireless links compared to raw data sharing.
In wireless networks, choosing the devices that share their learned model parameters is a significant practical problem, as the principle 'the more, the better' is not always suitable.
Over-the-air \ac{FL} device scheduling has been widely accepted as an efficient sub-optimal solution, which reduces the performance degradation caused by imperfect model aggregation~\cite{FL_device_scheduling_2023}.
In brief, device scheduling aims to identify the set of devices that contribute to the aggregated model in the most constructive way, while turning the rest off.
Finally, \ac{FL} and \ac{RL} can be used jointly to preserve data privacy and simultaneously adapt the behavior in real-time~\cite{FLRL_2020}.

\section{Conclusion}\label{sec:conclusion}

Radio sensing and \ac{ML} are two intelligence enablers of future wireless networks, which have attracted significant attention from the research community and standardization organizations.
Radio sensing aims to provide \ac{6G} wireless networks with a massive amount of raw data, which can be leveraged to extract the information—from type to location—about objects, \acp{TX}, and propagation environment. 
Meanwhile, \ac{ML}-based algorithms will ensure unprecedented accuracy, autonomy, and efficiency of the networks.
In this context, our tutorial offers guidelines and provides a comprehensive review of radio sensing capabilities as a crucial component of \ac{6G}.
For this purpose, we first reviewed the necessary background in wireless and \ac{ML}.
Then, we presented \ac{ML}-enhanced signal processing methods for radar-based sensing of passive objects, and their possible integration into future communication systems.
Furthermore, we detailed various aspects of \ac{ML}-enhanced passive \ac{EM} sensing, including standard spectrum sensing, signal classification, non-cooperative \ac{TX} localization, and channel charting. 
Additionally, we addressed several challenges and made assumptions about how radio sensing could be integrated into \ac{6G} networks. 
Based on this, we finally developed a vision of an \ac{AI}-powered multifunctional radio sensor network featuring radar, spectrum sensing, and signal classification capabilities which could facilitate autonomous real-time monitoring, decision making, and adaptation to complex and dynamic environments.
We hope that our work can serve as a starting point for novice researchers and engineers to better understand the state of the art, challenges, and possible research directions in the field of \ac{AI}-enhanced radio sensing for future \ac{6G} wireless networks.


\bibliography{IEEEabrv,literature_short.bib,refs_FAU}

\begin{thebibliography}{100}
\providecommand{\url}[1]{#1}
\csname url@samestyle\endcsname
\providecommand{\newblock}{\relax}
\providecommand{\bibinfo}[2]{#2}
\providecommand{\BIBentrySTDinterwordspacing}{\spaceskip=0pt\relax}
\providecommand{\BIBentryALTinterwordstretchfactor}{4}
\providecommand{\BIBentryALTinterwordspacing}{\spaceskip=\fontdimen2\font plus
\BIBentryALTinterwordstretchfactor\fontdimen3\font minus \fontdimen4\font\relax}
\providecommand{\BIBforeignlanguage}[2]{{%
\expandafter\ifx\csname l@#1\endcsname\relax
\typeout{** WARNING: IEEEtran.bst: No hyphenation pattern has been}%
\typeout{** loaded for the language `#1'. Using the pattern for}%
\typeout{** the default language instead.}%
\else
\language=\csname l@#1\endcsname
\fi
#2}}
\providecommand{\BIBdecl}{\relax}
\BIBdecl

\bibitem{LeCun2015}
Y.~LeCun, J.~Bengio, and G.~Hinton, ``Deep learning,'' \emph{Nature}, vol. 521, no. 7553, pp. 235--244, 2015.

\bibitem{yolo}
J.~Redmon, S.~Divvala, R.~Girshick, and A.~Farhadi, ``You only look once: Unified, real-time object detection,'' in \emph{Proc. IEEE Conf. on Computer Vision and Pattern Recognition (CVPR)}, 2016, pp. 779--788.

\bibitem{speech}
G.~Hinton \emph{et~al.}, ``Deep neural networks for acoustic modeling in speech recognition: The shared views of four research groups,'' \emph{IEEE Signal Process. Mag.}, vol.~29, no.~6, pp. 82--97, 2012.

\bibitem{ThesisMitola2000}
J.~Mitola, ``Cognitive radio: An integrated agent architecture for software defined radio,'' \emph{PhD dissertation}, 2000.

\bibitem{CR_overview2005}
S.~Haykin, ``Cognitive radio: brain-empowered wireless communications,'' \emph{IEEE J. Sel. Areas Commun.}, vol.~23, no.~2, pp. 201--220, 2005.

\bibitem{OSheaHoydis2017}
T.~O’Shea and J.~Hoydis, ``An introduction to deep learning for the physical layer,'' \emph{IEEE Trans. Cogn. Commun. Netw.}, vol.~3, no.~4, pp. 563--575, 2017.

\bibitem{tenBrinkHoydis2018}
S.~Dörner, S.~Cammerer, J.~Hoydis, and S.~ten Brink, ``Deep learning based communication over the air,'' \emph{IEEE J. Sel. Topics Signal Process.}, vol.~12, no.~1, pp. 132--143, 2018.

\bibitem{Bjornson2019}
E.~Björnson, L.~Sanguinetti, H.~Wymeersch, J.~Hoydis, and T.~L. Marzetta, ``Massive {MIMO} is a reality—what is next?: Five promising research directions for antenna arrays,'' \emph{Digit. Signal Process.}, vol.~94, pp. 3--20, 2019.

\bibitem{Letaief2019}
K.~B. Letaief, W.~Chen, Y.~Shi, J.~Zhang, and Y.-J.~A. Zhang, ``The roadmap to {6G}: {AI} empowered wireless networks,'' \emph{IEEE Commun. Mag.}, vol.~57, no.~8, pp. 84--90, 2019.

\bibitem{3GPP_data}
{3GPP TR 37.817}, ``Study on enhancement for data collection for {NR and EN-DC},'' \emph{V17.0.0}, 2022.

\bibitem{FNWF2023}
V.~Shatov, M.~Lübke, Y.~Su, and N.~Franchi, ``On integrated cooperative radio sensing for spatial electromagnetic analysis in {6G},'' in \emph{Proc. IEEE Future Networks World Forum (FNWF)}, 2023, pp. 1--8.

\bibitem{Lin2023_1}
X.~Lin, L.~Kundu, C.~Dick, and S.~Velayutham, ``Embracing {AI in 5G}-advanced toward {6G}: A joint {3GPP and O-RAN} perspective,'' \emph{IEEE Commun. Stand. Mag.}, vol.~7, no.~4, pp. 76--83, 2023.

\bibitem{Lin2023_2}
X.~Lin, ``Artificial intelligence in {3GPP 5G-A}dvanced: A survey,'' \emph{IEEE ComSoc Technology News}, 2023.

\bibitem{3GPP_ISAC}
{3GPP}, ``Feasibility study on integrated sensing and communication,'' \emph{Tech. Rep. 22.837 (Version 19.3.0)}, 2024.

\bibitem{AI_ISAC_2024}
N.~Wu \emph{et~al.}, ``{AI}-enhanced integrated sensing and communications: Advancements, challenges, and prospects,'' \emph{IEEE Commun. Mag.}, vol.~62, no.~9, pp. 144--150, 2024.

\bibitem{ISAC_DT_2024}
Z.~Wei \emph{et~al.}, ``Integrated sensing and communication driven digital twin for intelligent machine network,'' \emph{IEEE Internet Things Mag.}, vol.~7, no.~4, pp. 60--67, 2024.

\bibitem{Nokia_24}
\BIBentryALTinterwordspacing
Nokia. (2024) {6G} explained. [Online]. Available: \url{https://www.nokia.com/about-us/newsroom/articles/6g-explained/}
\BIBentrySTDinterwordspacing

\bibitem{Ericsson_24}
\BIBentryALTinterwordspacing
Ericsson. (2024) White paper: Co-creating a cyber-physical world. [Online]. Available: \url{https://www.ericsson.com/en/reports-and-papers/white-papers/co-creating-a-cyber-physical-world}
\BIBentrySTDinterwordspacing

\bibitem{Huawei_24}
\BIBentryALTinterwordspacing
T.~Wen, M.~Jianglei, Z.~Peiying, and C.~Yan. (2024) Co-creating a cyber-physical world. [Online]. Available: \url{https://www.huawei.com/en/huaweitech/publication/202401/ai-bridge-to-6g#}
\BIBentrySTDinterwordspacing

\bibitem{Nvidia_24}
\BIBentryALTinterwordspacing
E.~Obiodu, K.~Chowdhury, L.~Kundu, and X.~Lin. (2024) Boosting {AI}-driven innovation in {6G} with the {AI-RAN A}lliance, {3GPP}, and {O-RAN}. [Online]. Available: \url{https://developer.nvidia.com/blog/boosting-ai-driven-innovation-in-6g-with-the-ai-ran-alliance-3gpp\-and-o-ran/}
\BIBentrySTDinterwordspacing

\bibitem{cammerer2025sionnaresearchkitgpuaccelerated}
\BIBentryALTinterwordspacing
S.~Cammerer \emph{et~al.}, ``Sionna research kit: A {GPU}-accelerated research platform for {AI-RAN},'' 2025. [Online]. Available: \url{https://arxiv.org/abs/2505.15848}
\BIBentrySTDinterwordspacing

\bibitem{CR_AI_Survey2010}
A.~He \emph{et~al.}, ``A survey of artificial intelligence for cognitive radios,'' \emph{IEEE Trans. Veh. Technol.}, vol.~59, no.~4, pp. 1578--1592, 2010.

\bibitem{CR_AI_Survey2013}
M.~Bkassiny, Y.~Li, and S.~K. Jayaweera, ``A survey on machine-learning techniques in cognitive radios,'' \emph{IEEE Commun. Surv. Tutor.}, vol.~15, no.~3, pp. 1136--1159, 2013.

\bibitem{DeLima2021}
C.~De~Lima \emph{et~al.}, ``Convergent communication, sensing and localization in {6G} systems: An overview of technologies, opportunities and challenges,'' \emph{IEEE Access}, vol.~9, pp. 26\,902--26\,925, 2021.

\bibitem{ISAC_vision_Liu_2022}
F.~Liu \emph{et~al.}, ``Integrated sensing and communications: Toward dual-functional wireless networks for {6G} and beyond,'' \emph{IEEE J. Sel. Areas Commun.}, vol.~40, no.~6, pp. 1728--1767, 2022.

\bibitem{Demirhan2023}
U.~Demirhan and A.~Alkhateeb, ``Integrated sensing and communication for {6G}: Ten key machine learning roles,'' \emph{IEEE Commun. Mag.}, vol.~61, no.~5, pp. 113--119, May 2023.

\bibitem{JRC_Shatov}
V.~Shatov \emph{et~al.}, ``Joint radar and communications: Architectures, use cases, aspects of radio access, signal processing, and hardware,'' \emph{IEEE Access}, vol.~12, pp. 47\,888--47\,914, 2024.

\bibitem{respati2024}
M.~Ade Krisna~Respati and B.~M. Lee, ``A survey on machine learning enhanced integrated sensing and communication systems: Architectures, algorithms, and applications,'' \emph{IEEE Access}, vol.~12, pp. 170\,946--170\,964, 2024.

\bibitem{CognitiveRadio_ML_Survey2024}
N.~A. Khalek, D.~H. Tashman, and W.~Hamouda, ``Advances in machine learning-driven cognitive radio for wireless networks: A survey,'' \emph{IEEE Commun. Surv. Tutor.}, vol.~26, no.~2, pp. 1201--1237, 2024.

\bibitem{JammerDetection2024}
P.~Lohan, B.~Kantarci, M.~Amine~Ferrag, N.~Tihanyi, and Y.~Shi, ``From {5G} to {6G} networks: A survey on {AI}-based jamming and interference detection and mitigation,'' \emph{IEEE Open J. Commun. Soc.}, vol.~5, pp. 3920--3974, 2024.

\bibitem{Prelcic2024}
N.~González-Prelcic \emph{et~al.}, ``The integrated sensing and communication revolution for {6G}: Vision, techniques, and applications,'' \emph{Proc. IEEE}, vol. 112, no.~7, pp. 676--723, 2024.

\bibitem{Lu_ISAC_2024}
S.~Lu \emph{et~al.}, ``Integrated sensing and communications: Recent advances and ten open challenges,'' \emph{IEEE Internet Things J.}, vol.~11, no.~11, pp. 19\,094--19\,120, 2024.

\bibitem{SS_DL_Surv2023}
S.~N. Syed \emph{et~al.}, ``Deep neural networks for spectrum sensing: A review,'' \emph{IEEE Access}, vol.~11, pp. 89\,591--89\,615, 2023.

\bibitem{SpectrumSensingTrends2025}
F.~R.~V. Guimarães, J.~M.~B. da~Silva~Jr., C.~Casimiro~Cavalcante, G.~Fodor, M.~Bengtsson, and C.~Fischione, ``Machine learning for spectrum sharing: A survey,'' \emph{Foundations and Trends in Networking}, vol.~14, no. 1–2, p. 1–159, 2024.

\bibitem{SurvISACin6G2024}
D.~Wen, Y.~Zhou, X.~Li, Y.~Shi, K.~Huang, and K.~B. Letaief, ``A survey on integrated sensing, communication, and computation,'' \emph{IEEE Commun. Surv. Tutor.}, pp. 1--1, 2024.

\bibitem{DiSAC2025}
E.~C. Strinati \emph{et~al.}, ``Toward distributed and intelligent integrated sensing and communications for {6G} networks,'' \emph{IEEE Wirel. Commun.}, vol.~32, no.~1, pp. 60--67, 2025.

\bibitem{DISAC}
------, ``Distributed intelligent integrated sensing and communications: The {6G-DISAC} approach,'' in \emph{Proc. Joint European Conf. on Netw. and Commun. \& 6G Summit (EuCNC/6G Summit)}, 2024, pp. 392--397.

\bibitem{principles_of_mobile_communication}
G.~L. Stuber, \emph{Principles of Mobile Communication}, 4th~ed.\hskip 1em plus 0.5em minus 0.4em\relax Springer Nature, 2017.

\bibitem{goldsmith}
A.~Goldsmith, \emph{Wireless Communications}.\hskip 1em plus 0.5em minus 0.4em\relax Cambridge University Press, 2005.

\bibitem{rappaport}
T.~Rappaport, \emph{\BIBforeignlanguage{English (US)}{Wireless communications: Principles and practice}}, 2nd~ed., ser. Prentice Hall communications engineering and emerging technologies series.\hskip 1em plus 0.5em minus 0.4em\relax Prentice Hall, 2002.

\bibitem{Pradankla2022}
\BIBentryALTinterwordspacing
S.~Padakandla, ``A survey of reinforcement learning algorithms for dynamically varying environments,'' \emph{ACM Comput. Surv.}, vol.~54, no.~6, Jul. 2021. [Online]. Available: \url{https://doi.org/10.1145/3459991}
\BIBentrySTDinterwordspacing

\bibitem{CNN_basics}
Y.~LeCun and Y.~Bengio, \emph{Convolutional networks for images, speech, and time series}.\hskip 1em plus 0.5em minus 0.4em\relax Cambridge, MA, USA: MIT Press, 1998, p. 255–258.

\bibitem{Transformer}
A.~Vaswani \emph{et~al.}, ``Attention is all you need,'' in \emph{Proc. of the 31st Adv. Neural Inf. Process. (NeurIPS)}.\hskip 1em plus 0.5em minus 0.4em\relax Curran Associates Inc., 2017, p. 6000–6010.

\bibitem{5gchannelmodel}
{ETSI 3rd Generation Partnership Project (3GPP)}, ``{Study on channel model for frequencies from 0.5 to 100 GHz (3GPP TR 38.901 version 16.1.0 Release 16)},'' {ETSI}, Tech. Rep., 2020.

\bibitem{burkhardt2014quadriga}
F.~Burkhardt, S.~Jaeckel, E.~Eberlein, and R.~Prieto-Cerdeira, ``{QuaDRiGa}: a {MIMO} channel model for land mobile satellite,'' in \emph{Proc. 8th European Conf. on Antennas and Propag. (EuCAP 2014)}.\hskip 1em plus 0.5em minus 0.4em\relax IEEE, 2014, pp. 1274--1278.

\bibitem{Hoydis2023}
J.~Hoydis \emph{et~al.}, ``Sionna {RT}: Differentiable ray tracing for radio propagation modeling,'' in \emph{Proc. IEEE Globecom Workshops (GC Wkshps)}, 2023, pp. 317--321.

\bibitem{Remcom}
Remcom, ``{Wireless InSite},'' \url{http://www.remcom.com/wireless-insite}.

\bibitem{Alkhateeb2019}
A.~Alkhateeb, ``{DeepMIMO}: A generic deep learning dataset for millimeter wave and massive {MIMO} applications,'' in \emph{Proc. Information Theory and Applications Workshop (ITA)}, San Diego, CA, Feb 2019, pp. 1--8.

\bibitem{nist_nextg_website}
\BIBentryALTinterwordspacing
{National Institute of Standards and Technology (NIST)}, ``{NIST NextG Channel Model Alliance},'' 2025, accessed: 2025-02-13. [Online]. Available: \url{https://nextg.nist.gov/}
\BIBentrySTDinterwordspacing

\bibitem{cvnns}
N.~Benvenuto and F.~Piazza, ``On the complex backpropagation algorithm,'' \emph{IEEE Trans. Signal Process.}, vol.~40, no.~4, pp. 967--969, 1992.

\bibitem{shlezinger2023model}
N.~Shlezinger, J.~Whang, Y.~C. Eldar, and A.~G. Dimakis, ``Model-based deep learning,'' \emph{Proc. IEEE}, vol. 111, no.~5, pp. 465--499, 2023.

\bibitem{jiang2024isac}
W.~Jiang, D.~Ma, Z.~Wei, Z.~Feng, P.~Zhang, and J.~Peng, ``{ISAC-NET}: Model-driven deep learning for integrated passive sensing and communication,'' \emph{IEEE Trans. Commun.}, vol.~72, no.~8, pp. 4692--4707, 2024.

\bibitem{Eshraghian2023}
J.~K. Eshraghian \emph{et~al.}, ``Training spiking neural networks using lessons from deep learning,'' \emph{IEEE Proc.}, vol. 111, no.~9, pp. 1016--1054, Sep. 2023.

\bibitem{chen2023neuromorphic}
J.~Chen, N.~Skatchkovsky, and O.~Simeone, ``Neuromorphic integrated sensing and communications,'' \emph{IEEE Wirel. Commun. Lett.}, vol.~12, no.~3, pp. 476--480, 2023.

\bibitem{thomae2019}
R.~S. Thoma \emph{et~al.}, ``Cooperative passive coherent location: A promising 5g service to support road safety,'' \emph{IEEE Communications Magazine}, vol.~57, no.~9, pp. 86--92, 2019.

\bibitem{Xiong2023}
Y.~Xiong, F.~Liu, Y.~Cui, W.~Yuan, T.~X. Han, and G.~Caire, ``On the fundamental tradeoff of integrated sensing and communications under gaussian channels,'' \emph{{IEEE} Trans. Inf. Theory}, vol.~69, no.~9, pp. 5723--5751, 2023.

\bibitem{Yang2024a}
L.~Yang, Y.~Wei, Z.~Feng, Q.~Zhang, and Z.~Han, ``Deep reinforcement learning-based resource allocation for integrated sensing, communication, and computation in vehicular network,'' \emph{IEEE Trans. Wirel. Commun.}, vol.~23, no.~12, pp. 18\,608--18\,622, Dec. 2024.

\bibitem{Liu2025}
C.~Liu, M.~Xia, J.~Zhao, H.~Li, and Y.~Gong, ``Optimal resource allocation for integrated sensing and communications in internet of vehicles: A deep reinforcement learning approach,'' \emph{IEEE Trans. Veh. Technol.}, vol.~74, no.~2, pp. 3028--3038, Feb. 2025.

\bibitem{muth2024loss}
C.~Muth, B.~Geiger, D.~G. Gaviria, and L.~Schmalen, ``Loss design for single-carrier joint communication and neural network-based sensing,'' in \emph{Proc. 27th Int. Workshop on Smart Antennas (WSA)}, 2024, pp. 131--137.

\bibitem{MateosRamos2021}
J.~M. Mateos-Ramos \emph{et~al.}, ``End-to-end learning for integrated sensing and communication,'' in \emph{Proc. {IEEE} Int. Conf. Commun. (ICC)}, 2022.

\bibitem{muth2023autoencoder}
C.~Muth and L.~Schmalen, ``Autoencoder-based joint communication and sensing of multiple targets,'' in \emph{Proc. 26th Int. ITG Workshop on Smart Antennas and 13th Conf. on Systems, Communications, and Coding (WSA {\&} SCC)}, 2023, pp. 1--6.

\bibitem{fontanesi2024deep}
G.~Fontanesi \emph{et~al.}, ``A deep-{NN} beamforming approach for dual function radar-communication {THz UAV},'' \emph{IEEE Trans. Veh. Technol.}, vol.~74, no.~1, pp. 746--760, 2025.

\bibitem{Nguyen2023}
N.~T. Nguyen, L.~V. Nguyen, N.~Shlezinger, Y.~C. Eldar, A.~L. Swindlehurst, and M.~Juntti, ``Joint communications and sensing hybrid beamforming design via deep unfolding,'' 2023.

\bibitem{Yang2024}
X.~Yang, R.~Zhang, D.~Zhai, F.~Liu, R.~Du, and T.~X. Han, ``Constellation design for integrated sensing and communication with random waveforms,'' \emph{IEEE Trans. Wirel. Commun.}, vol.~23, no.~11, pp. 17\,415--17\,428, 2024.

\bibitem{geiger2025jointshaping}
B.~Geiger, F.~Liu, S.~Lu, A.~Rode, and L.~Schmalen, ``Joint optimization of geometric and probabilistic constellation shaping for {OFDM-ISAC} systems,'' in \emph{Proc. IEEE 5th Int. Symp. on Joint Commun. \& Sensing (JC\&S)}, 2025, pp. 1--6.

\bibitem{Trees2002}
H.~L. van Trees, \emph{{Optimum Array Processing: Part IV of Detection, Estimation, and Modulation Theory}}.\hskip 1em plus 0.5em minus 0.4em\relax Wiley, 2002.

\bibitem{jiang2019end}
W.~Jiang, A.~M. Haimovich, and O.~Simeone, ``End-to-end learning of waveform generation and detection for radar systems,'' in \emph{Proc. 53rd Asilomar Conference on Signals, Systems, and Computers}, Pacific Grove, CA, USA, 2019, pp. 1672--1676.

\bibitem{major2019vehicle}
B.~Major \emph{et~al.}, ``Vehicle detection with automotive radar using deep learning on range-azimuth-doppler tensors,'' in \emph{Proc. IEEE/CVF Int. Conf. on Computer Vision Workshop (ICCVW)}, 2019, pp. 924--932.

\bibitem{Dong2024}
H.~Dong and O.~B. Akan, ``Debrisense: Terahertz-based integrated sensing and communications (isac) for debris detection and classification in the internet of space (ios),'' 2024.

\bibitem{Tosi2025}
\BIBentryALTinterwordspacing
P.~Tosi, S.~Schieler, M.~Henninger, S.~Semper, and S.~Mandelli, ``Benchmarking cfar and cnn-based peak detection algorithms in isac under hardware impairments,'' 2025. [Online]. Available: \url{https://arxiv.org/abs/2505.10969}
\BIBentrySTDinterwordspacing

\bibitem{Schieler2025}
S.~Schieler, S.~Semper, and R.~Thomä, ``Wireless propagation parameter estimation with convolutional neural networks,'' \emph{International Journal of Microwave and Wireless Technologies}, p. 1–8, 2025.

\bibitem{Beuster2023}
J.~Beuster \emph{et~al.}, ``Measurement testbed for radar and emitter localization of uav at 3.75 ghz,'' in \emph{2023 17th European Conf. on Antennas and Propag. (EuCAP)}, 2023, pp. 1--5.

\bibitem{redmon2016lookonceunifiedrealtime}
\BIBentryALTinterwordspacing
J.~Redmon, S.~Divvala, R.~Girshick, and A.~Farhadi, ``You only look once: Unified, real-time object detection,'' 2016. [Online]. Available: \url{https://arxiv.org/abs/1506.02640}
\BIBentrySTDinterwordspacing

\bibitem{schielerEstimation2025}
S.~Schieler, S.~Semper, and R.~Thomä, ``Wireless propagation parameter estimation with convolutional neural networks (under review),'' pp. 1--5, 2025.

\bibitem{richterEstimation2005}
A.~Richter, \emph{\BIBforeignlanguage{en}{Estimation of radio channel parameters: models and algorithms}}.\hskip 1em plus 0.5em minus 0.4em\relax Ilmenau: ISLE, 2005.

\bibitem{semper2024}
S.~Semper, J.~Naviliat, J.~Gedschold, M.~Döbereiner, S.~Schieler, and R.~S. Thomä, ``Distributed computing and model-based estimation for integrated communications and sensing: A roadmap,'' \emph{IEEE Open J. Commun. Soc.}, vol.~5, pp. 6279--6290, 2024.

\bibitem{viberg91detection}
M.~Viberg, B.~Ottersten, and T.~Kailath, ``Detection and estimation in sensor arrays using weighted subspace fitting,'' \emph{IEEE Trans. Signal Process.}, vol.~39, no.~11, pp. 2436--2449, 1991.

\bibitem{stoica90maximum}
P.~Stoica and K.~Sharman, ``Maximum likelihood methods for direction-of-arrival estimation,'' \emph{IEEE Trans. Acoustics, Speech, and Signal Process.}, vol.~38, no.~7, pp. 1132--1143, 1990.

\bibitem{sharman89genetic}
K.~Sharman and G.~McClurkin, ``Genetic algorithms for maximum likelihood parameter estimation,'' in \emph{Proc. Int. Conf. on Acoustics, Speech, and Signal Process.}, 1989, pp. 2716--2719 vol.4.

\bibitem{zheng2024deep}
S.~Zheng \emph{et~al.}, ``Deep learning-based {DOA} estimation,'' \emph{IEEE Trans. Cogn. Commun. Netw.}, vol.~10, no.~3, pp. 819--835, 2024.

\bibitem{ma2022deep}
Y.~Ma, Y.~Zeng, and S.~Sun, ``A deep learning based super resolution {DoA} estimator with single snapshot {MIMO} radar data,'' \emph{IEEE Trans. Veh. Technol.}, vol.~71, no.~4, pp. 4142--4155, 2022.

\bibitem{chen2024sdoa}
P.~Chen, Z.~Chen, L.~Liu, Y.~Chen, and X.~Wang, ``Sdoa-net: An efficient deep-learning-based {DOA} estimation network for imperfect array,'' \emph{IEEE Trans. Instrum. Meas.}, vol.~73, pp. 1--12, 2024.

\bibitem{zhu2020two}
W.~Zhu, M.~Zhang, P.~Li, and C.~Wu, ``Two-dimensional {DOA} estimation via deep ensemble learning,'' \emph{IEEE Access}, vol.~8, pp. 124\,544--124\,552, 2020.

\bibitem{chen2022robust}
D.~Chen, S.~Shi, X.~Gu, and B.~Shim, ``Robust {DoA} estimation using denoising autoencoder and deep neural networks,'' \emph{IEEE Access}, vol.~10, pp. 52\,551--52\,564, 2022.

\bibitem{yu2023deep}
J.~Yu and Y.~Wang, ``Deep learning-based multipath {DoAs} estimation method for mmwave massive {MIMO} systems in low {SNR},'' \emph{IEEE Transactions Veh. Technol.}, vol.~72, no.~6, pp. 7480--7490, 2023.

\bibitem{shmuel2023subspacenet}
D.~H. Shmuel, J.~P. Merkofer, G.~Revach, R.~J. G.~v. Sloun, and N.~Shlezinger, ``Subspacenet: Deep learning-aided subspace methods for {DoA} estimation,'' \emph{IEEE Trans. Veh. Technol.}, pp. 1--15, 2024.

\bibitem{merkofer2024damusic}
J.~P. Merkofer, G.~Revach, N.~Shlezinger, T.~Routtenberg, and R.~J.~G. van Sloun, ``{DA-MUSIC}: Data-driven {DoA} estimation via deep augmented {MUSIC} algorithm,'' \emph{IEEE Trans. Veh. Technol.}, vol.~73, no.~2, pp. 2771--2785, 2024.

\bibitem{chatelier2025physically}
B.~Chatelier \emph{et~al.}, ``Physically parameterized differentiable music for doa estimation with uncalibrated arrays,'' \emph{IEEE Int. Conf. on Commun.}, 2025.

\bibitem{Barthelme2021}
A.~Barthelme and W.~Utschick, ``A machine learning approach to {DoA} estimation and model order selection for antenna arrays with subarray sampling,'' \emph{IEEE Trans. Signal Process.}, vol.~69, pp. 3075--3087, 2021.

\bibitem{schielerGridFree2024}
S.~Schieler, S.~Semper, R.~Faramarzahangari, M.~Döbereiner, C.~Schneider, and R.~Thomä, ``Grid-free harmonic retrieval and model order selection using convolutional neural networks,'' in \emph{Proc. 18th European Conf. on Antennas and Propag. (EuCAP)}, 2024, pp. 1--5.

\bibitem{schielerCNNMeasurement}
S.~Schieler, S.~Semper, C.~Schneider, and R.~Thomä, ``Measurement-based evaluation of cnn-based detection and estimation for {ISAC} systems,'' in \emph{Proc. IEEE Int. Radar Conf.}, 2025, pp. 1--5.

\bibitem{zyweck96radar}
A.~Zyweck and R.~Bogner, ``Radar target classification of commercial aircraft,'' \emph{IEEE Trans. Aerosp. Electron. Syst.}, vol.~32, no.~2, pp. 598--606, 1996.

\bibitem{sorowka15pedestrian}
P.~Sorowka and H.~Rohling, ``Pedestrian classification with 24 {GHz} chirp sequence radar,'' in \emph{Proc. 16th Int. Radar Symp. (IRS)}, 2015, pp. 167--173.

\bibitem{gupta2021target}
S.~Gupta, P.~K. Rai, A.~Kumar, P.~K. Yalavarthy, and L.~R. Cenkeramaddi, ``Target classification by mmwave {FMCW} radars using machine learning on range-angle images,'' \emph{IEEE Sens. J.}, vol.~21, no.~18, pp. 19\,993--20\,001, 2021.

\bibitem{ulrich2021deep}
M.~Ulrich, C.~Gläser, and F.~Timm, ``Deepreflecs: Deep learning for automotive object classification with radar reflections,'' in \emph{Proc. IEEE Radar Conf. (RadarConf21)}, 2021, pp. 1--6.

\bibitem{patel2019deep}
K.~Patel, K.~Rambach, T.~Visentin, D.~Rusev, M.~Pfeiffer, and B.~Yang, ``Deep learning-based object classification on automotive radar spectra,'' in \emph{Proc. IEEE Radar Conf. (RadarConf)}, 2019, pp. 1--6.

\bibitem{mateos2024model}
J.~M. Mateos-Ramos, C.~Häger, M.~F. Keskin, L.~L. Magoarou, and H.~Wymeersch, ``Model-based end-to-end learning for multi-target integrated sensing and communication under hardware impairments,'' \emph{IEEE Trans. Wirel. Commun.}, pp. 2574--2589, 2025.

\bibitem{mateos2024semi}
J.~M. Mateos-Ramos, B.~Chatelier, C.~Häger, M.~F. Keskin, L.~Le~Magoarou, and H.~Wymeersch, ``Semi-supervised end-to-end learning for integrated sensing and communications,'' in \emph{Proc. IEEE Int. Conf. on Machine Learning for Commun. and Netw. (ICMLCN)}, 2024, pp. 132--138.

\bibitem{hu2023radnet}
K.~Hu, X.~Hu, L.~Qi, G.~Lu, Y.~Zhong, and Y.~Han, ``Radnet: A radar detection network for target detection using {3D} range-angle-doppler tensor,'' in \emph{Proc. IEEE 19th Int. Conf. on Automation Science and Engineering (CASE)}, 2023, pp. 1--6.

\bibitem{Benyahia2022squeeze}
Z.~Benyahia, M.~Hefnawi, M.~Aboulfatah, E.~Abdelmounim, and T.~Gadi, ``Squeezenet-based range, angle, and doppler estimation for automotive {MIMO} radar systems,'' in \emph{Proc. Int. Conf. on Intelligent Systems and Computer Vision (ISCV)}, 2022, pp. 1--5.

\bibitem{Schieler2023}
\BIBentryALTinterwordspacing
S.~Schieler, S.~Semper, R.~Faramarzahangari, C.~Schneider, and R.~S. Thomae, ``4d joint harmonic retrieval and model order estimation with convolutional neural networks,'' in \emph{Proc. 5th International Conference on Advances in Signal Processing and Artificial Intelligence ({ASPAI}' 2023)}.\hskip 1em plus 0.5em minus 0.4em\relax {IFSA} Publishing, S. L., 2023, pp. 204--208. [Online]. Available: \url{\url{https://sensorsportal.com/ASPAI_2023/ASPAI_2023_Conference_Proceedings.pdf}}
\BIBentrySTDinterwordspacing

\bibitem{Schieler2022}
S.~Schieler, S.~Semper, R.~Faramarzahangari, M.~Döbereiner, C.~Schneider, and R.~Thomä, ``Grid-free harmonic retrieval and model order selection using deep convolutional neural networks,'' 2022.

\bibitem{garcia19trajectory}
{\'A}.~F. Garc{\'\i}a-Fern{\'a}ndez and L.~Svensson, ``Trajectory {PHD} and {CPHD} filters,'' \emph{IEEE Trans. Signal Process.}, vol.~67, no.~22, pp. 5702--5714, 2019.

\bibitem{vo19multi}
B.-N. Vo and B.-T. Vo, ``A multi-scan labeled random finite set model for multi-object state estimation,'' \emph{IEEE Trans. Signal Process.}, vol.~67, no.~19, pp. 4948--4963, 2019.

\bibitem{xia2019multi}
Y.~Xia, K.~Granstr{\"o}m, L.~Svensson, {\'A}.~F. Garc{\'\i}a-Fern{\'a}ndez, and J.~L. Williams, ``Multi-scan implementation of the trajectory poisson multi-bernoulli mixture filter,'' \emph{Journal of Advances in Information Fusion}, vol.~14, no.~2, pp. 213--235, 2019.

\bibitem{mahler2007statistical}
R.~Mahler, \emph{Statistical multisource-multitarget information fusion}.\hskip 1em plus 0.5em minus 0.4em\relax Artech, 2007.

\bibitem{liu2022learning}
C.~Liu \emph{et~al.}, ``Learning-based predictive beamforming for integrated sensing and communication in vehicular networks,'' \emph{IEEE J. Sel. Areas Commun.}, vol.~40, no.~8, pp. 2317--2334, 2022.

\bibitem{wang2023deep}
Z.~Wang and V.~W. Wong, ``Deep learning for {ISAC}-enabled end-to-end predictive beamforming in vehicular networks,'' in \emph{Proc. IEEE Int. Conf. on Commun.}, 2023, pp. 5713--5718.

\bibitem{demirhan2022radar}
U.~Demirhan and A.~Alkhateeb, ``Radar aided {6G} beam prediction: Deep learning algorithms and real-world demonstration,'' in \emph{Proc. IEEE Wirel. Commun. and Netw. Conf. (WCNC)}, 2022, pp. 2655--2660.

\bibitem{pinto2023deep}
J.~Pinto, G.~Hess, W.~Ljungbergh, Y.~Xia, H.~Wymeersch, and L.~Svensson, ``Deep learning for model-based multiobject tracking,'' \emph{IEEE Trans. Aerosp. Electron. Syst.}, vol.~59, no.~6, pp. 7363--7379, 2023.

\bibitem{gruber1967approach}
M.~Gruber, \emph{An approach to target tracking}.\hskip 1em plus 0.5em minus 0.4em\relax MIT Lincoln Laboratory, 1967.

\bibitem{julier1997new}
S.~J. Julier and J.~K. Uhlmann, ``New extension of the kalman filter to nonlinear systems,'' in \emph{Signal processing, sensor fusion, and target recognition VI}, vol. 3068.\hskip 1em plus 0.5em minus 0.4em\relax Spie, 1997, pp. 182--193.

\bibitem{gordon1993novel}
N.~J. Gordon, D.~J. Salmond, and A.~F. Smith, ``Novel approach to nonlinear/non-gaussian bayesian state estimation,'' in \emph{IEE proceedings F (radar and signal processing)}, vol. 140, no.~2.\hskip 1em plus 0.5em minus 0.4em\relax IET, 1993, pp. 107--113.

\bibitem{revach2022kalmannet}
G.~Revach, N.~Shlezinger, X.~Ni, A.~L. Escoriza, R.~J.~G. van Sloun, and Y.~C. Eldar, ``{KalmanNet}: Neural network aided kalman filtering for partially known dynamics,'' \emph{IEEE Trans. Signal Process.}, vol.~70, pp. 1532--1547, 2022.

\bibitem{ni2024adaptive}
X.~Ni, G.~Revach, and N.~Shlezinger, ``Adaptive kalmannet: Data-driven {Kalman} filter with fast adaptation,'' in \emph{Proc. IEEE Int. Conf. Acoust. Speech Signal Process. (ICASSP)}, 2024, pp. 5970--5974.

\bibitem{dahan2024uncertainty}
Y.~Dahan, G.~Revach, J.~Dunik, and N.~Shlezinger, ``Uncertainty quantification in deep learning based kalman filters,'' in \emph{Proc. IEEE Int. Conf. Acoust. Speech Signal Process. (ICASSP)}, 2024, pp. 13\,121--13\,125.

\bibitem{nuri2024learning}
I.~Nuri and N.~Shlezinger, ``Learning flock: Enhancing sets of particles for multi\~{} sub-state particle filtering with neural augmentation,'' \emph{arXiv preprint arXiv:2408.11348}, 2024.

\bibitem{mitra07gesture}
S.~Mitra and T.~Acharya, ``Gesture recognition: A survey,'' \emph{IEEE Transactions on Systems, Man, and Cybernetics, Part C (Applications and Reviews)}, vol.~37, no.~3, pp. 311--324, 2007.

\bibitem{na19deep}
N.~Lu, Y.~Wu, L.~Feng, and J.~Song, ``Deep learning for fall detection: Three-dimensional {CNN} combined with {LSTM} on video kinematic data,'' \emph{IEEE J. Biomed. Health Inform.}, vol.~23, no.~1, pp. 314--323, 2019.

\bibitem{mukhopadhyay15wearable}
S.~C. Mukhopadhyay, ``Wearable sensors for human activity monitoring: A review,'' \emph{IEEE Sens. J.}, vol.~15, no.~3, pp. 1321--1330, 2015.

\bibitem{qi2023resource}
W.~Qi, R.~Zhang, J.~Zhou, H.~Zhang, Y.~Xie, and X.~Jing, ``A resource-efficient cross-domain sensing method for device-free gesture recognition with federated transfer learning,'' \emph{IEEE Trans. Green Commun. Netw}, vol.~7, no.~1, pp. 393--400, 2023.

\bibitem{du2020three}
H.~Du, T.~Jin, Y.~Song, Y.~Dai, and M.~Li, ``A three-dimensional deep learning framework for human behavior analysis using range-doppler time points,'' \emph{IEEE Geosci. Remote Sens. Lett.}, vol.~17, no.~4, pp. 611--615, 2020.

\bibitem{bhat2023gesture}
N.~N. Bhat, R.~Berkvens, and J.~Famaey, ``Gesture recognition with mmwave {Wi-Fi} access points: Lessons learned,'' in \emph{Proc. IEEE 24th Int. Symp. on a World of Wireless, Mobile and Multimedia Networks (WoWMoM)}, 2023, pp. 127--136.

\bibitem{zhao2022angle}
Y.~Zhao, A.~Yarovoy, and F.~Fioranelli, ``Angle-insensitive human motion and posture recognition based on {4D} imaging radar and deep learning classifiers,'' \emph{IEEE Sens. J.}, vol.~22, no.~12, pp. 12\,173--12\,182, 2022.

\bibitem{Hoydis2023a}
J.~Hoydis \emph{et~al.}, ``Learning radio environments by differentiable ray tracing,'' \emph{IEEE Trans. on Machine Learning in Commun. and Netw.}, vol.~2, pp. 1527--1539, 2024.

\bibitem{bouget2021fusion}
V.~Bouget, D.~B{\'e}r{\'e}ziat, J.~Brajard, A.~Charantonis, and A.~Filoche, ``Fusion of rain radar images and wind forecasts in a deep learning model applied to rain nowcasting,'' \emph{Remote Sensing}, vol.~13, no.~2, p. 246, 2021.

\bibitem{aidin2021deep}
A.~Ferdowsi and D.~Whitefield, ``Deep learning for rain fade prediction in satellite communications,'' in \emph{Proc. IEEE Globecom Workshops (GC Wkshps)}, 2021, pp. 1--6.

\bibitem{Helal2022}
S.~Helal, H.~Sarieddeen, H.~Dahrouj, T.~Y. Al-Naffouri, and M.-S. Alouini, ``Signal processing and machine learning techniques for terahertz sensing: An overview,'' \emph{{IEEE} Signal Process. Mag.}, vol.~39, no.~5, pp. 42--62, sep 2022.

\bibitem{ThomaeDallmann2023}
R.~Thomä and T.~Dallmann, ``Distributed {ISAC} systems – multisensor radio access and coordination,'' in \emph{Proc. 20th European Radar Conf. (EuRAD)}, 2023, pp. 351--354.

\bibitem{Zhuang2023}
Z.~Zhuang, D.~Wen, Y.~Shi, G.~Zhu, S.~Wu, and D.~Niyato, ``Integrated sensing-communication-computation for over-the-air edge {AI} inference,'' 2023.

\bibitem{MENG2020115}
\BIBentryALTinterwordspacing
T.~Meng, X.~Jing, Z.~Yan, and W.~Pedrycz, ``A survey on machine learning for data fusion,'' \emph{Information Fusion}, vol.~57, pp. 115--129, 2020. [Online]. Available: \url{https://www.sciencedirect.com/science/article/pii/S1566253519303902}
\BIBentrySTDinterwordspacing

\bibitem{Wideband_SS_CR_2008}
Z.~Quan, S.~Cui, A.~H. Sayed, and H.~V. Poor, ``Wideband spectrum sensing in cognitive radio networks,'' in \emph{IEEE Int. Conf. Commun.}, 2008, pp. 901--906.

\bibitem{CognitiveRadarOverview_2019}
S.~Z. Gurbuz, H.~D. Griffiths, A.~Charlish, M.~Rangaswamy, M.~S. Greco, and K.~Bell, ``An overview of cognitive radar: Past, present, and future,'' \emph{IEEE Aerosp. Electron. Syst. Mag.}, vol.~34, no.~12, pp. 6--18, 2019.

\bibitem{LBT_2016}
C.~K. Kim, C.~S. Yang, and C.~G. Kang, ``Adaptive listen-before-talk {(LBT)} scheme for {LTE} and {Wi-Fi} systems coexisting in unlicensed band,'' in \emph{Proc. 13th IEEE Annual Consumer Commun. \& Netw. Conf. (CCNC)}, 2016, pp. 589--594.

\bibitem{Martone2021}
A.~Martone and M.~Amin, ``A view on radar and communication systems coexistence and dual functionality in the era of spectrum sensing,'' \emph{Digit. Signal Process.}, vol. 119, 12 2021.

\bibitem{EarlySS2008}
X.-L. Zhu, Y.-A. Liu, W.-W. Weng, and D.-M. Yuan, ``Channel sensing algorithm based on neural networks for cognitive wireless mesh networks,'' in \emph{Proc. 4th Int. Conf. on Wirel. Commun., Netw. and Mobile Comput.}, 2008, pp. 1--4.

\bibitem{SS_CNN_China_2020}
S.~Zheng, S.~Chen, P.~Qi, H.~Zhou, and X.~Yang, ``Spectrum sensing based on deep learning classification for cognitive radios,'' in \emph{China Commun.}, 2020, pp. 138--148.

\bibitem{maxminSpecrumSensing}
Y.~Zeng and Y.-C. Liang, ``Eigenvalue-based spectrum sensing algorithms for cognitive radio,'' \emph{IEEE Trans. Commun.}, vol.~57, no.~6, pp. 1784--1793, 2009.

\bibitem{Robinson_ICMLCN2024}
C.~P. Robinson, D.~Uvaydov, S.~D’Oro, and T.~Melodia, ``Deepsweep: Parallel and scalable spectrum sensing via convolutional neural networks,'' in \emph{Proc. IEEE Int. Conf. on Machine Learning for Commun. and Netw. (ICMLCN)}, 2024, pp. 505--510.

\bibitem{SpectrumTransformer2024}
W.~Zhang, Y.~Wang, X.~Chen, Z.~Cai, and Z.~Tian, ``Spectrum transformer: An attention-based wideband spectrum detector,'' \emph{Trans. Wirel. Commun.}, vol.~23, no.~9, pp. 12\,343--12\,353, 2024.

\bibitem{SS_DeepSense_Melodia2021}
D.~Uvaydov, S.~D’Oro, F.~Restuccia, and T.~Melodia, ``Deepsense: Fast wideband spectrum sensing through real-time in-the-loop deep learning,'' in \emph{Proc. IEEE Conf. on Computer Commun. (INFOCOM)}, 2021, pp. 1--10.

\bibitem{GAN_SS_2018}
K.~Davaslioglu and Y.~E. Sagduyu, ``Generative adversarial learning for spectrum sensing,'' in \emph{Proc. IEEE Int. Conf. on Commun. (ICC)}, 2018, pp. 1--6.

\bibitem{CNN_Spectrum_Sensing_2022}
L.~Cai, K.~Cao, Y.~Wu, and Y.~Zhou, ``Spectrum sensing based on spectrogram-aware {CNN} for cognitive radio network,'' \emph{IEEE Wirel. Commun. Lett.}, vol.~11, no.~10, pp. 2135--2139, 2022.

\bibitem{Wymeersch2009_IEEEProc}
H.~Wymeersch, J.~Lien, and M.~Z. Win, ``Cooperative localization in wireless networks,'' \emph{Proc. IEEE}, vol.~97, no.~2, pp. 427--450, 2009.

\bibitem{HiddenNode2008}
W.~Zhang, R.~K. Mallik, and K.~B. Letaief, ``Cooperative spectrum sensing optimization in cognitive radio networks,'' in \emph{Proc. IEEE Int. Conf. on Commun. (ICC)}, 2008, pp. 3411--3415.

\bibitem{LSTM_SS_2023}
D.~Janu, K.~Singh, S.~Kumar, and S.~Mandia, ``Hierarchical cooperative {LSTM}-based spectrum sensing,'' \emph{IEEE Commun. Lett.}, vol.~27, no.~3, pp. 866--870, 2023.

\bibitem{Coop_Unsupervised_SS_2022}
N.~A. Khalek and W.~Hamouda, ``Intelligent spectrum sensing: An unsupervised learning approach based on dimensionality reduction,'' in \emph{Proc. IEEE Int. Conf. on Commun. (ICC)}, 2022, pp. 171--176.

\bibitem{SpectrumSensing2024}
W.~Zhang, Y.~Wang, X.~Chen, L.~Liu, and Z.~Tian, ``Collaborative learning based spectrum sensing under partial observations,'' \emph{IEEE Trans. Cogn. Commun. Netw.}, pp. 1--1, 2024.

\bibitem{AMC_AISBETT_1987}
J.~Aisbett, ``Automatic modulation recognition using time domain parameters,'' \emph{Signal Processing}, vol.~13, no.~3, pp. 323--328, 1987.

\bibitem{ACM_old_features_1992}
J.~Reichert, ``Automatic classification of communication signals using higher order statistics,'' in \emph{Proc. IEEE Int. Conf. Acoust. Speech Signal Process. (ICASSP}, vol.~5, 1992, pp. 221--224 vol.5.

\bibitem{Dobre2007}
O.~Dobre, A.~Abdi, and W.~Su, ``Survey of automatic modulation classification techniques: Classical approaches and new trends,'' \emph{Communications, IET}, vol.~1, pp. 137 -- 156, 2007.

\bibitem{DNN_ACM_features_2016}
B.~Kim, J.~Kim, H.~Chae, D.~Yoon, and J.~W. Choi, ``Deep neural network-based automatic modulation classification technique,'' in \emph{Proc. Int. Conf. ICT Converg. (ICTC)}, 2016, pp. 579--582.

\bibitem{BD1943}
A.~K. Bhattacharyya, ``On a measure of divergence between two statistical populations defined by their probability distributions,'' \emph{Bulettin of the Calcutta Mathematical Society}, no.~35, pp. 99--109, 1943.

\bibitem{BD_AMC}
M.~H. Shah and X.~Dang, ``Novel feature selection method using {B}hattacharyya distance for neural networks based automatic modulation classification,'' \emph{IEEE Signal Process. Lett.}, vol.~27, pp. 106--110, 2020.

\bibitem{EarlyAMC2005}
A.~Fehske, J.~Gaeddert, and J.~Reed, ``A new approach to signal classification using spectral correlation and neural networks,'' in \emph{Proc. 1st IEEE Int. Symp. on New Frontiers in Dynamic Spectrum Access Netw.}, 2005, pp. 144--150.

\bibitem{DecisionTree2000}
A.~Swami and B.~Sadler, ``Hierarchical digital modulation classification using cumulants,'' \emph{IEEE Trans. Commun.}, vol.~48, no.~3, pp. 416--429, 2000.

\bibitem{RandomForest2017}
K.~Triantafyllakis, M.~Surligas, G.~Vardakis, and S.~Papadakis, ``Phasma: An automatic modulation classification system based on random forest,'' in \emph{Proc. IEEE Int. Symp. Dyn. Spectr. Access Netw. (DySPAN)}, 2017, pp. 1--3.

\bibitem{SVM2004}
H.~Gang, L.~Jiandong, and L.~Donghua, ``Study of modulation recognition based on {HOC}s and {SVM},'' in \emph{Proc. IEEE Veh. Technol. Conf. (VTC-Spring)}, vol.~2, 2004, pp. 898--902 Vol.2.

\bibitem{JRC_classification2025}
V.~Shatov, A.~Vagollari, Y.~Su, W.~Gerstacker, N.~Franchi, and M.~Lübke, ``Supervised learning of wireless signal waveforms in the era of radar-communications convergence,'' \emph{Under Review}, 2025.

\bibitem{Beyond5G_AMC_2020}
A.~P. Hermawan, R.~R. Ginanjar, D.-S. Kim, and J.-M. Lee, ``{CNN}-based automatic modulation classification for beyond {5G} communications,'' \emph{IEEE Commun. Lett.}, vol.~24, no.~5, pp. 1038--1041, 2020.

\bibitem{WaveformClass2021}
Y.~Shi, K.~Davaslioglu, Y.~E. Sagduyu, W.~C. Headley, M.~Fowler, and G.~Green, ``Deep learning for {RF} signal classification in unknown and dynamic spectrum environments,'' in \emph{Proc. IEEE Int. Symp. on Dynamic Spectrum Access Netw. (DySPAN)}, 2019, pp. 1--10.

\bibitem{Karagiannidis2024}
T.~K. Oikonomou \emph{et~al.}, ``{CNN}-based automatic modulation classification under phase imperfections,'' \emph{IEEE Wirel. Commun. Lett.}, vol.~13, no.~5, pp. 1508--1512, 2024.

\bibitem{AMC_Kumar2024}
A.~Kumar, M.~S. Chaudhari, and S.~Majhi, ``Automatic modulation classification for {OFDM} systems using bi-stream and attention-based {CNN-LSTM} model,'' \emph{IEEE Commun. Lett.}, pp. 1--5, 2024.

\bibitem{Dobre2024}
R.~Ding, F.~Zhou, Q.~Wu, C.~Dong, Z.~Han, and O.~A. Dobre, ``Data and knowledge dual-driven automatic modulation classification for {6G} wireless communications,'' \emph{IEEE Trans. Wirel. Commun.}, vol.~23, no.~5, pp. 4228--4242, 2024.

\bibitem{STFT}
F.~Hlawatsch and G.~Boudreaux-Bartels, ``Linear and quadratic time-frequency signal representations,'' \emph{IEEE Sig. Process. Mag.}, vol.~9, no.~2, pp. 21--67, 1992.

\bibitem{Vagollari2023}
A.~Vagollari, M.~Hirschbeck, and W.~Gerstacker, ``An end-to-end deep learning framework for wideband signal recognition,'' \emph{IEEE Access}, vol.~11, pp. 52\,899--52\,922, 2023.

\bibitem{FFT_AMC_Comparison2022}
H.-K. Le, V.-P. Hoang, V.-S. Doan, M.-T. Hoang, and N.~P. Dao, ``Performance analysis of convolutional neural networks with different window functions for automatic modulation classification,'' in \emph{Proc. Int. Conf. ICT Converg. (ICTC)}, 2022, pp. 153--157.

\bibitem{Zhang_infocom_2021}
W.~Zhang, M.~Feng, M.~Krunz, and A.~Hossein Yazdani~Abyaneh, ``Signal detection and classification in shared spectrum: A deep learning approach,'' in \emph{Proc. IEEE Conf. on Comput. Commun.}, 2021, pp. 1--10.

\bibitem{Jam_factory_floor_2023}
S.~Dinh-Van, T.~M. Hoang, B.~B. Cebecioglu, D.~S. Fowler, Y.~K. Mo, and M.~D. Higgins, ``A defensive strategy against beam training attack in {5G} mmwave networks for manufacturing,'' \emph{IEEE Trans. Inf. Forensics Secur.}, vol.~18, pp. 2204--2217, 2023.

\bibitem{DT_spectrum2024}
A.~Schosser, F.~Burmeister, P.~Schulz, M.~D. Khursheed, S.~Ma, and G.~Fettweis, ``Advancing spectrum anomaly detection through digital twins,'' \emph{IEEE Commun. Mag.}, pp. 1--7, 2024.

\bibitem{Co_AMC_legacy_1_2010}
J.~L. Xu, W.~Su, and M.~Zhou, ``Distributed automatic modulation classification with multiple sensors,'' \emph{IEEE Sens. J.}, vol.~10, no.~11, pp. 1779--1785, 2010.

\bibitem{Co_AMC_legacy_2_2013}
O.~Ozdemir, R.~Li, and P.~K. Varshney, ``Hybrid maximum likelihood modulation classification using multiple radios,'' \emph{IEEE Commun. Lett.}, vol.~17, no.~10, pp. 1889--1892, 2013.

\bibitem{Co_AMC_legacy_cumulants_2017}
J.~Zhang, D.~Cabric, F.~Wang, and Z.~Zhong, ``Cooperative modulation classification for multipath fading channels via expectation-maximization,'' \emph{IEEE Trans. Wirel. Commun.}, vol.~16, no.~10, pp. 6698--6711, 2017.

\bibitem{Rajendran2018}
S.~Rajendran, W.~Meert, D.~Giustiniano, V.~Lenders, and S.~Pollin, ``Deep learning models for wireless signal classification with distributed low-cost spectrum sensors,'' \emph{IEEE Trans. Cogn. Commun. Netw.}, vol.~4, no.~3, pp. 433--445, 2018.

\bibitem{DL_CoAMC_2020}
Y.~Wang, J.~Wang, W.~Zhang, J.~Yang, and G.~Gui, ``Deep learning-based cooperative automatic modulation classification method for {MIMO} systems,'' \emph{IEEE Trans. Veh. Technol.}, vol.~69, no.~4, pp. 4575--4579, 2020.

\bibitem{Collab_AMC_2020}
Y.~Wang \emph{et~al.}, ``Distributed learning for automatic modulation classification in edge devices,'' \emph{IEEE Wirel. Commun. Lett.}, vol.~9, no.~12, pp. 2177--2181, 2020.

\bibitem{OFDM_class_features_collab2024}
Y.~Chen, J.~He, W.~Jiang, Y.~Zhang, S.~Huang, and Z.~Feng, ``Toward collaborative and channel-robust automatic modulation classification for {OFDM} signals,'' \emph{IEEE Wirel. Commun. Lett.}, vol.~13, no.~11, pp. 3187--3191, 2024.

\bibitem{Reinf_CSS_2024}
A.~Gao \emph{et~al.}, ``Attention enhanced multi-agent reinforcement learning for cooperative spectrum sensing in cognitive radio networks,'' \emph{IEEE Trans. Veh. Technol.}, vol.~73, no.~7, pp. 10\,464--10\,477, 2024.

\bibitem{Fontaine2020}
J.~Fontaine, A.~Shahid, R.~Elsas, A.~Seferagic, I.~Moerman, and E.~De~Poorter, ``Multi-band sub-{GHz} technology recognition on {NVIDIA’s Jetson Nano},'' in \emph{Proc. IEEE 92nd Veh. Technol. Conf. (VTC2020-Fall)}, 2020, pp. 1--7.

\bibitem{Vagollari2021}
A.~Vagollari, V.~Schram, W.~Wicke, M.~Hirschbeck, and W.~Gerstacker, ``Joint detection and classification of {RF} signals using deep learning,'' in \emph{Proc. IEEE 93rd Veh. Technol. Conf. (VTC2021-Spring)}, 2021, pp. 1--7.

\bibitem{Rad_Com_Coex_Classif_2020}
G.~Kong, M.~Jung, and V.~Koivunen, ``Waveform classification in radar-communications coexistence scenarios,'' in \emph{Proc. 2020 IEEE Global Commun. Conf. (GLOBECOM)}, 2020, pp. 1--6.

\bibitem{RadComClass2024Latest}
T.~Huynh-The, V.-P. Hoang, J.-W. Kim, M.-T. Le, and M.~Zeng, ``Wavenet: Towards waveform classification in integrated radar-communication systems with improved accuracy and reduced complexity,'' \emph{IEEE Internet Things J.}, vol.~11, no.~14, pp. 25\,111--25\,123, 2024.

\bibitem{RML16_OShea}
T.~J. O'Shea, J.~Corgan, and T.~C. Clancy, ``Convolutional radio modulation recognition networks,'' in \emph{Proc. Int. Conf. Eng. Appl. Neural Netw.}\hskip 1em plus 0.5em minus 0.4em\relax Springer, 2016, pp. 213--226.

\bibitem{RadioML_Dataset_2018}
T.~J. O’Shea, T.~Roy, and T.~C. Clancy, ``Over-the-air deep learning based radio signal classification,'' \emph{IEEE J. Sel. Top. Signal Process.}, vol.~12, no.~1, pp. 168--179, 2018.

\bibitem{AMC_Dataset2023}
V.~Sathyanarayanan, P.~Gerstoft, and A.~E. Gamal, ``{RML22}: Realistic dataset generation for wireless modulation classification,'' \emph{IEEE Wirel. Commun.}, vol.~22, no.~11, pp. 7663--7675, 2023.

\bibitem{WidebandDataset2021}
N.~West, T.~O’Shea, and T.~Roy, ``A wideband signal recognition dataset,'' in \emph{Proc. IEEE Workshop Signal Process. Adv. Wirel. Commun. (SPAWC)}, 2021, pp. 6--10.

\bibitem{dataset_fraunhofer2022}
J.~Wicht, U.~Wetzker, and V.~Jain, ``{Spectrogram Data Set for Deep-Learning-Based {RF} Frame Detection},'' \emph{Data}, vol.~7, no.~12, 2022.

\bibitem{SpectrumSensingDataset2023}
C.~Tassie \emph{et~al.}, ``Detection of co-existing {RF} signals in {CBRS} using {ML}: Dataset and {API}-based collection testbed,'' \emph{IEEE Commun. Mag.}, vol.~61, no.~9, pp. 82--88, 2023.

\bibitem{deepmtl}
C.~Zhan, M.~Ghaderibaneh, P.~Sahu, and H.~Gupta, ``{DeepMTL}: Deep learning based multiple transmitter localization,'' in \emph{Proc. IEEE 22nd Int. Symp. on a World of Wireless, Mobile and Multimedia Networks (WoWMoM)}, 2021, pp. 41--50.

\bibitem{deeptxfinder}
A.~{Zubow}, S.~{Bayhan}, P.~{Gaw{\l}owicz}, and F.~{Dressler}, ``{DeepTxFinder}: Multiple transmitter localization by deep learning in crowdsourced spectrum sensing,'' in \emph{Proc. 29th Int. Conf. on Computer Communications and Networks (ICCCN)}, 2020, pp. 1--8.

\bibitem{ml_fingerprinting_survey}
N.~Singh, S.~Choe, and R.~Punmiya, ``Machine learning based indoor localization using {Wi-Fi RSSI} fingerprints: An overview,'' \emph{IEEE Access}, vol.~9, pp. 127\,150--127\,174, 2021.

\bibitem{experimental_dnnrss}
I.~Bizon, Z.~Li, A.~Nimr, M.~Chafii, and G.~P. Fettweis, ``Experimental performance of blind position estimation using deep learning,'' in \emph{Proc. IEEE Global Commun. Conf. (GLOBECOM)}, 2022, pp. 4553--4557.

\bibitem{picm}
I.~Bizon, A.~Nimr, G.~Fettweis, and M.~Chafii, ``Indoor positioning using correlation based signal analysis and convolutional neural networks,'' in \emph{Proc. 19th Int. Symp. on Wirel. Commun. Syst. (ISWCS)}, 2024, pp. 1--6.

\bibitem{MultiTaskLearning2022}
Y.~Zhang and Q.~Yang, ``A survey on multi-task learning,'' \emph{IEEE Trans. Knowl. Data Eng.}, vol.~34, no.~12, pp. 5586--5609, 2022.

\bibitem{Miloshevski2024}
L.~Milosheski, M.~Mohorčič, and C.~Fortuna, ``Spectrum sensing with deep clustering: Label-free radio access technology recognition,'' \emph{IEEE Open J. Commun. Soc.}, vol.~5, pp. 4746--4763, 2024.

\bibitem{Shatov_ISWCS2024}
V.~Shatov, N.~Shanin, B.~Perner, Y.~Su, N.~Franchi, and M.~Lübke, ``{LSTM-DNN}-based joint wireless signal waveform classification and blind transmitter localization,'' in \emph{Proc. Int. Symp. on Wirel. Commun. Systems (ISWCS)}, 2024.

\bibitem{DualTaskLearning2025}
V.~Shatov, N.~Shanin, T.~Veihelmann, B.~Perner, N.~Franchi, and M.~Lübke, ``Dual-task supervised learning for distributed spectrum monitoring applications,'' in \emph{Proc. Joint European Conf. on Netw. and Commun. \& 6G Summit (EuCNC/6G Summit)}, 2025.

\bibitem{weighted_multitask}
T.~Gong \emph{et~al.}, ``A comparison of loss weighting strategies for multi task learning in deep neural networks,'' \emph{IEEE Access}, vol.~7, pp. 141\,627--141\,632, 2019.

\bibitem{studer_cc}
C.~Studer, S.~Medjkouh, E.~G{\"{o}}n{\"{u}}ltas, T.~Goldstein, and O.~Tirkkonen, ``Channel charting: Locating users within the radio environment using channel state information,'' \emph{IEEE Access}, 2018.

\bibitem{stephan2024angle}
P.~Stephan, F.~Euchner, and S.~ten Brink, ``Angle-delay profile-based and timestamp-aided dissimilarity metrics for channel charting,'' \emph{IEEE Trans. Commun.}, vol.~72, no.~9, pp. 5611--5625, 2024.

\bibitem{euchner2024uncertainty}
F.~Euchner, P.~Stephan, and S.~ten Brink, ``Uncertainty-aware dimensionality reduction for channel charting with geodesic loss,'' in \emph{Asilomar Conf. on Signals, Systems, and Computers}, 2024.

\bibitem{dichasus2021}
F.~Euchner, M.~Gauger, S.~D\"orner, and S.~ten Brink, ``{A Distributed Massive MIMO Channel Sounder for "Big CSI Data"-driven Machine Learning},'' in \emph{Proc. 25th Int. ITG Workshop on Smart Antennas (WSA)}, 2021.

\bibitem{MultiModalSensing_Cheng_2024}
X.~Cheng \emph{et~al.}, ``Intelligent multi-modal sensing-communication integration: Synesthesia of machines,'' \emph{IEEE Commun. Surv. Tutor.}, vol.~26, no.~1, pp. 258--301, 2024.

\bibitem{MMMTL_2021}
R.~Hu and A.~Singh, ``{UniT}: Multimodal multitask learning with a unified transformer,'' in \emph{IEEE/CVF Int. Conf. on Computer Vision (ICCV)}, 2021, pp. 1419--1429.

\bibitem{6Gvision_2020}
W.~Saad, M.~Bennis, and M.~Chen, ``A vision of {6G} wireless systems: Applications, trends, technologies, and open research problems,'' \emph{IEEE Netw.}, vol.~34, no.~3, pp. 134--142, 2020.

\bibitem{Schuster2023}
A.~Schuster, T.~Reissland, and R.~Weigel, ``Hybrid sensing and communications receiver front-end for mobile communications applications,'' in \emph{Proc. IEEE MTT-S Latin America Microw. Conf. (LAMC)}, 2023.

\bibitem{antennas2023}
G.-B. Wu \emph{et~al.}, ``A universal metasurface antenna to manipulate all fundamental characteristics of electromagnetic waves,'' \emph{Nat. Commun.}, no.~14, 2023.

\bibitem{60GHz_metasurface}
A.~Jabbar \emph{et~al.}, ``60 {GHz} programmable dynamic metasurface antenna ({DMA}) for next-generation communication, sensing, and imaging applications: From concept to prototype,'' \emph{IEEE Open J. Antennas Propag.}, vol.~5, no.~3, pp. 705--726, 2024.

\bibitem{Hol_MIMO_6G}
C.~Huang \emph{et~al.}, ``Holographic {MIMO} surfaces for {6G} wireless networks: Opportunities, challenges, and trends,'' \emph{IEEE Wirel. Commun.}, vol.~27, no.~5, pp. 118--125, 2020.

\bibitem{Open6GHubTestbed2024}
B.~Nuss \emph{et~al.}, ``Flexible and scalable broadband massive {MIMO} testbed for joint communication and sensing applications,'' in \emph{Proc. European Wireless}, 2024, pp. 1--6.

\bibitem{TUI_Testbed2024}
M.~Engelhardt \emph{et~al.}, ``Accelerating innovation in {6G} research: Real-time capable {SDR} system architecture for rapid prototyping,'' \emph{IEEE Access}, vol.~12, pp. 118\,718--118\,732, 2024.

\bibitem{MultipleAccess6G_2024}
B.~Clerckx \emph{et~al.}, ``Multiple access techniques for intelligent and multifunctional {6G}: Tutorial, survey, and outlook,'' \emph{Proc. IEEE}, pp. 1--48, 2024.

\bibitem{RL_MAC_2012}
Y.~Chu, P.~D. Mitchell, and D.~Grace, ``Aloha and {Q}-learning based medium access control for wireless sensor networks,'' in \emph{Int. Symp. Wirel. Commun. Syst. (ISWCS)}, 2012, pp. 511--515.

\bibitem{FL_device_scheduling_2023}
A.~Bereyhi, A.~Vagollari, S.~Asaad, R.~R. Müller, W.~Gerstacker, and H.~V. Poor, ``Device scheduling in over-the-air federated learning via matching pursuit,'' \emph{IEEE Trans. Signal Process.}, vol.~71, pp. 2188--2203, 2023.

\bibitem{FLRL_2020}
H.~Wang, Z.~Kaplan, D.~Niu, and B.~Li, ``Optimizing federated learning on non-iid data with reinforcement learning,'' in \emph{IEEE Conf. on Comput. Commun. (INFOCOM)}, 2020, pp. 1698--1707.

\end{thebibliography}
\bibliographystyle{IEEEtran}

\end{document}